%% file: paper.tex
\pdfoutput=1
\documentclass[cernpreprint,texlive=2011,txfonts,texmf,UKenglish]{atlasdoc}

\usepackage[subfigure]{atlaspackage}
\usepackage{atlasphysics}

\usepackage{epstopdf}
\usepackage{afterpage}
\usepackage{mathrsfs}
\usepackage{xspace} 
\usepackage{units}
\usepackage{authblk}
\usepackage{euscript}
\usepackage{upgreek}
\usepackage{rotating}
\usepackage{graphicx}

\hypersetup{colorlinks,breaklinks}
\hypersetup{linkcolor=blue,citecolor=blue,filecolor=black,urlcolor=blue}
\hypersetup{pdftitle={ATLAS draft cover},pdfauthor={The ATLAS Collaboration}}

\input{jetetmisssymbols.tex}

\input{boostedsymbols.tex}

\usepackage{pifont}

\newcommand{\abg}{afterglow$_{\rm BIB}$}
\newcommand{\agpp}{afterglow$_{\rm pp}$}

\newcommand{\expfor}[2]{$#1\!\times\! 10^{#2}$}

\newcommand{\beq}{\begin{equation}} 
\newcommand{\eeq}{\end{equation}} 

\usepackage{url}
\PreprintIdNumber{CERN-EP-2016-029}

\AtlasTitle{Beam-induced and cosmic-ray backgrounds observed in the ATLAS detector during the LHC 2012 proton-proton running period}
\AtlasJournal{JINST}

\AtlasAbstract{
  This paper discusses various observations on beam-induced and cosmic-ray backgrounds in the ATLAS detector during the LHC 2012 proton-proton run.
  Building on published results based on 2011 data, the correlations between background and residual pressure of the beam vacuum are revisited. 
  Ghost charge evolution over 2012 and its role for backgrounds are evaluated. New methods to monitor ghost charge 
  with beam-gas rates are presented and observations of LHC abort gap population by ghost charge are discussed in detail.
  Fake jets from colliding bunches and from ghost charge are analysed with improved methods, showing that ghost charge in individual
  radio-frequency buckets of the LHC can be resolved.  Some results of two short periods of dedicated cosmic-ray background data-taking are shown;
  in particular cosmic-ray muon induced fake jet rates are compared to Monte Carlo simulations and to the fake jet rates from beam background.
  A thorough analysis of a particular LHC fill, where abnormally high background was observed, is presented. Correlations between
  backgrounds and beam intensity losses in special fills with very high $\beta^*$ are studied. 

\vspace{3mm}


}

\begin{document}

\maketitle
\tableofcontents
\clearpage

\input{introduction}

\input{lhcatlas}

\input{ncb}
\input{bcm}
\input{fillpattern}

\input{afterglow}

\input{beamgas}

\input{ghostmon}

\input{fakejets}
\input{ghost}

\input{special}

\input{run213816}

\input{highbeta}
\input{25ns}

\input{conclusions}

\appendix
\clearpage
\input{app_TS3fitsBCM}

\section*{Acknowledgments}

This paper required frequent consultation with LHC experts from several areas and we are especially grateful to  G.\,Bregliozzi,  
E.\,Shaposhnikova and J.\,Wenninger for useful discussions and invaluable advice.

\input{Acknowledgements}


\printbibliography

\clearpage

\include{atlas_authlist}

\end{document}

%% file: jetetmisssymbols.tex
\newcommand{\antikt}{anti-$k_{t}$}
\newcommand{\Antikt}{Anti-$k_{t}$}

%% file: introduction.tex
\section{Introduction}

In 2012 the Large Hadron Collider (LHC) increased its beam energy to 4\,\TeV{} and 
raised beam intensities with respect to 2011. Bunch intensities up to \expfor{1.6}{11} protons 
were routinely reached. 

The high-luminosity experiments at the LHC are designed to cope with intense background from $pp$ collision debris,
compared to which the usual levels of beam-induced backgrounds (BIB) are negligible. 
When beam intensities and energies increase, the risk for adverse beam conditions, that could 
compromise the performance of inner detectors, grows. In order to rapidly recognise and
mitigate such conditions, a thorough understanding of background sources and observables is 
necessary. The main purpose of this paper is to contribute to this knowledge.

The experience accumulated during the 2011 operation, in measuring and monitoring beam 
backgrounds \cite{backgroundpaper2011}, allowed optimisation of the beam structure 
and analysis procedures to reach better sensitivity for background observables.
In this paper, a summary of the main observations on BIB and cosmic-ray 
backgrounds (CRB), as well as ghost collision rates, in the ATLAS detector is presented. The topics 
cover a variety of background types in different experimental conditions encountered in 2012.

The operational conditions of the LHC, relevant for this analysis, are first explained, 
followed by a short discussion of the background detection and triggering methods.
Subsequent sections are devoted to detailed analyses, starting with re-establishing 
the correlation between vacuum quality and BIB in the ATLAS inner detector region, refining 
the analysis already performed on 2011 data. 

Improved methods to monitor ghost collisions, i.e. protons (ghost charge) in nominally 
empty radio-frequency (RF) buckets of the LHC colliding with nominal intensity 
bunches in the other beam (unpaired bunches), are introduced and used to separate beam backgrounds 
in unpaired bunches into collision, beam-gas and random noise components. Implications
for correcting luminosity measurements using unpaired bunch backgrounds are discussed.

The most significant non-collision background for physics searches comes from fake jets created by radiative energy
losses of BIB or CRB muons in the calorimeters. Although the rate of such fake jets is low
they still can form a non-negligible background in searches for rare physics processes.
The sources, rates and characteristics of fake jets are discussed in detail and it is shown 
that in normal physics conditions the dominant fake jet background comes from BIB, although 
the fraction of fake jets from CRB increases towards higher apparent $\pt$.

With the new, more sensitive, analysis methods it is possible to detect and quantify the BIB from 
ghost charge, despite the very low rate. Several unexpected observations are discussed, the most significant being the dominant role of 
the LHC momentum cleaning as a source of BIB. 
In particular, it is shown that BIB from ghost charge in the LHC abort gap can be detected by ATLAS. 
The effects of the LHC abort gap cleaning mechanism on these backgrounds
are evaluated and it is shown that the abort gap is repopulated within about one minute when the cleaning
is switched off.
Although the levels of BIB from ghost charge are tiny, they can be observed clearly also
as fake jets, which are significantly out of time. In searches for rare long-lived particles,
such special backgrounds need to be rigorously removed.

The 2012 operation included two fills with very special characteristics. The first was a normal
fill, but following a magnet quench close to ATLAS. This quench caused local outgassing and resulted 
in very high backgrounds at the start of the fill. A detailed analysis of these backgrounds provides 
insights into the conditioning process of the beam pipe surface.

The other fill of interest used special optics for forward-physics experiments. The fill had
very low beam intensity and luminosity, but large loss spikes due to repeated tightening of the betatronic beam cleaning. 
These conditions allow detail studies of correlations between losses at the LHC collimators and backgrounds 
seen in ATLAS. Although the optics was very different from high-luminosity operation, the methods developed and
results obtained motivate similar tests during LHC Run-2 with normal optics but special low-intensity beams.

At the very end of 2012, three fills were dedicated to studies of operation with 25\,ns bunch spacing,
which is the baseline condition for LHC Run-2.
Although only one of these fills was of sufficient length and intensity for backgrounds analysis, the observations 
will provide a useful point of comparison between the end of LHC Run-1 and the start-up after the long shutdown.

%% file: lhcatlas.tex
\section{The LHC and the ATLAS detector}

\newcommand{\AtlasCoordFootnote}{%
ATLAS uses a right-handed coordinate system with its origin at the nominal interaction point (IP)
in the centre of the detector and the $z$-axis along the beam pipe.
The $x$-axis points from the IP to the centre of the LHC ring,
and the $y$-axis points upwards.
Cylindrical coordinates $(r,\phi)$ are used in the transverse plane, 
$\phi$ being the azimuthal angle around the $z$-axis.
The pseudorapidity is defined in terms of the polar angle $\theta$ as $\eta = -\ln \tan(\theta/2)$.
Angular distance is measured in units of $\Delta R \equiv \sqrt{(\Delta\eta)^{2} + (\Delta\phi)^{2}}$.}

The LHC accelerator and the ATLAS detector are described in references\,\cite{lhcpaper} and \cite{atlaspaper}, respectively.
Only a concise summary is given here, focusing 
on aspects relevant for the 2012 background analysis.

\subsection{The LHC collider}
\label{sect:lhc}

The features of the LHC, relevant to background, have been described 
in reference\,\cite{backgroundpaper2011} and remained largely the same for the 2012 operation.

The RF of the LHC is 400.79\,MHz and the revolution time 88.9244\,$\upmu{\rm s}$, which means there 
are 35640 RF buckets that can accommodate particles. 
Nominally, only every tenth bucket can be filled and groups of ten buckets
are identified with Bunch Crossing IDentifiers (BCID), which take values in the range 1--3564.

In order to be able to safely eject the full-energy LHC beam, an abort gap of slightly more than 3\,$\upmu{\rm s}$ is left in the
bunch pattern to fully accommodate the rise-time of the beam extraction magnet. For the safety of the LHC it is
imperative that the amount of ghost charge in the abort gap, especially its early part, does not become too large.

\begin{figure}
  \centering
  {\includegraphics[width=0.9\textwidth]{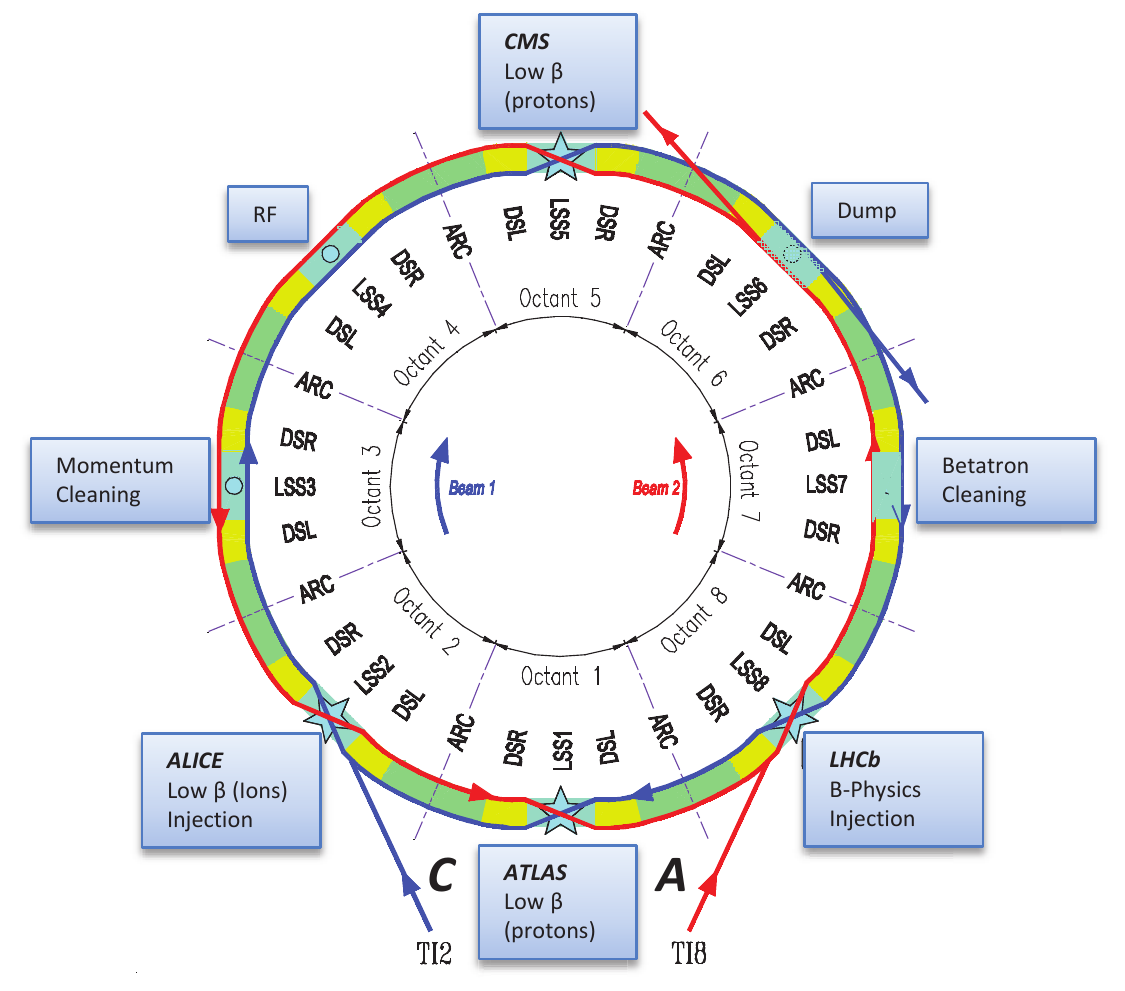}}            
  \caption{The general layout of the LHC\,\protect\cite{lhcdesignreport}. The dispersion suppressors (DSL and DSR) are 
sections between the straight section and the regular arc. In this paper they are considered to be
part of the arc, for simplicity. LSS denotes the Long Straight Section -- roughly 500\,m
long parts of the ring without net bending. All insertions (experiments, cleaning, dump, RF)
are located in the middle of these sections. Beams are injected through transfer lines TI2 and TI8.
The ATLAS convention of labelling sides by `A' and `C' is indicated.}
  \label{fig:layout}
\end{figure}

The LHC layout, shown in figure\,\ref{fig:layout}, comprises eight arcs, which are joined by Long Straight Sections (LSS) of slightly more than 250\,m
half-length. Each LSS houses an Interaction Region (IR) in its middle, the ATLAS detector being located in IR1.
The LHC beam cleaning equipment is situated in IR3 (momentum cleaning) and IR7 (betatron cleaning),
i.e. two octants away from ATLAS for beam-2 and beam-1, respectively. The role of the beam cleaning
is to intercept the primary and secondary beam halo, but some protons escape, forming a tertiary halo.\footnote{The definition of 
the halo hierarchy is related to table\,\ref{tab:apertures}. The primary collimators intercept the primary beam halo, but some protons scatter out and form the
secondary halo which is intercepted by the secondary collimators, which scatter out some tertiary halo that ends up on the tertiary collimators.}
In order to intercept this component and to provide local protection against accidental beam losses, tertiary 
collimators (TCT\footnote{The naming convention of LHC machine elements is described in reference\,\protect\cite{lhcdesignreport}.}) 
are placed about 150\,m from the experiments on the incoming beams. 
The aperture settings of collimators, with respect to the nominal normalised emittance of 3.5\,$\upmu{\rm m}$,
are listed in table\,\ref{tab:apertures}. It can be seen that the  momentum cleaning collimators in IR3 are much more open 
than those of the betatron cleaning in IR7. Since most of the cleaning takes place in IR7, it has much more efficient absorbers
than IR3 so that per intercepted proton there is more leakage of cleaning debris from the latter.

\begin{table}
\centering
\begin{tabular}{c|c|c|c}
$\beta^*$ & TCP in IR7 (in IR3) & TCS in IR7 (in IR3) & TCT in IR1,5 (in IR 2,8) \\ \hline
0.6\,m    & 4.3 (12) & 6.3 (15.6) & 9.0 (12) \\ \hline
1000\,m   & 2.0$^\dagger$ (5.9$^\dagger$) & 6.3 (15.6) & 17 (26) \\ \hline
\end{tabular}
\caption{Apertures of primary (TCP), secondary (TCS) and tertiary (TCT) collimators in units of nominal betatronic $\sigma$, corresponding to a normalised emittance 
of 3.5\,$\upmu{\rm m}$\,\protect\cite{collimatorPaper}.
Settings for normal high-luminosity optics ($\beta^*=0.6$\,m) and the special high-$\beta^*$ fill ($\beta^*=1000$\,m), discussed in section\,\protect\ref{sect:highbeta}, are given.
$^\dagger$For the 1000\,m optics, the given TCP settings varied and the values listed correspond to the tightest settings.}
\label{tab:apertures}
\end{table}

The inner triplet of quadrupoles, providing the final focus, operates at 1.9\,K and 
extends from $z=23\,$m to $z=\,54$\,m. It is equipped with a perforated beam-screen, operated at 20\,K, in order to
protect the superconducting coils from synchrotron radiation, electron cloud effects and resistive heating by the image currents of the passing beam. 
The perforation allows for residual gas to condense on the cold bore of the coils. This cryo-pumping effect is responsible
for the very low pressure reached in the cold sections of the LHC\,\cite{cryopumping}.

The residual pressure close to the experiment is monitored by several vacuum gauges of Penning and ionisation types.
The 2011 background analysis revealed that the BIB seen in ATLAS at small radius is correlated with 
the pressure measured at 22\,m. Further gauges are at 58\,m (on accelerator side of inner triplet), 150\,m (close to the TCT) and $\sim$250\,m (at exit of the
arc) from the IP. All of these were considered in the analysis but finally only the
22\,m and 58\,m readings were found to show a correlation with the observed backgrounds.
The gauge at 58\,m is located in a short warm section without Non-Evaporative Getter coating\,\cite{neg-benvenuti}. Electron-cloud
formation was discovered to be a problem in this region and already in 2011 solenoids
were placed around the beam-pipe in order to suppress electron multipacting.
The solenoids were found to be efficient and remained operational throughout 2012.

The inner triplet quadrupole absorber (TAS) is another machine element of importance for background formation. 
It is a 1.8\,m long copper block located at $z=19\,$m from the interaction point (IP) with a 17\,mm radius aperture for the 
beam. While the TAS provides a shielding effect against beam backgrounds, high-energy 
particles impinging on it can initiate showers that are sufficiently penetrating to partially leak through.

A few thousand beam loss monitors (BLM) are distributed all around the LHC ring in order to monitor beam losses 
and to initiate a protective beam dump in case of a severe anomaly. The time resolution of a BLM is limited by its electronics 
to about 40\,$\upmu{\rm s}$, so it cannot be used to determine loss rates of individual bunches. Since the BLMs are located
around very different machine elements, with different internal shielding, their response with respect to one lost proton is
not uniform. Thus, without detailed response simulations, the BLMs cannot be used to compare the losses in two different locations. 
They serve mainly to give information about the time development of losses on a given accelerator element. The loss-rates on 
the TCT would be most interesting for background studies.
Unfortunately, in this location, the BLMs are subject to intense debris from the collisions, so during physics 
operation they have no sensitivity to halo losses.

Another beam monitoring system is the Longitudinal Density Monitor (LDM), which is used to 
measure the population in each RF bucket by synchrotron light emission\,\cite{ldmreference}. The system
has sufficiently good time resolution and charge sensitivity to detect ghost charge in individual RF buckets 
with an intensity several orders of magnitude below the nominal bunch intensity of $\sim\! 10^{11}$\,protons/bunch. 
In 2012 the system was operational only in some LHC fills and elaborate calibration and background subtraction
had to be developed in order to extract the signal.

The beam intensity is measured by two devices of which only one, the fast beam current 
transformer (FBCT), provides intensity information bunch by bunch. Where appropriate, the
intensity values provided by the FBCT have been used to normalise backgrounds.

Normal data-taking happens in the STABLE BEAMS mode. This is preceded by phases called
FLAT TOP, when beams have reached full energy,  SQUEEZE, when the optics at the interaction points 
is changed to provide the low $\beta^*$ focusing for physics\footnote{The $\beta$-function determines the variation 
of the beam envelope around the ring and depends on the focusing properties of the magnetic lattice. Details can be found in reference\,\protect\cite{betaref}.} 
and ADJUST, during which the beams are brought into collision. 
Together FLAT-TOP and SQUEEZE last typically 20\,minutes. During this time the beams remain separated, which provides particularly clean 
conditions for background monitoring.

\subsection{The ATLAS detector}
\label{sect:atlas}

ATLAS is a general purpose detector at the LHC with almost
$4\pi$ coverage. It is optimised to study proton-proton collisions
at the highest possible energies, but has capabilities also for
heavy-ion and very forward physics. The ATLAS inner detector is
housed inside a solenoid which produces a 2\,T axial field. It is surrounded 
by calorimeters and a muon spectrometer based on a toroidal magnet configuration.
The calorimeters extend up to a pseudorapidity $|\eta|=4.9$, where 
$\eta=-\ln \tan(\theta/2)$, with $\theta$ being the polar angle with respect to 
the nominal LHC beam-line in the beam-2 direction.
Charged particle tracks are measured by the inner detector in the
range $|\eta|<2.5$.
In the right-handed ATLAS coordinate system, with its origin at the 
nominal IP, the azimuthal angle $\phi$ is measured with respect to 
the  $x$-axis, which points towards the centre of the LHC ring.  
As shown in figure\,\ref{fig:layout}, side A of ATLAS is defined as the side of the incoming, clockwise, 
LHC beam-1 while the side of the incoming beam-2 is labelled C. The coding of LHC machine 
elements uses letters L (left of IR, when viewed from ring centre) for 
ATLAS side A and, correspondingly, R for side C.
The $z$-axis in the ATLAS coordinate system points from C to A, i.e. along the beam-2 direction.
The most relevant ATLAS subdetectors for the analyses presented in
this paper are the Beam Conditions Monitor (BCM)\,\cite{bcm},  LUCID, the calorimeters, 
and the Pixel detector.

The BCM detector consists of 4 diamond modules ($8\times 8\,{\rm mm}^2$ active area) on 
each side of the IP at a $z$-distance of 1.84\,m from the IP
and a mean radius of $r=5.5$\,cm from the beam-line, corresponding to $|\eta|=4.2$.
The modules on each side are arranged in a cross, i.e.
two in the vertical plane and two in the horizontal plane. The individual modules will be referred to
as Ax-, Cy+, etc. where the first letter refers to the side according to ATLAS convention, the
second letter to the azimuth and the sign is according to the ATLAS coordinate 
system. 

The LUCID detector was introduced as a dedicated luminosity monitor. It consists of
16  Cherenkov tubes per side, each connected to its own photomultiplier (PMT), situated in the forward 
region at a distance $z=18.3$\,m from the IP, giving a pseudorapidity coverage of $5.6<|\eta|<6.0$.
In 2012 LUCID was  
operated without gas in the tubes, most of the time. In this configuration, only the Cherenkov light from the
quartz-window of a PMT was used for particle detection.  

The Pixel detector consists of three barrel layers at mean radii of 
50.5\,mm, 88.5\,mm and 122.5\,mm. All layers have a half-length of
400\,mm, giving an $|\eta|$-coverage out to 1.9 and 2.7 for the outermost 
and innermost layers, respectively. In each layer the modules are 
slightly tilted with respect to the tangent. The pixel size in the barrel modules 
is $r\phi \times z = 50\times 400\,\upmu{\rm  m}$. 
The full coverage, with three points per track, is extended to $|\eta|=2.5$ by three endcap 
pixel disks.

A high-granularity liquid-argon (LAr) electromagnetic calorimeter with lead as absorber material
covers the pseudorapidity range $|\eta|<1.5$ in the barrel region. The half-length of the LAr barrel is
3.2\,m and it extends from $r=1.5$\,m to $r=2$\,m.
The hadronic calorimetry in the region $|\eta|<1.7$ is provided by a scintillator-tile calorimeter (TileCal), extending from
$r=2.3$\,m to $r=4.3$\,m with a half-length of 8.4\,m.
Hadronic endcap calorimeters (HEC) based on LAr technology cover the range $1.5 < |\eta| < 3.2$. The absorber materials 
are iron and copper, respectively.
The calorimetry coverage is extended by the Forward Calorimeter up to $|\eta| = 4.9$.
All calorimeters provide nanosecond timing resolution.

%% file: ncb.tex
\section{Non-collision backgrounds}
\label{sec:ncb}

The non-collision backgrounds (NCB) are defined to include CRB and BIB.
The main sources of the latter are\,\cite{backgroundpaper2011, roderick-beam-background}:
\begin{itemize}
\item Inelastic beam-gas events in the LSS or the adjacent arc. Simulations indicate that contributions 
from up to about 500\,m away from the IP can be seen\,\cite{nikolai-elastic}.
\item Beam losses on limiting apertures. The contributions to the experiments come predominantly from losses on the TCTs, which in the normal optics are the smallest 
apertures in the vicinity of ATLAS.
\item Elastic beam-gas scattering around the ring. According to simulations\,\cite{nikolai-elastic,roderick-beam-background}, the scattered 
protons are intercepted by the beam cleaning insertions or the TCTs. In the latter case their effect adds to the halo losses on the TCTs.
\end{itemize} 
Beam-gas events, within about $\pm$50\,m from the IP, can spray secondary particles on the ATLAS inner detectors, but it is unlikely that they
reach large radii and give signals in, e.g., the barrel calorimeters. The fake jets due to BIB, which are a major subject of this paper, are 
caused by high-energy muons produced as a consequence of proton interactions with residual gas or machine elements far enough from the
IP to allow for the high-energy muons to reach the radii of the calorimeters. Radiative energy losses of these muons in 
calorimeter material, if large enough, are reconstructed as jets and can form a significant background to certain physics searches\,\cite{mono-jet-paper}. 
A characteristic feature of the high-energy muon component of BIB is that, due to the bending in the dipole magnets of the LHC, it is 
predominantly in the horizontal plane. 

The CRB is entirely due to high-energy muons. These can penetrate the 60\,m thick overburden and reach the experiment. Just like the
BIB muons, the CRB muons can create fake jets in the calorimeters by radiative energy losses and thereby 
introduce backgrounds to physics searches\,\cite{R-hadron-paper}.

A significant part of this paper is devoted to studies of ghost charge. The definition
adopted here is to call ghost charge all protons outside the RF buckets housing nominally filled 
bunches.\footnote{This is slightly different from reference\,\protect\cite{lumipaper2011}, where the charge in a nominally empty RF-bucket, but 
within $\pm$12.5\,ns of a colliding bunch, is referred to as `satellite bunch'. For this paper such a differentiation has no significance 
and is omitted for simplicity.}
There are two different mechanisms which lead to ghost bunch formation.\footnote{When the bunched structure of the ghost charge is significant, 
the term ghost bunch will be used.} 
\begin{itemize}
\item Ghost bunch formation in the injectors: Of particular interest are ghost bunches formed in
the Proton Synchrotron (PS) during the generation of the LHC bunch structure. In the PS a complicated
multiple splitting scheme\,\cite{blas97} is applied on the bunches injected from the Booster (PSB). The result of
this is to split a single PSB bunch into six bunches, separated by 50\,ns. If any of the protons injected from the
PSB do not fall into a PS bucket, the protons spilling over might be captured in an otherwise empty bucket and undergo 
the same splitting. In this case six ghost bunches with a 50\,ns spacing will be formed. 
Similar spill-over can occur in the injection from the PS into the Super Proton Synchrotron (SPS). In this case
ghost bunches with a 5\,ns spacing can be formed. If the production mechanism is of significance in a given context,
these bunches will be referred to as {\em injected ghost bunches}.
\item De-bunching in the LHC: In the course of a fill, a small fraction of the protons develop large enough
momentum deviations to leave their initial bucket\,\cite{Elena-Epac2004}. These escaped protons can drift in the LHC 
for tens of minutes and complete several turns more than their starting bucket before being intercepted by
the beam cleaning\,\cite{Elena-Epac2004, Elena}. Due to their relatively long lifetime these de-bunched protons can be assumed to be rather
uniformly distributed. Some of the de-bunched charge can be re-captured by the RF forming 
ghost bunches all round the ring. Thus the {\em de-bunched ghost charge} maintains an imprint of the bucket structure. 
\end{itemize}

%% file: bcm.tex
\section{Background monitoring methods}
\label{sect:bcmlucid}

In the ATLAS first level (L1) trigger, the LHC bunches are grouped according to their different characteristics 
into bunch groups (BG). Two of these groups are of particular importance for beam background analysis: 
{\em unpaired isolated} and {\em unpaired non-isolated}.
In the first of these groups the requirement is to have no bunch in the other beam within 150\,ns, while the 
second group includes those unpaired bunches which fail to fulfil this isolation requirement.
The timing of the central trigger is such that the collision time ($t=0$) of two filled bunches falls into the middle 
of the BCID.
When reference to an empty BCID is made in this paper, it means any BCID without a bunch in either beam.

During each LHC fill ATLAS data-taking is subdivided into Luminosity Blocks (LB), typically 60\,s in duration but some  
can be as short as 10\,s. While recorded events carry an exact time-stamp, trigger rates and luminosity data
are recorded only as averages over a LB.

\subsection{Background monitoring with the BCM}
\label{sect:bcm}

Throughout LHC Run-1, the BCM was the primary device in
ATLAS to monitor beam backgrounds. Gradually, the full capabilities and optimal usage
of the detector were explored and allowed for refinement of some of the results obtained
on 2011 data\,\cite{backgroundpaper2011} and augmenting these with new studies. In this section 
some aspects relevant to the BCM data, taken in 2012, are presented in detail.

\subsubsection*{BCM time resolution}

The time resolution of the BCM is measured to be of the order of 0.5\,ns. In the readout the 25\,ns duration 
of a BCID is subdivided into 64 bins, each 390.625\,ps wide. For each
recorded event, the entire vector of 64 bins is stored, allowing the exact arrival time
and duration of the signal to be determined. The bins are aligned such that the nominal collision time falls 
into bin 27, i.e. about 2\,ns before the centre of the readout interval. More details about the readout windows are
given in appendix\,\ref{app:TS3fitsBCM}.

\subsubsection*{BCM triggers and luminosity data}

In 2012 the BCM detector provided two different L1 triggers:
\begin{itemize}
\item L1\_BCM\_AC\_CA is a background-like `coincidence', i.e. requires an early 
hit in upstream\footnote{Upstream and downstream are defined with respect to the beam direction.} 
detectors associated with an in-time hit in downstream  detectors. The
window widths are the same as in 2011: the early window at $-6.25\,\pm\,2.73$\,ns
and the in-time window at $+6.25\,\pm\,2.73$\,ns, where the IP-passage of the bunch
is at 0.

\item L1\_BCM\_Wide is a trigger designed to select collision events by requiring a coincidence
of hits on both sides. Until the third technical stop (TS3) of the LHC in mid-September, the
window setting of 2011 was used, i.e. the window was open from 0.39\,ns until 8.19\,ns
following the collision. Since this alignment did not seem optimal for a nominal 
signal arrival at 6.1\,ns, TS3 was used to realign it with the in-time window 
of the BCM\_AC\_CA trigger. 
A detailed discussion of the consequences of this realignment and reduction of window width is given in
appendix\,\ref{app:TS3fitsBCM}.
\end{itemize}

Another significant modification implemented during TS3 was to combine all four BCM modules per side of ATLAS
into one read-out driver (ROD), where previously two independent RODs had each served a pair of modules.
Consequently the L1\_BCM\_Wide rates recorded prior to TS3 have to be doubled to be comparable with post-TS3 trigger
rates, as shown in appendix\,\ref{app:TS3fitsBCM} .

The combination of all modules into a single ROD also affected the BCM\_AC\_CA rates,
but the effect is less obvious. The correction factor for pre-TS3 rates, 
derived in appendix\,\ref{app:TS3fitsBCM}, is 1.2. 

The rates per BCID 
were recorded in a special monitoring database averaged over 300\,s. For some of the per-BCID rate studies presented 
in this paper this integration time proved too long and recourse to the luminosity
data from the BCM had to be made.
These data, dedicated for luminosity measurement with the BCM, are available as LB-averages, 
i.e. with a typical time resolution of 60\,s, for each BCID independently.
As for the BCM triggers, the luminosity data are based on a hit in any of the four modules
on one side. The time-window to accept events for the luminosity algorithms 
is 12.5\,ns, starting at the nominal collision time.
For the purpose of this paper only the single-side rates are relevant and will
be denoted as BCM-TORx.\footnote{This notation is used to be consistent with the terminology
used for luminosity measurements, where the ``TOR'' denotes a logical OR of all four modules on one side.
The ``x'' stands for either ``A'' or ``C''.} 
By comparing BCM-TORx rates  
in colliding bunches, the difference in efficiency, including acceptance, between sides A and C was found to be  $<$1\%. 
Since those data are recorded independently for each side, it is not possible to reconstruct 
the background-like timing pattern. Consequently these data are most useful for unpaired bunches in conditions 
where the BIB is high compared to other signals.
Such cases will be encountered in sections\,\ref{sect:run213816} and \ref{sect:highbeta}. 
Recourse to the BCM luminosity data will also be made in section\,\ref{sect:ghostlumi} when describing a 
new method to disentangle ghost collisions from noise and BIB. 

\subsubsection*{BCM data quality}

A noise of unidentified origin appeared in A-side BCM modules on 27 October evening. The noise lasted 
until the morning of 26 November and constituted a 
significant increase in the level of random hits and will be clearly visible in many plots 
in this paper. The noise period has been excluded from analyses where it would have
influenced the result. In trend plots over the year the period is included, but highlighted and
in most cases should be ignored.

A few LHC fills, predominantly early in the year, were affected by
various types of data quality problems, mostly loss of trigger synchronisation or
beam intensity information. These fills have been removed from the analyses and
the trend plots. 

\subsection{Using LUCID luminosity data in background analysis}
\label{sect:lucid}

The LUCID data are recorded by the luminosity data-acquisition software, independently of
the ATLAS trigger and are available per BCID and LB, typically with very good statistics.

The LUCID detector is not as fast as the BCM and suffers more from long-lived collision debris,
which will be discussed in section\,\ref{sect:afterglow}.
Unlike in 2011, the 2012 operation had two specific cases where LUCID proved very 
useful to detect backgrounds. These were a normal fill with abnormally high background  and
the high-$\beta^*$ fill with very low luminosity and sparse bunch pattern, discussed in 
sections\,\ref{sect:run213816} and\,\ref{sect:highbeta}, respectively.

\begin{figure}[t]
\centering 
\includegraphics[width=0.7\textwidth]{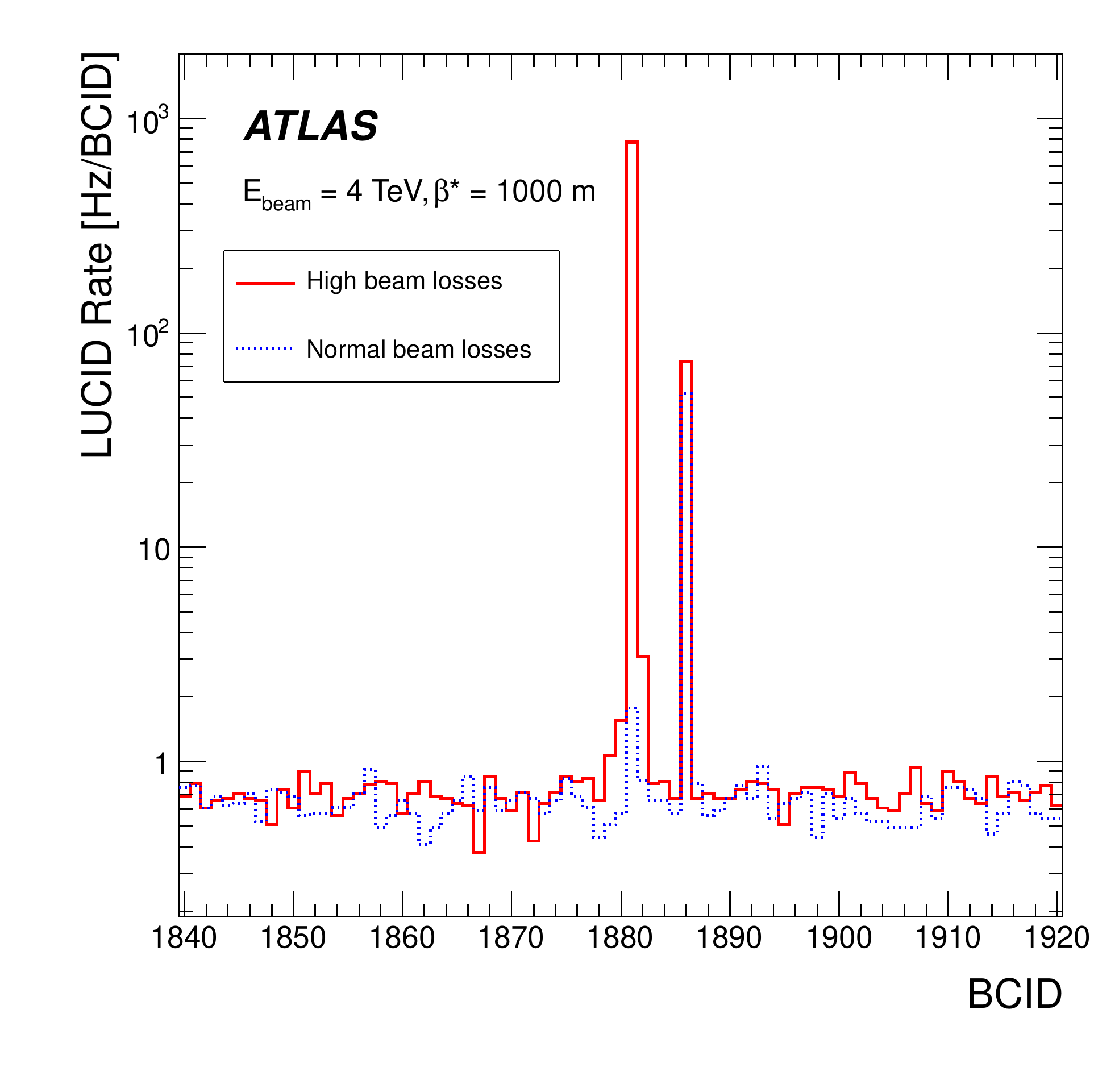}
\caption{Background and luminosity signal seen by LUCID in the high-$\beta^*$ fill (discussed in section\,\protect\ref{sect:highbeta}). 
Both data-sets average over three ATLAS luminosity blocks, i.e. about 180\,s.}
\label{fig:lucidearly}
\end{figure}

The use of LUCID in those special cases is made possible by its large distance from the IP, which allows the
separation of the background hits from the luminosity signal.
The usable signal comes from the background associated with the incoming bunch, observed in the upstream LUCID. 
The incoming bunch passes the upstream detector about 60\,ns before the actual collision which means that
the background from that bunch appears five BCIDs before the luminosity signal, twice the time-of-flight between the IP and 
LUCID. This is illustrated in figure\,\ref{fig:lucidearly}, using data from the high-$\beta^*$ fill
with very low luminosity and only two colliding bunches.
The normal luminosity signal is seen in BCID 1886 and is of comparable size during high beam losses and in normal conditions. The
background signal appears five BCID earlier and peaks in BCID 1881. In normal conditions it is much smaller
than the luminosity peak, but during high losses it can become very prominent. 
Unlike the BCM-TORx signal, this early LUCID signal has practically no luminosity contamination
even for paired bunches and lends itself very well to monitoring of the beam background of bunches 
with a long empty gap preceding them.

\subsection{Using calorimeter jets in background analysis}
\label{sec:calojets}

In the analysis of 2011 backgrounds jets proved to be a useful tool to study characteristics of NCB\,\cite{backgroundpaper2011}. 
A jet trigger L1\_J10 was defined in the 2012 data-taking for 
unpaired bunches in order to select BIB events with fake jets. 
The L1\_J10 trigger fires on a transverse energy deposition above 10\,\GeV, calibrated at approximately the 
electromagnetic energy scale\,\cite{jes}, in an $\eta$--$\phi$ region 
with a width of about $\Delta\eta\times\Delta\phi=0.8\times0.8$ anywhere within $|\eta| < 3.0$ and, with reduced efficiency, up to $|\eta| = 3.2$.
A similar trigger with a 30\,\GeV{} transverse energy threshold, L1\_J30, was defined for recording CRB data.

In order to suppress instrumental backgrounds that are not due to BIB or CRB,
abrupt noise spikes are masked during the data reconstruction and efficiently
removed by the standard data quality requirements~\cite{Aad:2014una}. 

For an offline analysis of the recorded data,
the anti-$k_t$ jet algorithm~\cite{Cacciari:2008gp}
with a radius parameter $R=0.4$ is used to reconstruct jets from the energy
deposits in the calorimeters.
The inputs to this algorithm are topologically connected clusters of calorimeter cells\,\cite{jes}, seeded by cells with 
energy significantly above the measured noise. These topological clusters are calibrated at the electromagnetic energy scale, 
which measures the energy deposited by electromagnetic showers in the calorimeter.
The measured jet transverse momentum
is corrected for detector effects,  including the non-compensating character of the calorimeter, by weighting
energy deposits arising from electromagnetic and hadronic showers differently.   
In addition, jets are corrected for contributions from pileup,  
as described in reference~\cite{jes}. 
The minimum jet transverse momentum considered is 10\,\GeV.

The jet time is defined as the weighted average of the time of the calorimeter cell energy deposits in the jet, weighted by the square of the cell energies.
The calorimeter time is defined such that it is zero for the expected arrival of collision secondaries at the given location,
with respect to the event time recorded by the trigger.

\subsection{Pixel background tagger}
\label{sec:tagger}

In the course of the 2011 background analysis\,\cite{backgroundpaper2011}, an algorithm was developed to tag beam background events 
based on the presence of elongated clusters in the Pixel barrel layers. While collision products, emerging from events at the IP, create short
clusters at central pseudorapidities, the clusters due to BIB, with trajectories almost parallel to the beam pipe, create long clusters
at all $\eta$ in the Pixel barrel. At high $|\eta|$ the clusters from $pp$ collision products in the barrel modules also become long, so the tagging method has its best
discrimination power below $|\eta|\sim 1.5$. 

%% file: fillpattern.tex
\section{Data-taking Conditions}
\label{sect:datataking}

In 2012 the LHC collided protons at $\sqrt{s}=8\,\TeV$, i.e. with $4\,\TeV$ energy per beam.
Except for some special fills, the bunch spacing was 50\,ns which allowed for
slightly fewer than 1400 bunches per beam. The typical bunch intensity at the start of a fill varied 
between \expfor{1.2\!-\!1.7}{11}\,protons. Most of 2012 physics operation was at $\beta^*=0.6$\,m.
In order to avoid parasitic 
collisions a crossing half-angle of 145\,$\upmu{\rm rad}$ between the two proton beams was used.
The normalised emittance was typically around 2.5\,$\upmu{\rm m}$ which is well below the nominal value of 3.5\,$\upmu{\rm m}$.

The colliding bunches were grouped into trains with $2\!-\!4\times 36$ bunches, with a separation 
of 900\,ns between the trains to allow for the injection kicker rise time. Within these long trains the 36 bunch sub-trains 
were separated by nine empty BCIDs. 

Since backgrounds depend on beam conditions, they are expected to be different 
for various beam structures. 
The beam pattern of 2011, with 1331 colliding bunches, was used at the beginning of 2012 data-taking until
instabilities of the unpaired bunches\,\cite{evian2012} required finding a different filling scheme. The intermediate solution was 
a 1377 colliding bunch pattern with only three unpaired bunches per beam, which were all non-isolated according 
to the ATLAS standard definition. In order to retain one bunch in the {\em unpaired isolated} group, which is used primarily for 
background monitoring, it was decided to relax the isolation requirement from 150\,ns to 100\,ns. Continuing instabilities led the LHC 
to introduce temporarily a fill pattern with 1380 colliding bunches, leaving none unpaired. During this
period, lasting from 24 May until 5 June, there was no background monitoring capability.
After optimisation of the LHC machine parameters, the unpaired bunches could be reintroduced and
soon afterwards (from LHC fill 2734 onwards) an optimal fill pattern was implemented. 
This pattern with 1368 colliding bunches has a mini-train of six unpaired bunches per beam 
in the ideal location, immediately after the abort gap in odd BCIDs 1--11 and 13--23. 
In this pattern the first colliding train started in BCID 66 and the last colliding bunch before the abort gap was in BCID 3393.
Initially the beam-1 mini-train was first, but at the end of the year (24 November, from LHC fill 3319 onwards) the 
trains were swapped in order to disentangle possible systematic beam-1/beam-2 differences and effects caused by the 
relative order of the unpaired trains.

\begin{table}
\begin{center}
\begin{tabular}{l|c|c|c} 
Period & Colliding & Unpaired (non-)isolated per beam &  Dates \\ \hline
1   & 1331 &   (40)9      & 18.4. -- 19.5. \\ 
2   & 1377 &   (2)1       & 19.5. -- 24.5. \& 5.6. -- 15.6.\\ 
--- & 1380 &   (0)0       & 24.5. --  5.6 \\
3   & 1368 &   (3)3       & 15.6. -- 17.9.\\ 
4   & 1368 &   (3)3       &  29.9. -- 6.12.\\ \hline
\end{tabular}
\caption{Periods with different numbers of colliding and unpaired bunches.
Periods 3 and 4 are separated by TS3 during which changes to the BCM trigger logic were implemented. Since the pattern with 1380 colliding bunches had no unpaired bunches,
background monitoring in that period was not possible.}
\label{beam-periods}						     
\end{center}
\end{table}

All periods described above are listed in table\,\ref{beam-periods}.
The period with 1368 colliding bunches is divided into pre-TS3 and post-TS3 periods because the BCM trigger rates
are not directly comparable, as discussed in appendix\,\ref{app:TS3fitsBCM}.
In addition, there were several fills with  
special bunch structure early in 2012 and a few such fills appeared also later in the year. Many of these were dedicated fills,
e.g. for van der Meer scans, high-luminosity or 25\,ns tests. Since these fills do not share common
characteristics, they are not included in the trend plots presented in this paper.

At the high beam intensities reached in 2012 and with 50\,ns bunch spacing, outgassing from
the beam pipe becomes a significant issue for the vacuum quality. Special scrubbing fills at $450\,\GeV$
beam energy but high intensity were used in early 2012 to condition the beam pipe surfaces. 
Despite these special fills some vacuum conditioning most likely continued throughout the first months
of physics operation.

\begin{table}
\begin{center}
\begin{tabular}{l|c}
Characteristics & Dates \\ \hline
\multicolumn{2}{l}{BCM} \\ \hline
Before chromaticity changes & 15.6. -- 3.8. \\
From chromaticity changes until BCM Noise & 10.8. -- 27.10. \\
After BCM Noise & 26.11. -- 6.12. \\ \hline
\multicolumn{2}{l}{Jets} \\ \hline
Before chromaticity changes & 15.6. -- 3.8. \\
From chromaticity changes until unpaired swap & 10.8. -- 24.11. \\
After unpaired swap & 25.11. -- 6.12. \\ \hline
\end{tabular}
\caption{Break-points, which had significant influence on rates or
data quality for background monitoring based on BCM (top) and jets (bottom), during 2012 data taking 
with 1368 colliding bunches. Within a given period data from different fills should be comparable.} 
\label{breakpoints}						     
\end{center}
\end{table}

As will be seen, an event of significance for some analyses in this paper was when LHC changed 
chromaticity settings between 3--9 August. During this period, extending over several days, the LHC was optimising
performance with different chromaticities and a switch of octupole polarities on 7 August.

All the break-points marking the important changes during the data-taking with 1368 colliding bunches, leading to significant changes in 
background rates or influencing the data quality,  are summarised in Table\,\ref{breakpoints}. 
Periods 1 and 2 of Table\,\ref{beam-periods} are not mentioned there
since those early periods have very different bunch patterns and have to be treated
separately. No particular events, aside from the pattern changes, were identified during those periods.

%% file: afterglow.tex
\section{Afterglow}
\label{sect:afterglow}

The term afterglow was introduced in the context of the ATLAS luminosity analysis\,\cite{lumipaper2010} 
to describe signals caused by delayed tails of particle cascades following a $pp$ collision. 
The afterglow decreases rapidly over the first few BCIDs following a collision, but a long tail extends up to 
about 10\,$\upmu{\rm s}$, as a result of which significant afterglow buildup is observed in colliding trains, with 50\,ns bunch spacing.
In the region of the long tail, more than $\sim$100\,ns after the last paired bunch crossing, the afterglow hits 
appear without any time-structure and are thus indistinguishable from instrumental, or other, noise. 

In the rapidly dropping part immediately after the collision the distinction between 
prompt signal and afterglow is somewhat ambiguous. A natural definition for the afterglow from $pp$ collisions 
is to consider as prompt all hits with a delay less than the BCID half-width of 12.5\,ns, while the 
rest is being counted as afterglow.

\begin{figure}
\centering 
\includegraphics[width=0.7\textwidth]{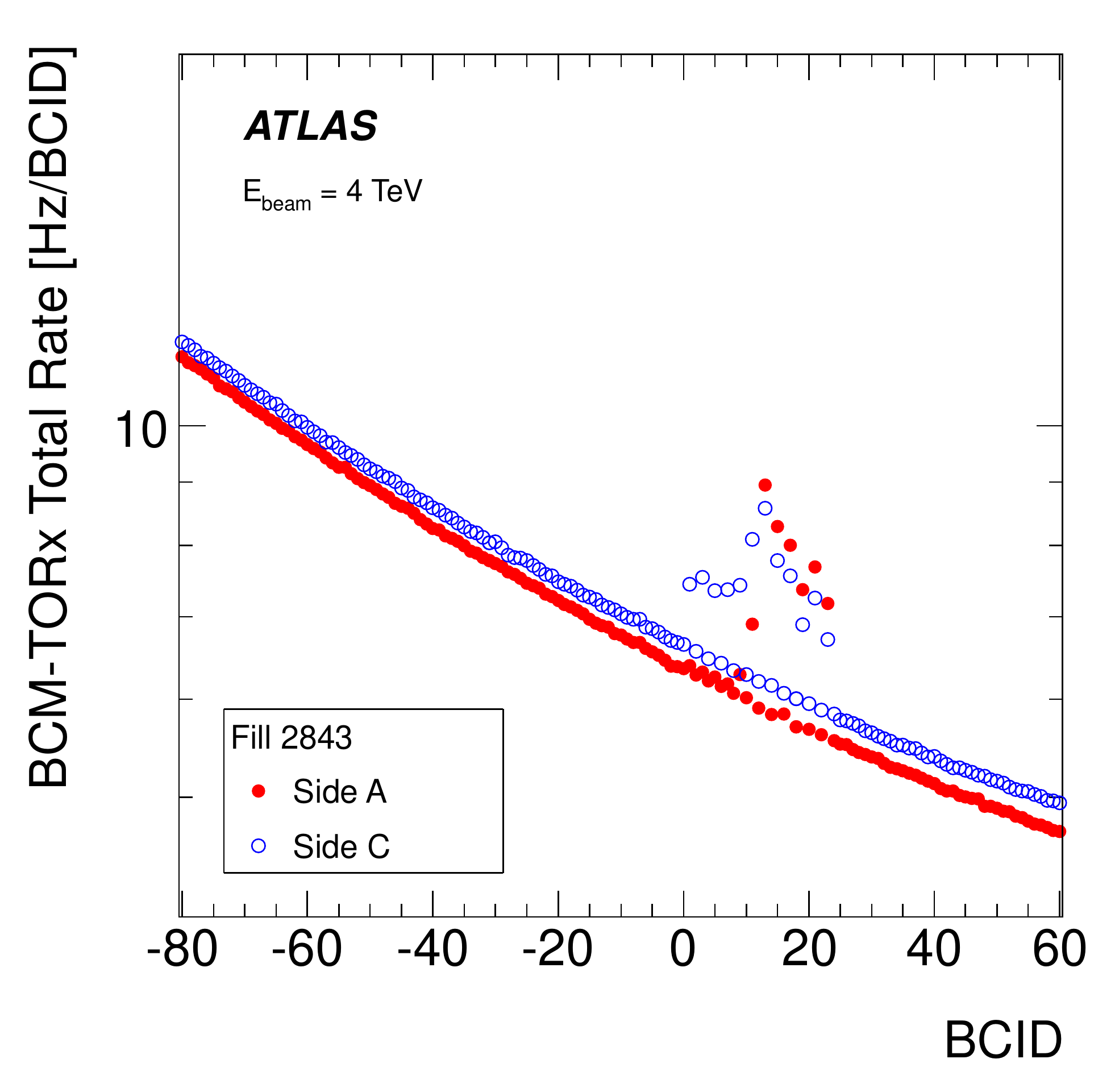}
\caption{
Afterglow distribution, as seen by the BCM-TORx algorithm, at the end of the abort gap and in the region of the unpaired bunches.
The signals associated with the
unpaired bunches are seen in odd BCIDs from 1 to 23. The asymmetry of the rate distribution in them will be discussed in 
section\,\protect\ref{sect:ghostlumi}. For the purpose of
plotting the BCID numbers in the abort gap have been re-mapped to negative values by BCID\,=\,BCID$_{\rm true}$\,-\,3564.}
\label{fig:ghostAGCorr}
\end{figure}

Figure~\ref{fig:ghostAGCorr} shows the afterglow tail created by the colliding trains before the abort gap in a normal physics 
fill and extending all the way to the unpaired bunches in odd BCIDs 1--23. 
Since the rates are much lower than one count per bunch crossing\footnote{With the revolution time of 89.9244\,$\upmu{\rm s}$, exactly 
one event per bunch crossing would result in a rate of 11245\,Hz/BCID}, the afterglow contribution to the unpaired bunches can be
removed by subtracting the rate in the preceding BCID from that in the BCID occupied by the unpaired bunch.

In analogy to the afterglow following $pp$ collisions, there should be delayed debris from a BIB 
event, which will be referred to as \abg{}. 
The component of the \abg{} which arrives within the same BCID as the unpaired bunch is particularly significant for
some observations to be discussed in this paper.
This part has a very non-uniform, rapidly dropping, time distribution within the BCID and cannot be easily subtracted.
The \abg{} level is a small fraction of the primary beam-gas rate, i.e. the rate 
seen as in-time hits in the downstream modules. However, since the \abg{} is correlated with the beam-gas events, 
there will be a bias towards primary beam-gas and \abg{} signal to form a collision-like
coincidence, i.e. provide real and apparent in-time hits on both sides of the IP. In order for this to happen the
afterglow has to arrive at the upstream detectors with a delay corresponding to the time-of-flight between the two BCM detector
arms, i.e. about $12.5\,$ns $\pm \Delta t$. Here, $\Delta t$ is the tolerance allowed by the coincidence trigger window. 
If the downstream signal is exactly in time, $\Delta t = 2.7\,{\rm ns}$ for L1\_BCM\_Wide after TS3 and
6.3\,ns for observing simultaneous BCM-TORx signals on both sides.

\begin{figure}
\centering
\mbox{
\subfigure[]{
\includegraphics[width=0.49\textwidth]{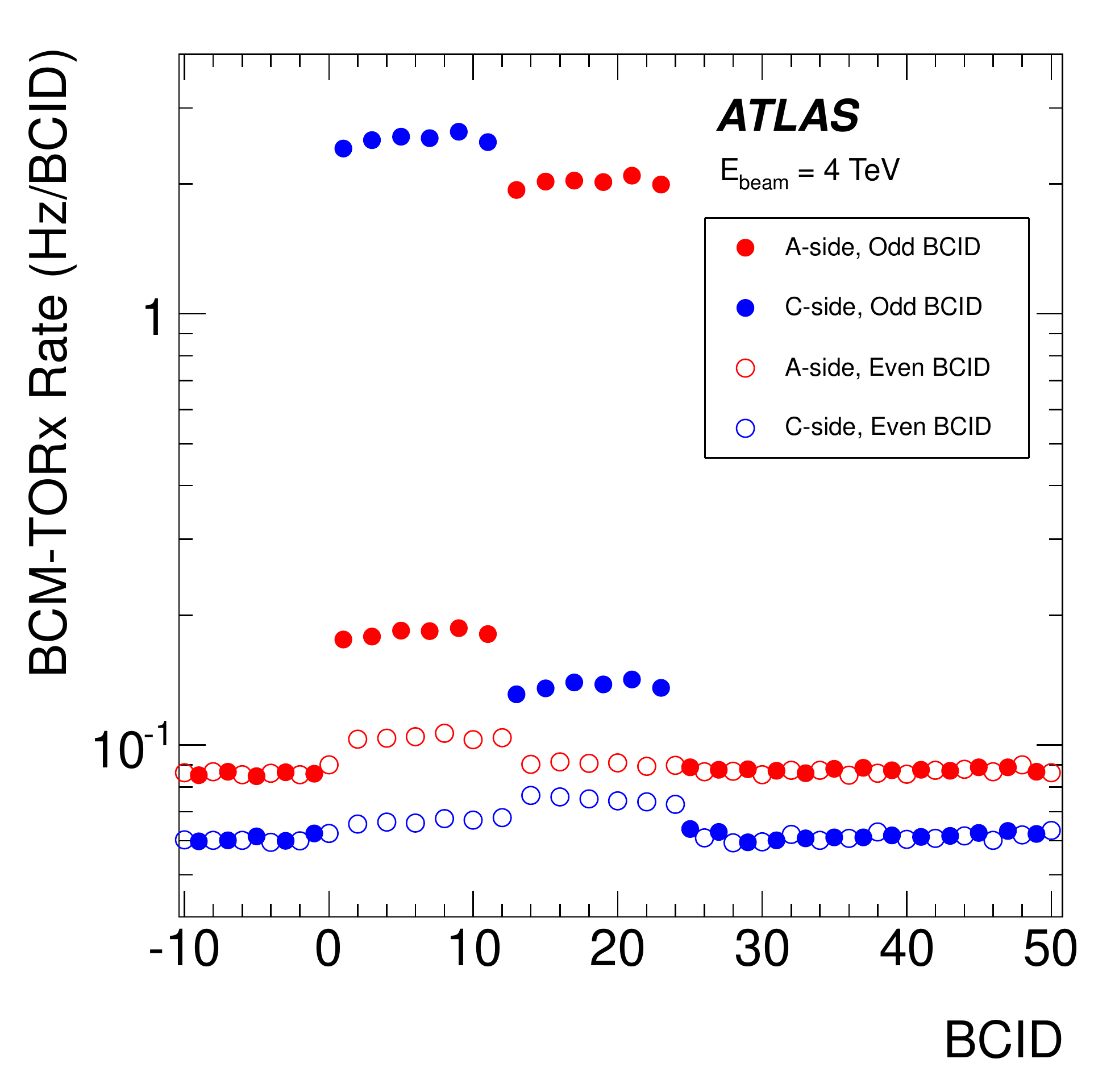}
\label{ghostsep}
}
\subfigure[]{
\includegraphics[width=0.49\textwidth]{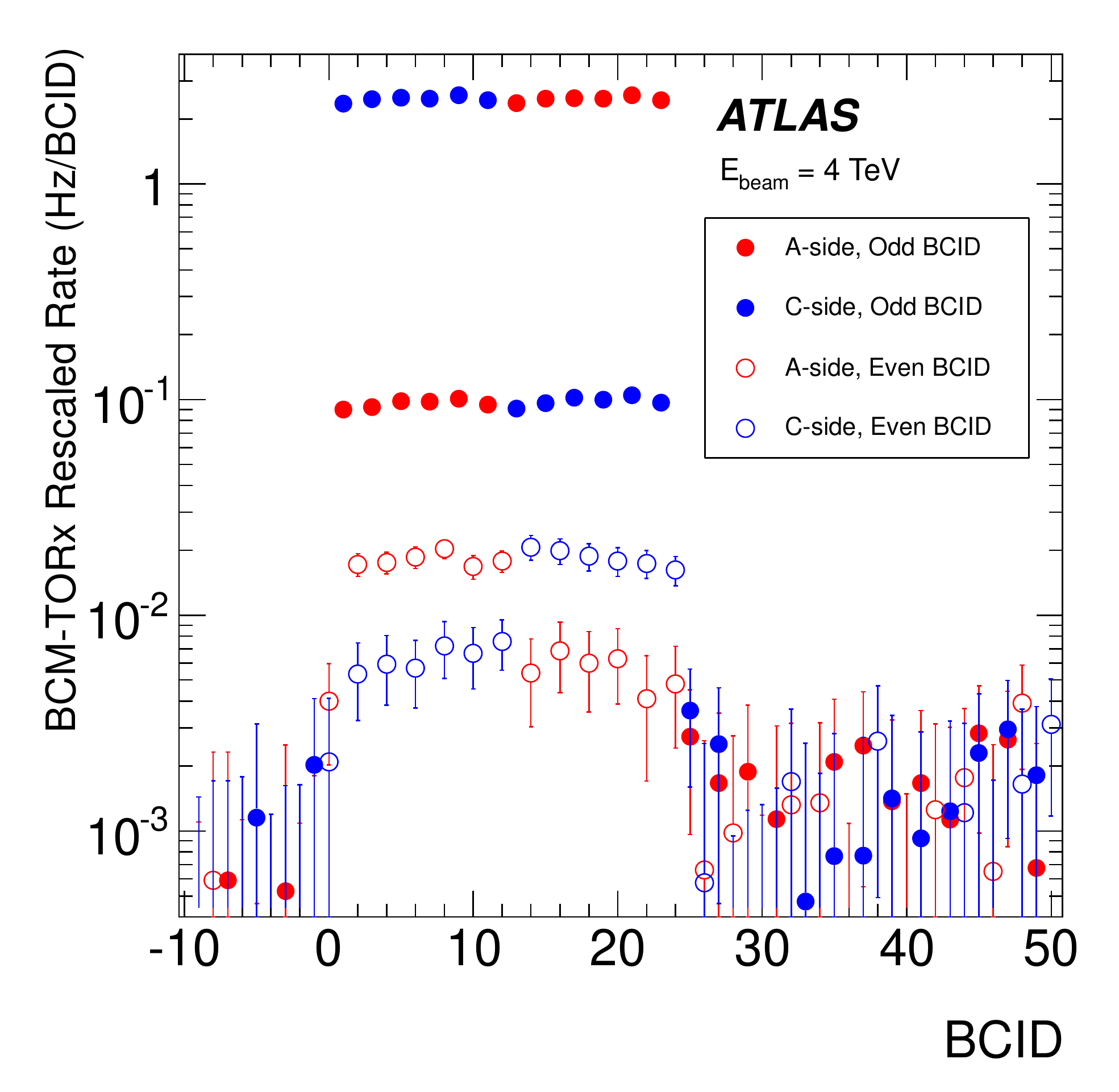}
\label{ghostsep-scaled}
}
}
\caption{Single sided BCM-TORx rates of unpaired and empty bunches during periods with separated full energy beams before (a) and 
after (b) noise subtraction and rescaling to match the primary beam-background rates of both beams, as explained in the text.
For clarity, signals in odd and even BCIDs are shown with different symbols. The unpaired bunches are all in odd BCIDs.}
\label{ghostsep-all}
\end{figure}

Since the \abg{} signal is very small, the overwhelming afterglow from the collisions prevents extracting it from stable-beam data.
However, around 20 minutes of data at the start of each fill is obtained during the FLAT-TOP and SQUEEZE phases, in which the full-energy
beams are separated and so there is no large afterglow from collisions. Figure\,\ref{ghostsep} shows 
the BCM-TORx rates for unpaired bunches and empty BCID around them, averaged over most fills with the 
1368 colliding bunch structure. Only fills until the start of the BCM noise period are included.
It is clearly seen that the background noise (BCIDs$<$1 and BCIDs$>$23) is low and BCID-independent. 

In addition to the primary beam-gas signals, there are additional smaller signals in the upstream modules.
These are the hits from \abg{}. The \abg{} signals are not easily identified in figure\,\ref{ghostsep} because
they are barely above the noise level, which is different for the two sides. Also, the primary 
beam-gas signals themselves disagree by about 28\%, i.e. much more than the efficiency difference of the two sides.
This discrepancy reflects a real difference in beam background, possibly due to systematically worse vacuum on
side A. These arguments motivate processing the data by first subtracting the background noise and then rescaling one
beam such that the primary BIB signals match. For each beam the noise level is determined as an average over 20 empty BCIDs 
preceding the first unpaired bunch. The rates in BCID 13--24 are multiplied by a factor of 1.28 in order to bring the primary BIB
signals on the two sides into agreement. 

The rates after these adjustments are shown in figure\,\ref{ghostsep-scaled}, where 
two levels of signal rate in odd BCIDs on sides A and C are seen to match each other. 
The perfect agreement of the second level, around 0.1\,Hz/BCID, after 
simply matching the primary BIB rates ($\sim$2.5\,Hz/BCID) on the two sides, means that it is proportional
to the primary BIB rate at a level of 3.9\% thereof, which leaves little doubt about its interpretation as \abg{} signals. 

In figure\,\ref{ghostsep-scaled} the factor 1.28 is applied also to the empty (even) BCIDs in the range 14--24 between beam-2 bunches. 
It is not evident if this is justified since the origin of the signals in them is not certain. A contribution from beam-1 ghost charge 
cannot be excluded. The statistical uncertainties on those points are large enough for the two sides to agree both with and without scaling. 

In order to avoid confusion between \abg{}, which is important only because of its correlation with BIB events, and the much more significant
level of afterglow from $pp$ collisions, the latter will be referred to as \agpp{}.

\begin{figure}[h!]
\centering 
\mbox{
\subfigure[]{
\includegraphics[width=0.49\textwidth]{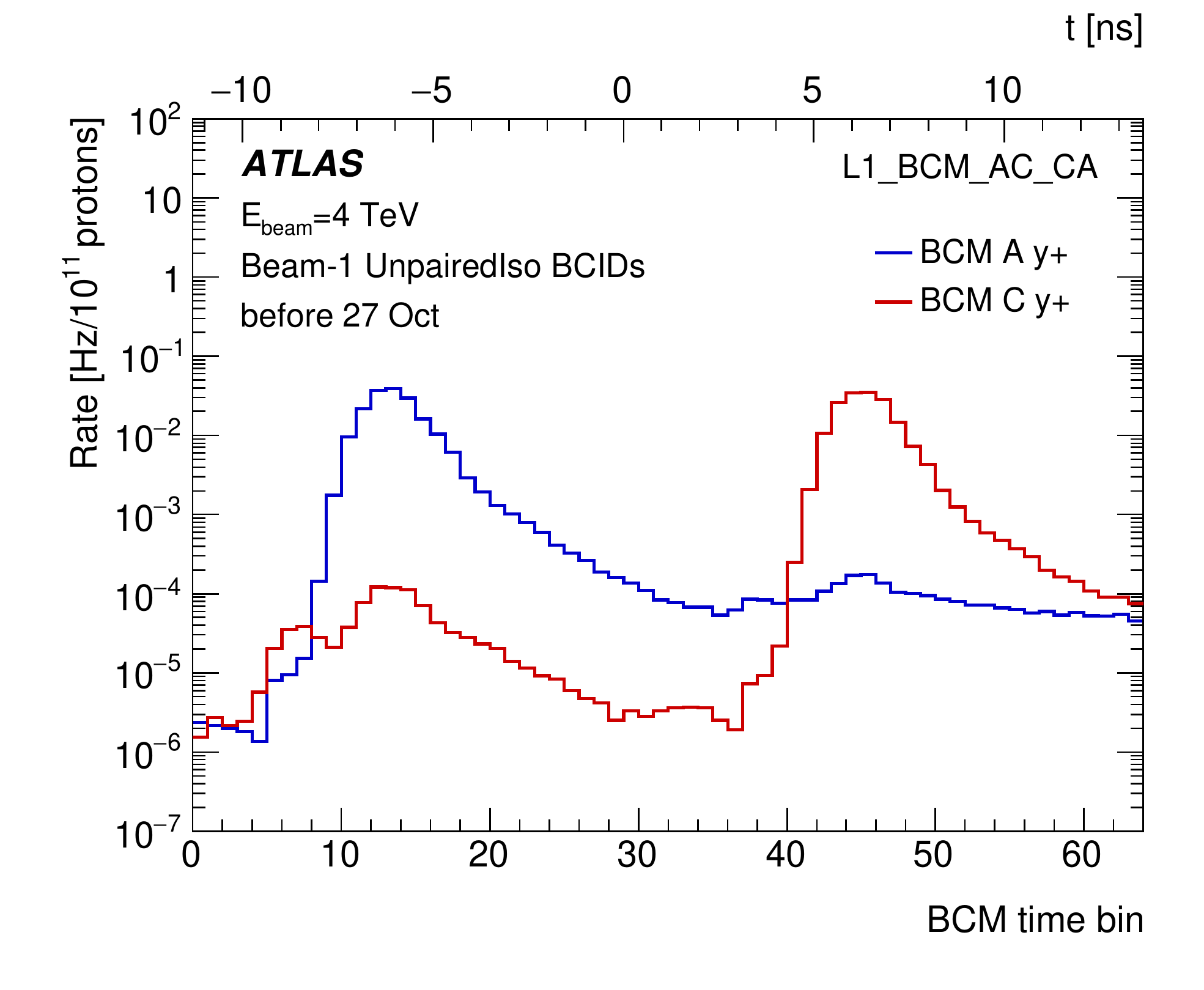}
\label{fig:bcmresponsea}
}
\subfigure[]{
\includegraphics[width=0.49\textwidth]{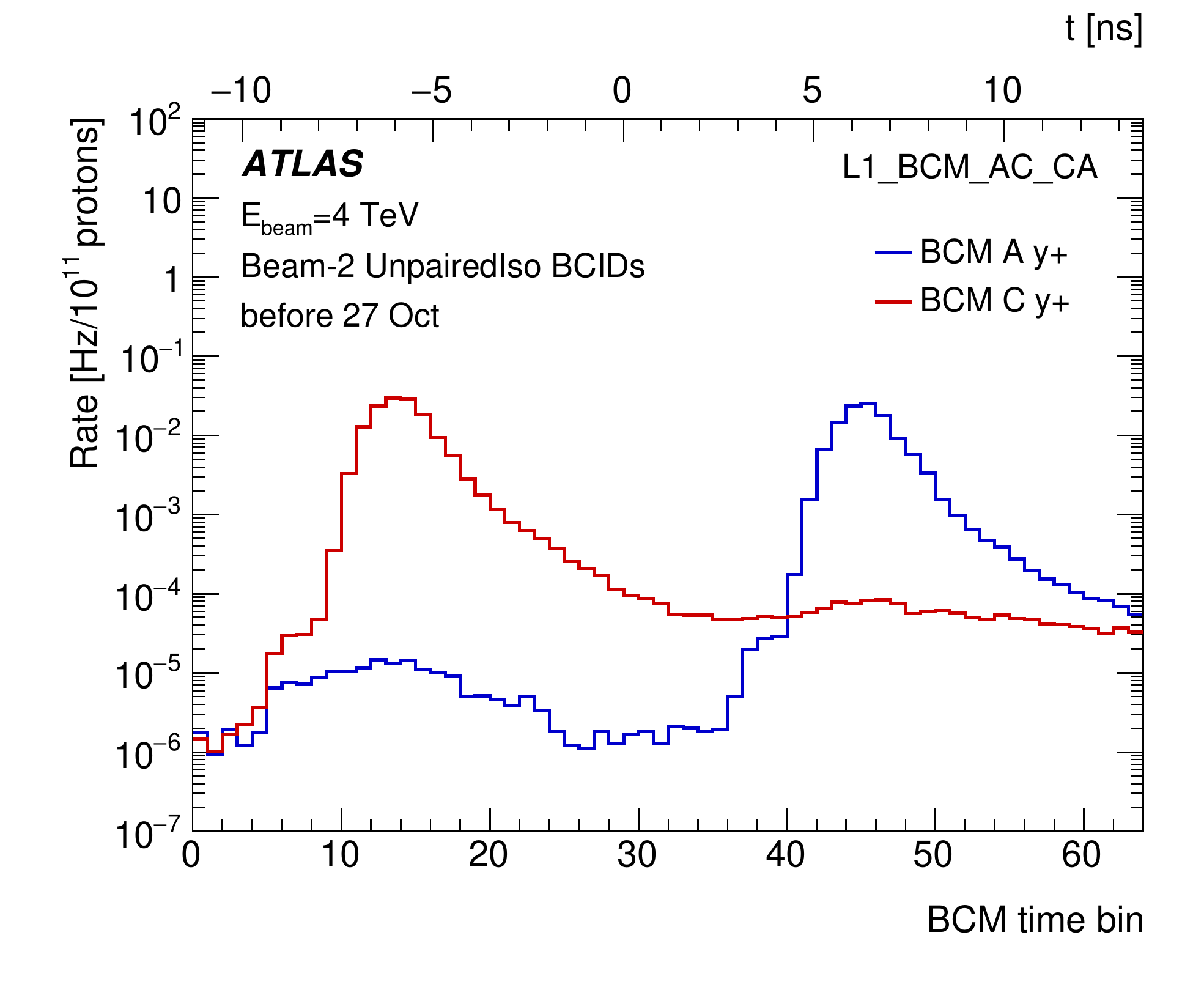}
\label{fig:bcmresponseb}
}
}
\caption{Response of the BCM $y+$ station in the BCIDs defined for beam-1 (a) and beam-2 (b) in the
events triggered by L1\_BCM\_AC\_CA\_UNPAIRED\_ISO.
Data with 1368 colliding bunches until the BCM noise period are used.}
\label{fig:bcmresponse}
\end{figure}

The excellent time resolution of the BCM signal and data acquisition system enabled the arrival time of the
background to be measured precisely. By exploiting the fine time binning of the recorded BCM signal, the existence of an
\abg{} tail was verified and its shape was studied in detail.
 Figure\,\ref{fig:bcmresponse} shows the time distributions 
in two BCM modules in events triggered by the L1\_BCM\_AC\_CA trigger on unpaired isolated bunches of either beam.
For beam-1 a pronounced early peak on the A-side is observed in figure\,\ref{fig:bcmresponsea}, corresponding to the background associated with
the incoming bunch. It is followed by a long tail, consistent with \abg{}. 
Beam-1 exits on the C-side and correspondingly the peak appears around bin 43, i.e. `in-time' with respect to the nominal collisions, 
again followed by a tail due to \abg{}. In figure\,\ref{fig:bcmresponseb} similar structures are seen for beam-2, but on the opposite sides.

However, there are also some entries consistent with early hits in downstream modules. A flat, or slowly falling,
pedestal is expected from noise and afterglow causing random triggers. But instead clear peak-like structures are seen in figure\,\ref{fig:bcmresponse},
especially in unpaired BCIDs of beam-1. These can be attributed to ghost bunches in the other beam, but a detailed discussion is
deferred to section\,\ref{sect:ghostBG}.

%% file: beamgas.tex
\section{Background monitoring}
\label{sect:beamgas}

\subsection{Beam-gas events}
\label{sect:vacuum}

The analysis of 2011 background data revealed a clear correlation of the
beam background as seen by the BCM and the residual gas pressure 
reported by the gauges at 22\,m. Using the data from a dedicated test fill 
without electron-cloud suppression by the solenoids at 58\,m, the 
contribution of the pressure at 58\,m to the observed background was estimated
to be 3--4\,\%\,\cite{backgroundpaper2011}.

In 2012 no such dedicated test was performed, but the contributions of various
vacuum sections were estimated by fitting the background (${\rm BIB}_{\rm BCM}$) data with a 
simple 3-parameter fit:

\begin{equation}
{\rm BIB}_{\rm BCM} =  A(f \cdot p22 + (1-f) \cdot p58) + b,
\label{eq:fit}
\end{equation}
where $p22$ and $p58$ are the pressures measured at 22\,m and 58\,m, respectively, and
$A$, $f$ and $b$ are free parameters.
The constant $b$ is introduced to take into account any background not correlated with the
two pressures included in the fit. Since ${\rm BIB}_{\rm BCM}$ is normalised by bunch intensity, 
the fit implies that also $b$ is assumed to be proportional to beam intensity, which 
is  a valid assumption if $b$ is due to beam-gas further upstream. However, if
a residual contribution comes from beam-halo losses or noise, it is not necessarily
proportional to intensity.

The pressure values given by the three gauges, available at 22\,m, were not always
consistent. If information from an individual gauge was not received for a short time interval, 
the gap was bridged by using the last available value, provided it was in the same fill and not older than 10 minutes.
Obviously erratic readings were rejected by requiring that the value was within a reasonable range ($10^{-11}$ -- $10^{-6}$\,mbar).\footnote{Pressures typically were in the range $10^{-10}$ -- $10^{-9}$\,mbar.}
The pressure was determined by first taking the average of the pair of readings closest to each other and including 
the value from the third gauge only if it did not deviate by more than a factor of three from this pair-average. Such
a three-gauge average was accepted in 97\% of luminosity blocks and in only 0.3\% of cases a two-gauge average was used.
The number of cases that all gauges deviated by more than a factor of three from each other negligible.
In 2.7\% of luminosity blocks no valid pressure data could be determined and the luminosity block was ignored.

\begin{figure}[t]
\centering 
\includegraphics[width=\textwidth]{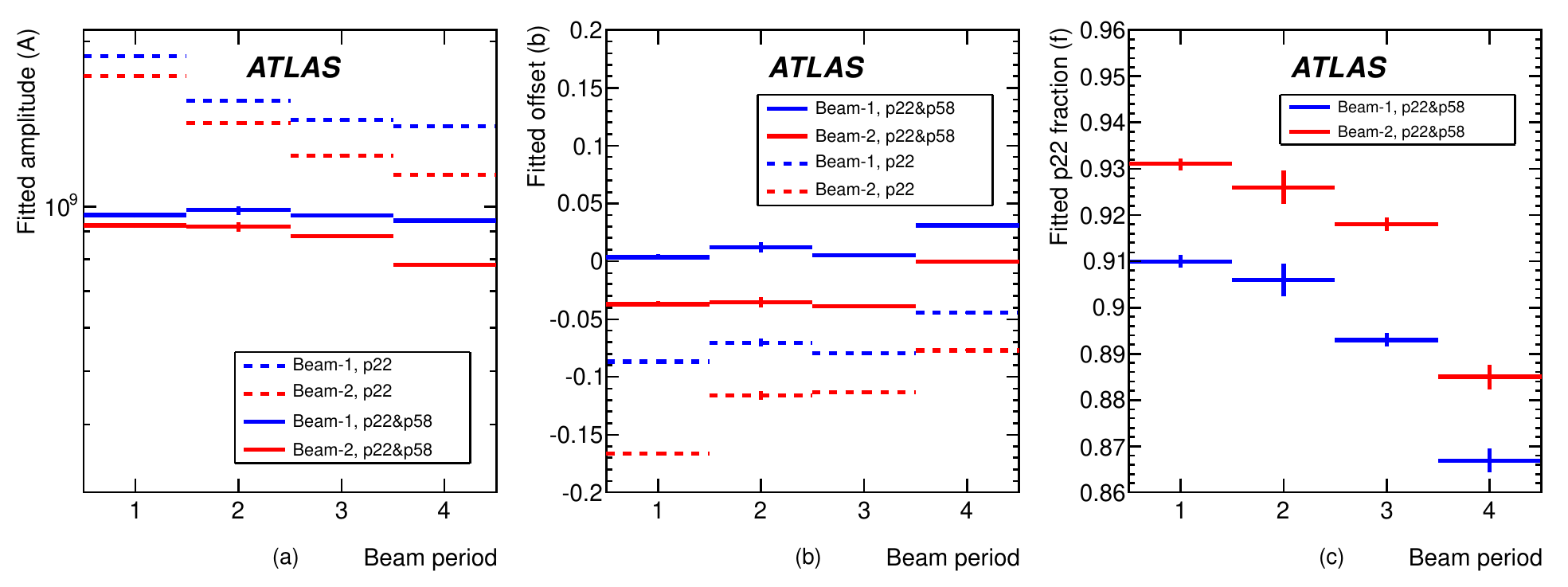}
\caption{Parameters $A$ (a), $b$ (b) and $f$ (c), resulting from the fit of equation\,\protect\ref{eq:fit} to data in periods 1--4, as detailed in table\,\protect\ref{beam-periods}.}
\label{fig:fitparams}
\end{figure}

Figure\,\ref{fig:fitparams} shows the obtained fit parameters for the periods listed in table\,\ref{beam-periods} using
either $p22$ only, i.e. $f=1.0$, or a combination of $p22$ and $p58$ in Eq.\,\ref{eq:fit}.
It can be seen that the value of $A$, which corresponds to the absolute normalisation, 
is systematically lower for beam-2, which implies that for the same measured pressure there is
less background from beam-2 than beam-1. The difference varies over the year, but is roughly
20\% during the operation with 1368 colliding bunches (periods 3 and 4).
The two sides of ATLAS are symmetric, so there is no obvious explanation why the backgrounds
should be different. However, this difference in the fitted $A$ is close to the 28\% that was derived 
from figure\,\ref{ghostsep-scaled}.

Ideally, the offset $b$ should reflect how well the pressures alone
describe ${\rm BIB}_{\rm BCM}$. If $b$ vanishes, it implies that
there is no additional source contributing significantly, while $b>0$ means that
not all sources are included in the fit. 
Using $p22$ only results in $b$-values which are negative by a significant amount.
These have no obvious physical interpretation and indicate that a linear fit using 
a single pressure is insufficient to describe the data. 
Using both, $p22$ and $p58$, clearly improves the model and results in $b$-values more 
consistent with zero.

The fraction $f$ indicates how large a role $p22$ plays in explaining the observed background. 
The values gradually decrease during the year, but remain in a band of $90\pm 4$\%, confirming 
the result for 2011 operation, that the background seen by the BCM is strongly correlated with the
pressure at 22\,m.

\begin{figure}[t]
\centering 
\includegraphics[width=\textwidth]{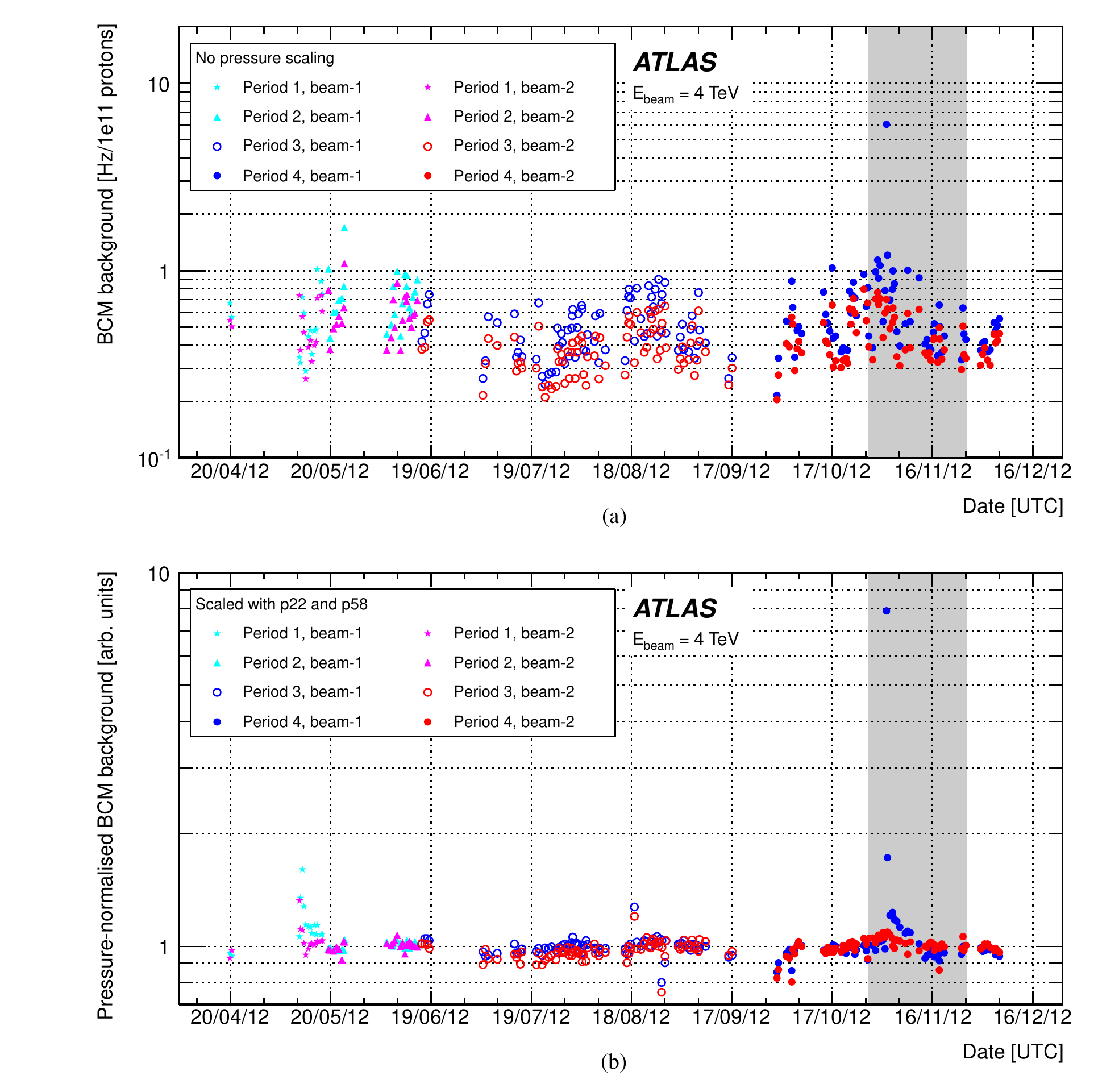}
\caption{BCM background rate for both beams normalised by bunch intensity (a) and after an
additional normalisation with the pressures at 22\,m and 58\,m (b) using Eq.\,\protect\ref{eq:fit} with the parameters as
shown in figure\,\protect\ref{fig:fitparams}. The shaded area indicates the period when the BCM was noisy.}
\label{fig:bcmAC-halo-all2012}
\end{figure}

Figure\,\ref{fig:bcmAC-halo-all2012} shows the intensity normalised BCM backgrounds for both beams
separately before and after scaling with the residual pressures using a fit with
$p22$ and $p58$ and parameter values shown in figure\,\ref{fig:fitparams}. It should be remarked, however, that a similar plot using only $p22$ for scaling
would be almost indistinguishable from figure\,\ref{fig:bcmAC-halo-all2012}(b). By construction,
the scaling with equation\,\ref{eq:fit} will result in an average of 1.0 in figure\,\ref{fig:bcmAC-halo-all2012}(b). 
The remarkable feature is that by this scaling the 
fill-to-fill scatter is almost entirely removed, despite the fact that the parameters are fitted over
four long periods, covering most of the year. 

The background estimates for beam-1 and beam-2, shown in figure\,\ref{fig:bcmAC-halo-all2012}(a), for the 
period with 1368 colliding bunches\footnote{One fill with abnormally high beam-1 background, 
which will be discussed later, is excluded.} lead to a beam-1/beam-2 ratio of $1.21\pm 0.17$, which is 
perfectly consistent with the difference  obtained before for the fitted parameter $A$.
This ratio is also consistent with the factor 1.28 derived from non-colliding beam data
in section\,\ref{sect:afterglow}, which suggests that vacuum conditions, i.e. beam-gas event 
rates, do not significantly change when beams are brought into collision. Since neither 
figure\,\ref{ghostsep-all}, nor figure\,\ref{fig:bcmAC-halo-all2012}(a), involves a pressure measurement, these 
good agreements suggest that the difference observed in parameter $A$ is not caused by different calibration of
the gauges, but by a real difference in backgrounds. One possibility is that the pressure
on side A has a different profile than that on side C and therefore the pressure gauges at 22\,m
do not have the same response in terms of average pressure in the region that contributes to the background.

%% file: ghostmon.tex
\subsection{Ghost collisions}
\label{sect:ghostlumi}

The rate of ghost collisions, i.e. $pp$-collisions in encounters of unpaired and ghost bunches, can be 
estimated from the rate of events recorded by the L1\_J10 and L1\_BCM\_Wide triggers in unpaired bunches.
An independent method is to use the luminosity data from the BCM, which are recorded at 
high rate independently of the ATLAS trigger. In particular, the single-sided BCM event counting, BCM-TORx,
as described in section\,\ref{sect:bcm} can be used. In the following, both methods will be presented 
and results compared.

\subsubsection*{Ghost collision rates from recorded events}
\label{sec:ghostreco}

From June 2012 onwards L1\_BCM\_Wide and L1\_J10 were both run without 
prescale on unpaired isolated and unpaired non-isolated bunches.
A significant fraction of the raw L1\_BCM\_Wide and  L1\_J10 trigger rates in unpaired bunches are 
due to accidental coincidences or, in the case of the latter, fake jets due to BIB muons. 
In this analysis ghost collisions are selected offline from the recorded event data
by requiring the presence of a reconstructed vertex in a volume consistent with the luminous region and with at 
least two associated tracks. In order to estimate the rates correctly, the vertex 
reconstruction efficiency has to be known. A method for estimating this from data for L1\_BCM\_Wide will be
presented. Luminosity blocks, or entire fills, that do not meet general data quality requirements are removed from the analysis.

\begin{figure}
\centering
\mbox{
\subfigure[]{
\includegraphics[width=0.47\textwidth]{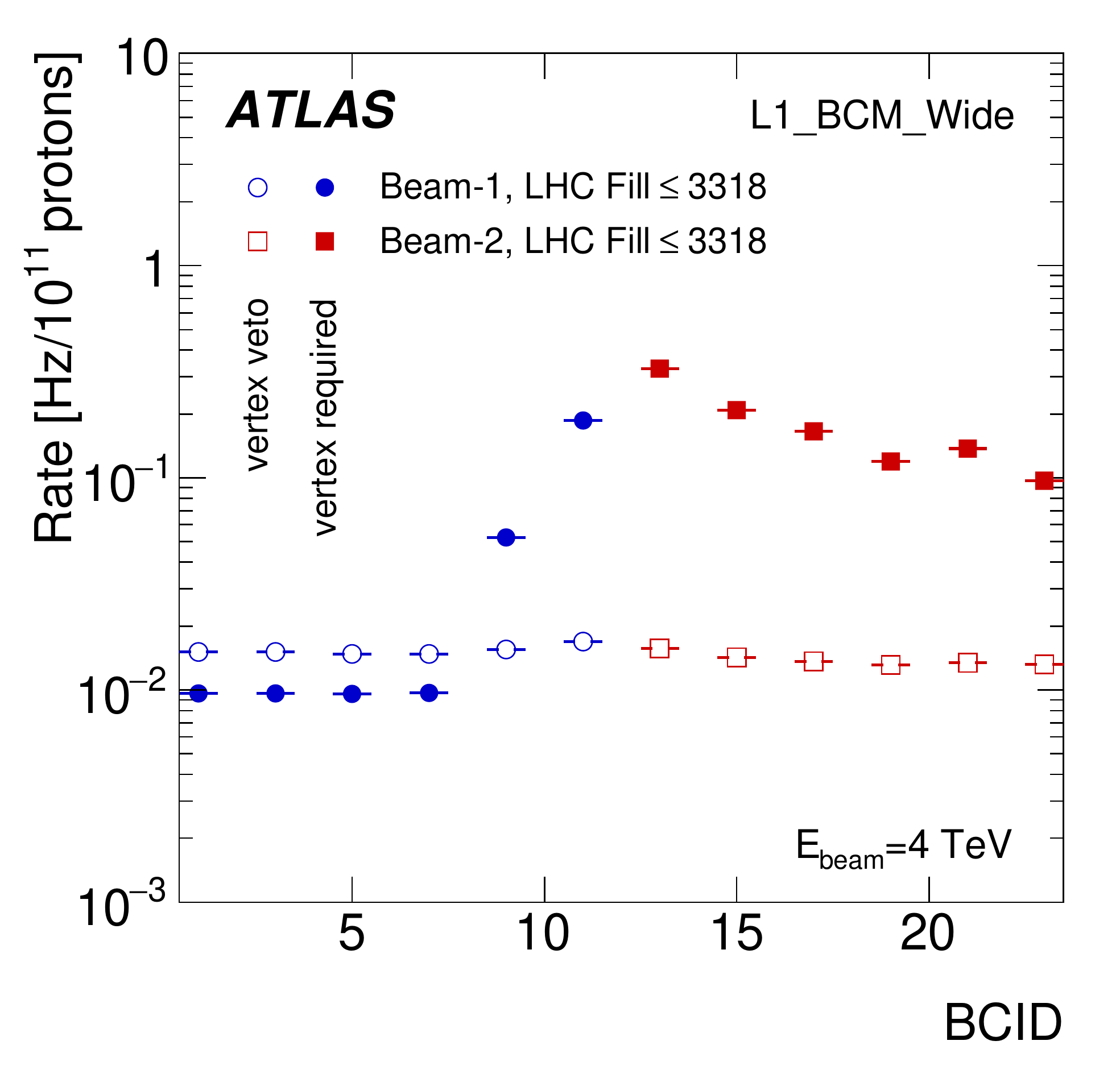}
\label{fig:bcmGhostColl}
}
\subfigure[]{
\includegraphics[width=0.47\textwidth]{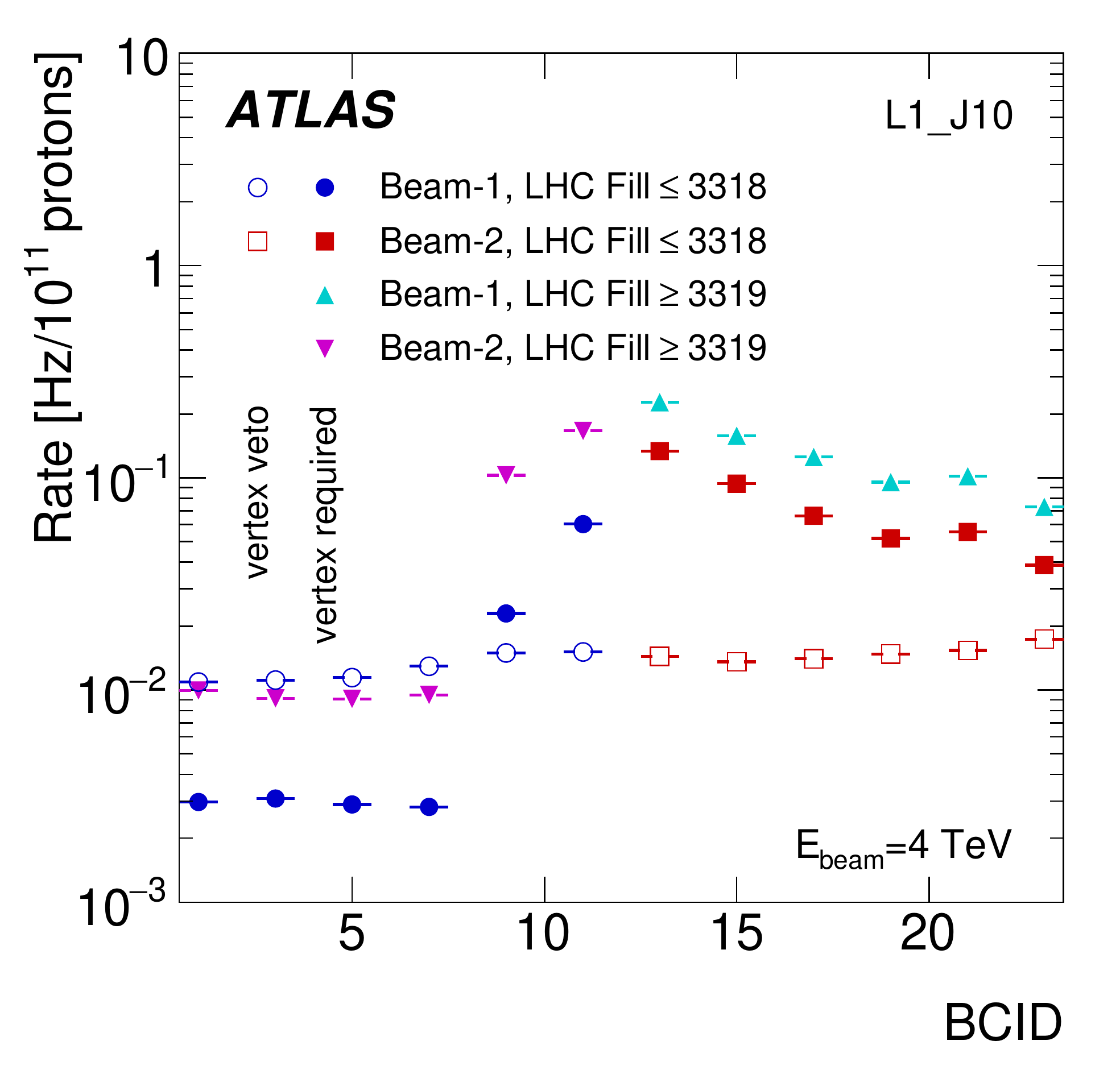}
\label{fig:j10GhostColl}
}
}
\caption{Left: rate of L1\_BCM\_Wide triggered events in unpaired BCIDs for the period prior to swapping the
unpaired bunches. The solid symbols show the rate after vertex requirement and the open symbols with
a vertex veto. Right: rate of L1\_J10 triggered events in the same period, and after swapping the unpaired trains. 
\label{fig:GhostColl}
}
\end{figure}

Figure\,\ref{fig:bcmGhostColl} shows the L1\_BCM\_Wide rates for events with and without a 
reconstructed vertex. The asymmetry of the ghost collision rate is striking. As expected,
the rate is much higher for lower isolation -- but only for the first unpaired train, i.e. beam-1.
For the second train, comprising unpaired bunches in beam-2, the rates remain high even for isolated bunches.
This asymmetry arises from the beam extraction from the PS, where the kicker is timed to the start of the
unpaired train, but will also extract any possible trailing ghost bunches. 
As a result the unpaired trains systematically have more intensity in trailing than heading injected ghost bunches. 
The L1\_J10 rates shown in figure\,\ref{fig:j10GhostColl} confirm this shape. The plots also show that the unpaired 
isolated definition of having no bunch within 3 BCID in the other beam is adequate for the first train, while it includes 
a non-negligible tail of ghost collisions in the second train. This feature is preserved also after swapping the
unpaired trains late in 2012, i.e. depends on the order of the trains and not on the beam.

\begin{figure}
\centering
\includegraphics*[width=\textwidth]{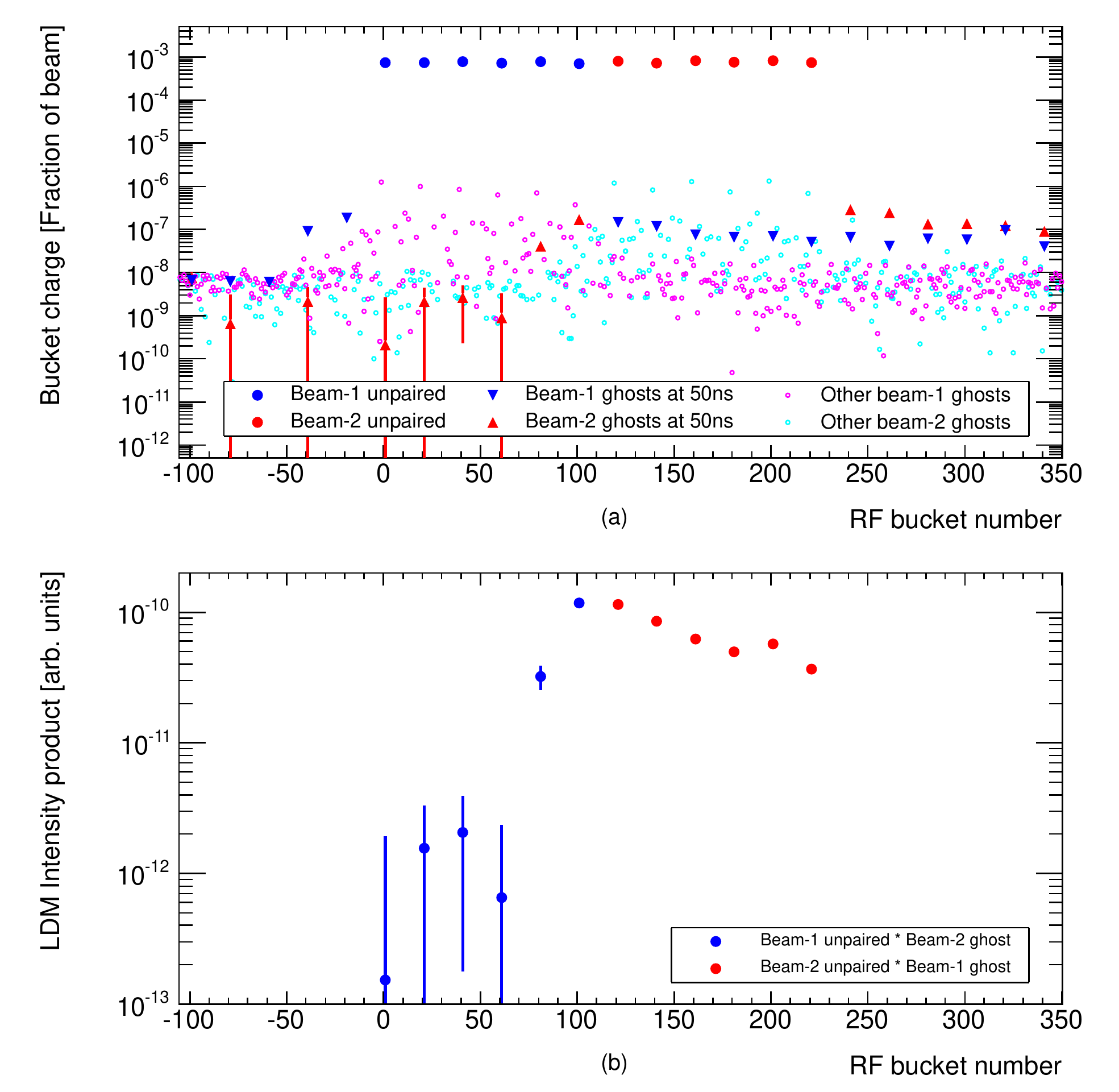}
\caption{Charge in individual RF buckets (a) as measured by the LDM in LHC fill 3005. The buckets corresponding to the 50\,ns 
beam structure are shown by the larger symbols. The signal of each bucket has been normalised by the sum of signals from all
buckets of the beam. The product (b) of the bucket charges in beam-1 and beam-2.  
}
\label{fig:ldmplot}
\end{figure}

The ghost collision rates seen in figure\,\ref{fig:GhostColl} are a product of ghost and unpaired bunch intensities in
colliding RF buckets. These quantities can be directly measured by the LDM and the results for one of the fills
when the system was operational are shown in figure\,\ref{fig:ldmplot}. Figure\,\ref{fig:ldmplot}(a) shows the charge in
individual buckets and confirms the higher intensity of trailing injected ghost bunches and their 50\,ns spacing. 
As explained in section\,\ref{sec:ncb}, this is due to spill-over in the PSB to PS injection. Figure\,\ref{fig:ldmplot}(b) shows the product of the 
bucket charges and largely confirms the shape seen in figure\,\ref{fig:GhostColl}. The small differences, especially for bucket 
101 (BCID 11), are likely to be explained by fill-to-fill differences, since figure\,\ref{fig:ldmplot} shows a single fill, while 
figure\,\ref{fig:GhostColl} is a long-term average. The point in RF-bucket 81 falls into the LDM trigger reset and had to be
estimated from the value in bucket 101 using the ratio of the corresponding positions in front of other bunch-trains.

The rates with vertex veto, shown in figure\,\ref{fig:GhostColl}, are almost BCID-independent, consistent with
their origin being random coincidences of hits -- or noise in the case of jets --  which are distributed uniformly in time. 
However, a barely visible hump in the centre (around BCID 11)  can be identified, which is due to real collisions where 
vertex reconstruction has failed.
Assuming that the probability of fake vertex reconstruction and the vertex reconstruction inefficiency are both small,
the latter can be estimated from the relative sizes of the small and the large peaks in figure\,\ref{fig:bcmGhostColl}.     
This yields an estimate of 1.1\,\% for the vertex reconstruction inefficiency. It must be emphasised, however, that 
this is the value with respect to collision events seen by the BCM and not a global inefficiency of ATLAS vertex reconstruction.
Furthermore, a comparison of the flat tails in figure\,\ref{fig:bcmGhostColl}, taking into account this inefficiency,
provides an estimate of 1.27 for the beam-1/beam-2 ratio of BIB, which is perfectly consistent with 
the factor 1.28 derived in section\,\ref{sect:afterglow} and $1.20\pm 0.17$ found in section\,\ref{sect:beamgas}.

\begin{figure}
\centering 
\includegraphics[width=\textwidth]{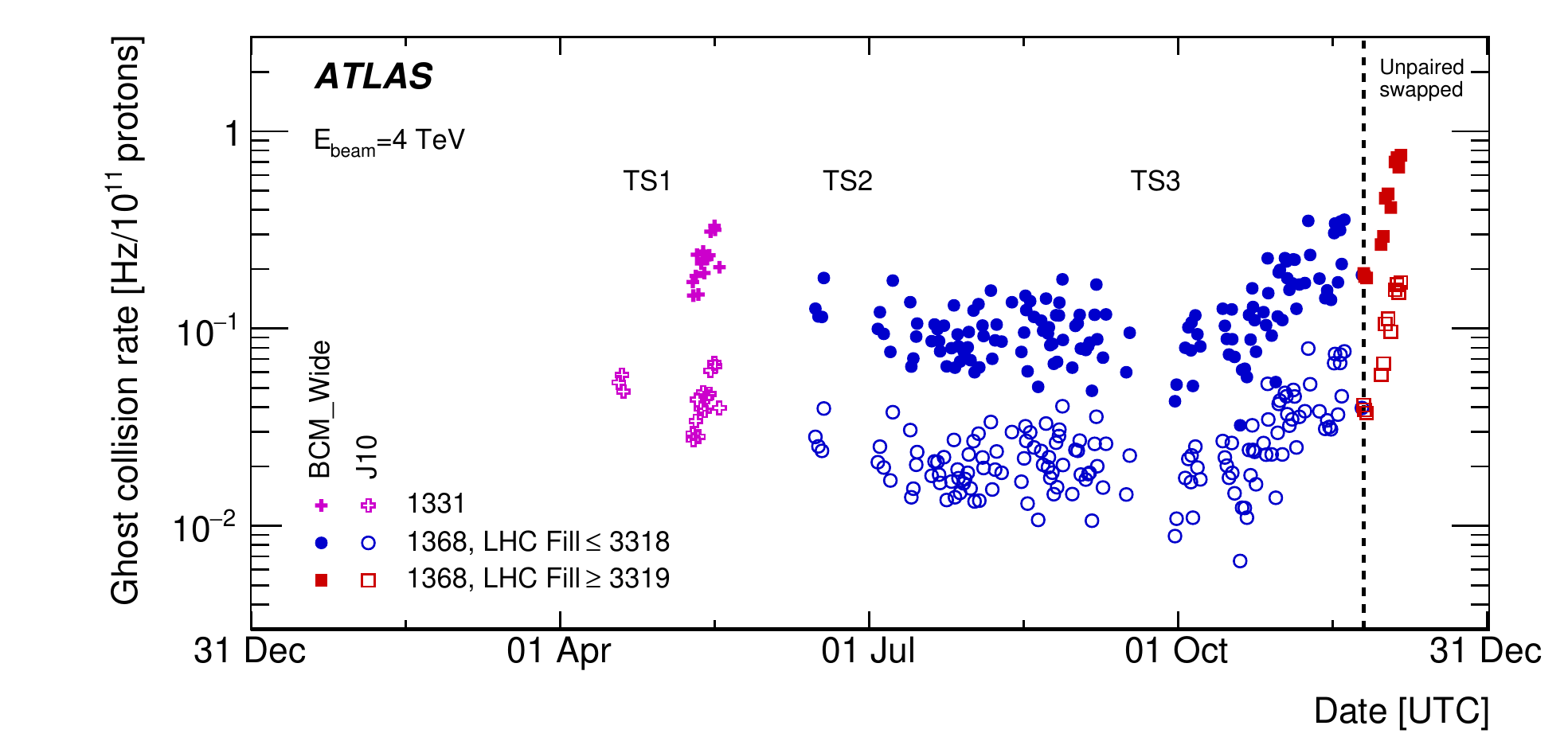}
\caption{Ghost collision rate in unpaired bunches in the 2012 data, estimated from events
triggered by L1\_BCM\_Wide and L1\_J10 triggers. In both cases the
presence of at least one reconstructed vertex is required. The early BCM points are 
not included, since the device was not properly timed in. The technical stops are indicated in the plot.
}
\label{fig:ghost_rate}
\end{figure}

The ghost collision rate during 2012 operation is shown in figure\,\ref{fig:ghost_rate}, where it can be 
seen that the rate is rather constant until TS3 but rises steeply thereafter. L1\_BCM\_Wide and L1\_J10 rates exhibit a 
similar rise, both in terms of shape and relative magnitude. Since the BCM and the calorimeters are totally independent 
and look at very different observables, it is not conceivable that the rise would be due to an increase
of some random contribution. Thus, the identical rise must mean that the intensity of injected ghost bunches, 
colliding with the unpaired bunches, increased rapidly after TS3.

\subsubsection*{Ghost collision rates from luminosity data}

The single-sided event rate, BCM-TORx, is composed of three contributions
\begin{equation}
{\rm Rate} = {\rm Collisions} + {\rm BIB} + {\rm Pedestal},
\label{eq:tor-rate}
\end{equation}
where the last term is defined to include both instrumental noise and \agpp{}. 
As shown in figure\,\ref{fig:ghostAGCorr} this pedestal can be estimated from the BCID before the unpaired bunch.

Prompt secondaries from upstream BIB events will not be counted in the upstream detector, since they arrive before the 
BCM-TORx window is open. A possible contribution of backscattering from beam-gas events to upstream detectors is
strongly suppressed by timing and the good vacuum close to the IP. The only process, besides $pp$-collisions, which can give 
in-time hits in upstream detectors is the \abg{}, discussed in section\,\ref{sect:afterglow}.
From this argumentation, and Eq.\,\ref{eq:tor-rate}, it follows that after pedestal subtraction, essentially all of the rate observed for 
unpaired bunches in the upstream BCM detector must be due to ghost collisions and \abg{}, the latter
being proportional to the primary BG signal seen in the downstream modules.

\begin{figure}
\centering 
\includegraphics[width=\textwidth]{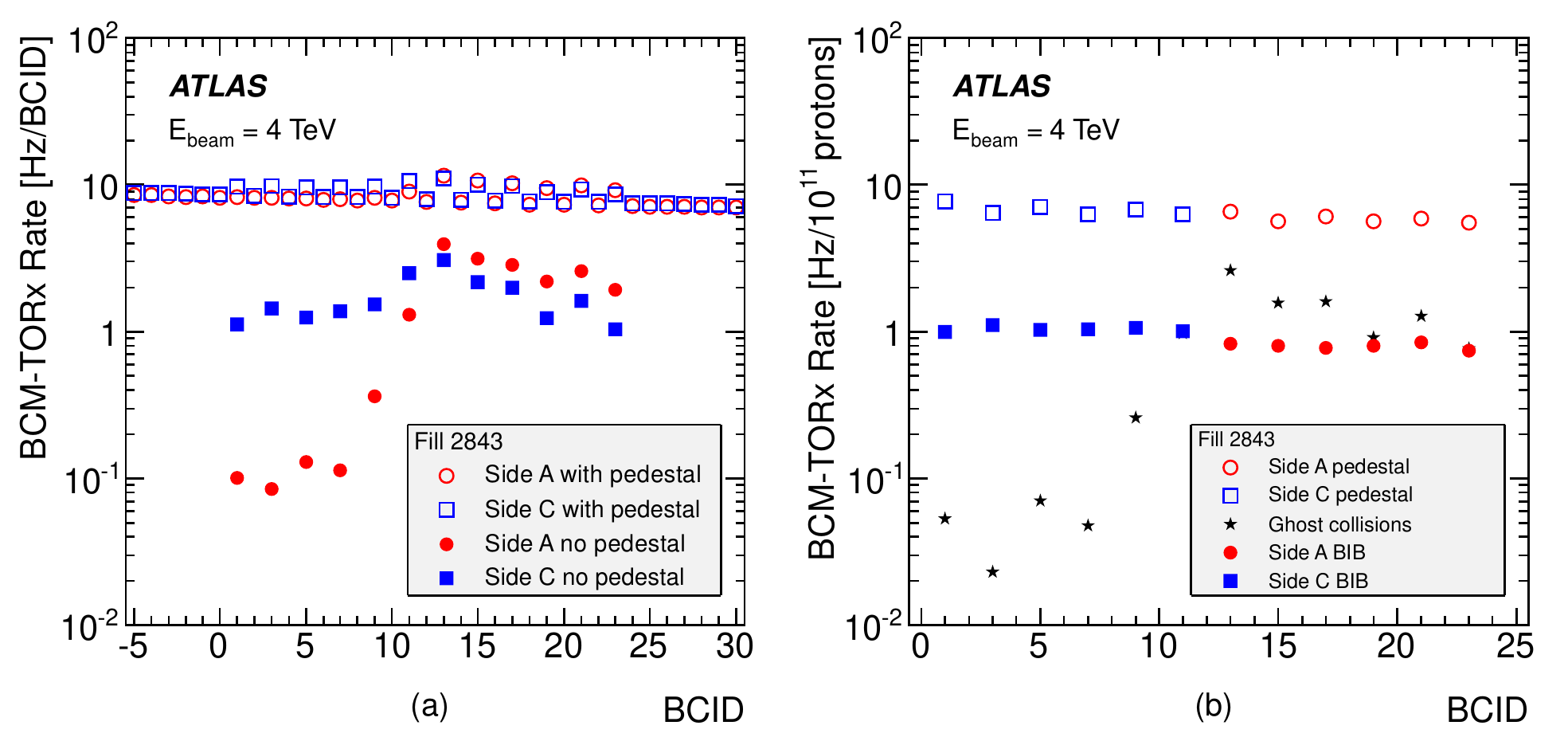}
\caption{The procedure of separating the background components in unpaired bunches. The open symbols in plot (a) show the
total single-sided rate seen by the BCM while the solid symbols show the data after subtraction of the background and
restricted to the BCID-range of the unpaired punches. Plot (b) shows the three background components separated, as explained in the text.
}
\label{newGhostRate}
\end{figure}

The procedure to separate the single-side BCM rate into its three components, and the results obtained, are illustrated in figure\,\ref{newGhostRate}. 
In figure\,\ref{newGhostRate}(a) the open symbols indicate that the raw BCM-TORx rate is dominated by the pedestal and the other contributions are barely visible.
A clear structure, resembling figure\,\ref{fig:bcmGhostColl} appears when the pedestal is subtracted, so that the data contain only beam-background and ghost collisions.
The downstream modules
are timed to see the primary BIB and ghost collision products emitted in
the direction of the unpaired bunch,  while the upstream modules see the ghost collision secondaries emitted in the direction of the ghost bunch and a contribution from \abg{}. In 
figure\,\ref{ghostsep-scaled} the latter was estimated to be a fraction $f=0.039$ of the primary BIB signal. 
Thus the rates seen in the upstream ($R_u$) and downstream ($R_d$) detectors can be written as:
\beq
R_u = f B + G
\label{eq:ru}
\eeq
and
\beq
R_d = B + G,
\label{eq:rd}
\eeq
where $B$ stands for the primary rate from BIB and $G$ for the rate from ghost collisions and the detection efficiencies are assumed to
be identical on both sides. The equations can be solved to
yield
\beq
B = \frac{R_d-R_u}{1-f} 
\label{eq:ghostEst1}
\eeq
and
\beq
G = \frac{R_u - f R_d}{1-f}.
\label{eq:ghostEst2}
\eeq
It is worth to note that the value of $f$ is small and in the limit $f\rightarrow 0$ the equations simplify to
$G=R_u$ and $B=R_d-R_u$, i.e. the upstream detector measures the ghost collision rate and the difference between
downstream and upstream detectors gives the rate due to BIB.
Figure\,\ref{newGhostRate}(b) shows all the background components separated.

\begin{figure}
\centering 
\includegraphics*[width=\textwidth]{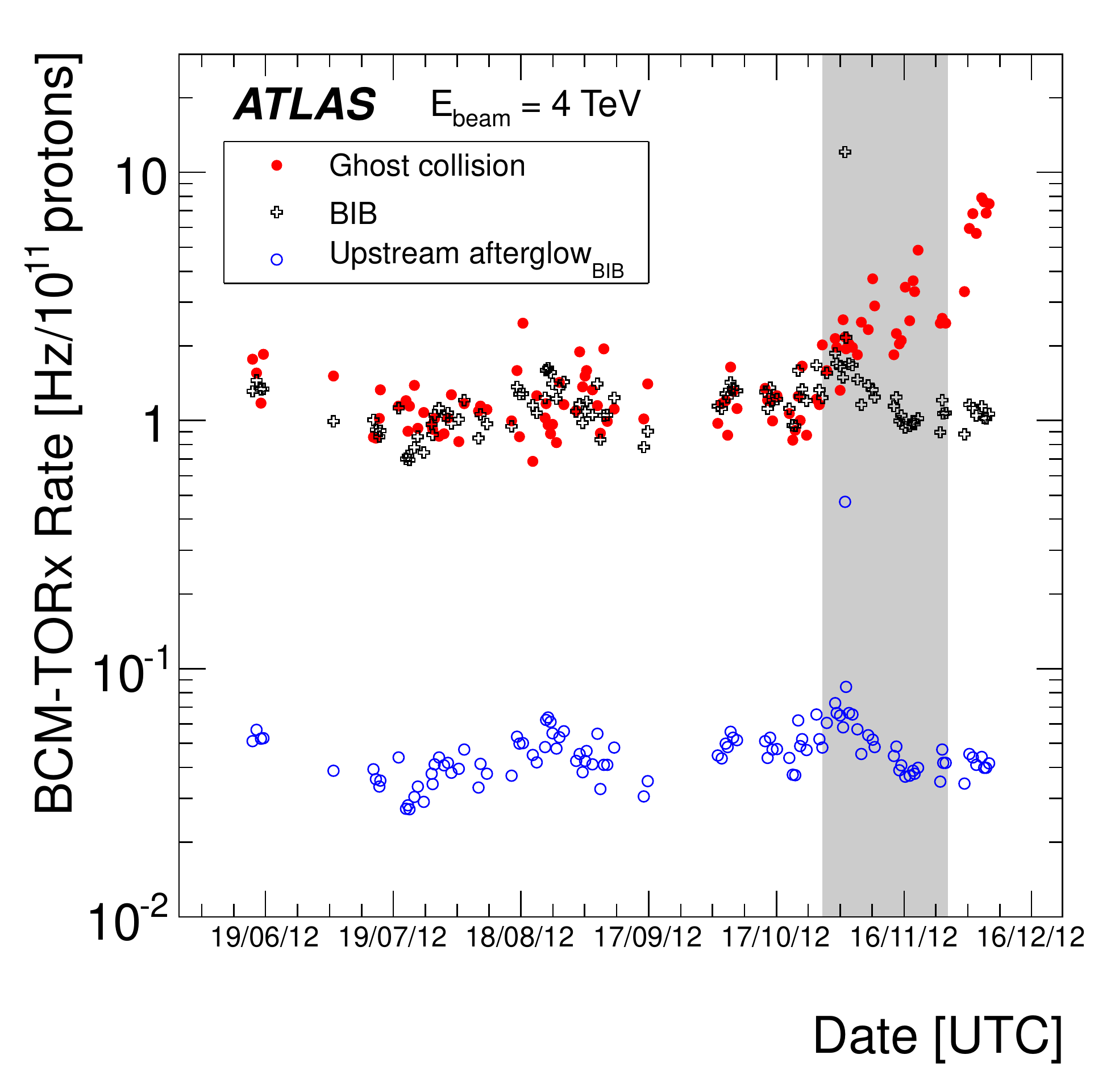}
\label{ghostVsUpsORCorrS}
\caption{The BCM-TORx rate in unpaired bunches due to BIB, \abg{} and ghost collisions for all 2012 fills with 1368 colliding bunches.
Only the first 100 LB ($\sim 100$\,minutes) of stable beams in each fill have been used in order to remove fill-length dependence.}
\label{fig:ghostVsUpsORCorrS}
\end{figure}

Figure\,\ref{fig:ghostVsUpsORCorrS} compares the BCM-TORx rates, i.e. the components of Eqs.\,\ref{eq:ru} and \ref{eq:rd}, due to ghost 
collisions ($G$), BIB ($B$) and \abg{} ($fB$) for all fills in 2012 which had 1368 colliding bunches. It can be seen that for most of the year, beam-gas 
and ghost collisions, i.e. genuine luminosity, contribute about the same amount to the BCM-TORx rate in unpaired bunches, but at 
the very end of the year there is a steep rise in the ghost collision rate and it becomes the dominant contribution. 
This result implies that if the rate seen in unpaired bunches is used as background correction to a BCM-based luminosity
measurement in a high-luminosity fill, a detailed decomposition, as described here, must be done in order to separate out the ghost collision contribution.

Having presented two different methods to monitor ghost collision rates, a comparison between the results remains to be done. The fake ghost collision
triggers, i.e. coincidences without a real collision, are a sum of many independent contributions and therefore provide the most sensitive basis for comparison. 

Assuming that $B$ in Eq.\,\ref{eq:ghostEst1} and the pedestal are uncorrelated between sides A and C, the random coincidence rates can 
be obtained by simple multiplication. The rate of these fake ghost collisions should be the same as that obtained from the event-by-event 
analysis after applying a vertex veto. 
Since the pedestal is uniformly distributed in time, the different window widths of the BCM\_Wide trigger and the luminosity 
trigger have to be taken into account, as detailed in appendix\,\ref{app:TS3fitsBCM}.

The problematic component of the non-collision L1\_BCM\_Wide rate is the \abg{}, because it does not fulfil the
requirement of being uncorrelated with $B$ on the other side. Unfortunately figure\,\ref{ghostsep-scaled} 
only determines the total \abg{} rate to be about 3.9\% of $B$ but does not give any information about the 
correlation between $B$ and \abg{} signals, i.e. how often the latter coincides with the former. 

If the L1\_BCM\_AC\_CA triggered sample were an unbiased subset of BIB events giving a signal in the downstream detector,
the correlation could be estimated as the fraction of L1\_BCM\_AC\_CA triggered events, which are also triggered by 
L1\_BCM\_Wide, but have no vertex.
However, the events selected by L1\_BCM\_AC\_CA are likely to be biased towards higher multiplicities with respect to BIB events 
giving only downstream hits. A higher multiplicity will also imply a higher likelihood to obtain 
a L1\_BCM\_Wide trigger due to an associated \abg{} hit. Thus the observed fraction of 1.3\% should be considered an upper limit. 

A better method to estimate the correlation is to require that the estimated rate agrees with that of events recorded 
by the L1\_BCM\_Wide trigger, after applying a vertex veto. The best match over all 2012 is obtained when 0.9\% of the 
downstream beam-gas hits are assumed to be in coincidence with an upstream \abg{} hit. 
Being slightly lower than the 1.3\,\%, this value is considered perfectly reasonable.

\begin{figure}
\centering 
\includegraphics*[width=\textwidth]{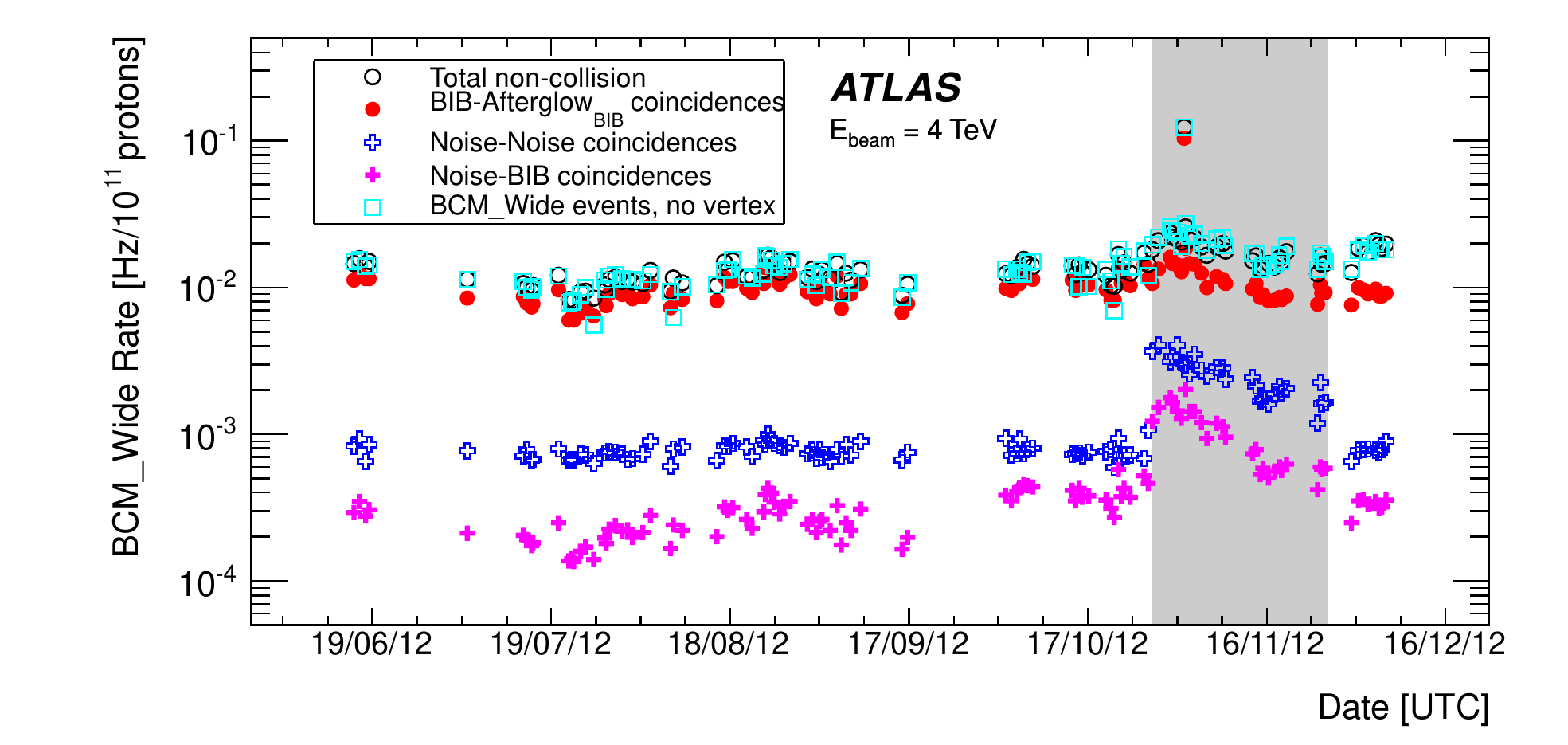}
\caption{The total non-collision rate in the BCM\_Wide trigger for 2012 LHC Fills with 1368 colliding bunches and the 
individual contributions to this rate. The open circles include a contribution of real ghost collisions due to the
1.1\% vertex inefficiency. The jump in the pedestal-related components is due to the appearance of the 
BCM noise (shaded area), which then gradually decreased. Only the first 100 LB ($\sim 100$\,minutes) of stable beams in each fill have been
used in order to remove fill-length dependence due to the decreasing afterglow.}
\label{fakeghostyear}
\end{figure}

When this coincidence fraction of 0.9\% is applied to all fills of 2012 with 1368 colliding bunches, the non-collision BCM\_Wide rate
estimates shown in figure\,\ref{fakeghostyear} are obtained. The plot is consistent with the assumption that most of this rate comes
from a coincidence formed by BIB and its associated \abg{}. 
The open circles in figure\,\ref{fakeghostyear} are the sum of the three other components shown
with the addition of 1.1\% of ghost collisions in order to take into account the vertex reconstruction inefficiency.
Inclusion of this small fraction of ghost collisions is needed to describe the rise towards the end of the year. 
An agreement of the open circles and open squares is enforced on average by the matching of the data, as described above.
This procedure, however, does not constrain agreement of the fill-to-fill fluctuations 
or the slight rise over the year. For both very good consistency between the two methods is observed.

%% file: fakejets.tex
\newcommand{\fch}{f_\textrm{ch}}
\newcommand{\fmax}{f_\textrm{max}}

\section{Fake jets}

Muons can emerge from the particle showers initiated by  beam gas interactions or  scattering
of the beam halo protons at limiting apertures of the LHC. Such BIB muons, with
energies potentially up to the \TeV{} range, may enter the ATLAS calorimeters and
deposit energy which is then reconstructed as a fake jet.
The simulations of BIB show that the high-energy muons leading to fake jets at radial
distances of $R>1$\,m, originate from the tertiary collimators or, in the case of beam-gas collisions, 
even further away~\cite{backgroundpaper2011,roderick-beam-background}.

Nearly every physics analysis in ATLAS requires good quality jets with the pile-up contribution
suppressed and non-collision backgrounds removed. In reference\,\cite{backgroundpaper2011} various sets of jet cleaning criteria
were proposed to identify NCB. Since then, these have been commonly used in ATLAS analyses. 
While calorimeter noise appears at a rate low enough to not significantly increase the trigger rate,
calorimeter hot cells, BIB and CRB-muons may take up a significant
fraction of the event recording bandwidth.
The number of events triggered by a fake jet and entering a given analysis, strongly depends on the event topology considered.
Single-jet selections are most likely to pick up a fake jet on top of minimum-bias collisions that happen
during every crossing of two nominal bunches.
Therefore, analyses with jets and missing transverse momentum (\MET{}) in the final state, such as the mono-jet analysis~\cite{mono-jet-paper},
crucially depend on an efficient jet-cleaning strategy.
More complicated topologies will include NCB only if it is in combination with other hard collision products.

This section reports on the observation of fake jets due to BIB muons in unpaired bunches as well as the ones extracted from collision data by inverting the jet cleaning selection.
In addition, the rates of fake jets due to CRB muons 
are compared to a dedicated Monte Carlo simulation. Finally the rates of fake jets due to BIB and CRB are compared as a function
the $\pt$ of the reconstructed jet.

\subsection{Fake jets in unpaired bunches}
\label{sec:jetunpaired}

Unpaired bunches provide a unique 
environment for studying fake jets due to BIB muons.
For such analysis, the recorded events triggered by L1\_J10 are selected from a dataset satisfying general data quality conditions, as described in section\,\ref{sec:calojets}.
Jets emerging from ghost collisions are efficiently suppressed by a vertex veto. 

\begin{figure}
\centering 
\mbox{
\subfigure[]{
\includegraphics[width=0.49\textwidth]{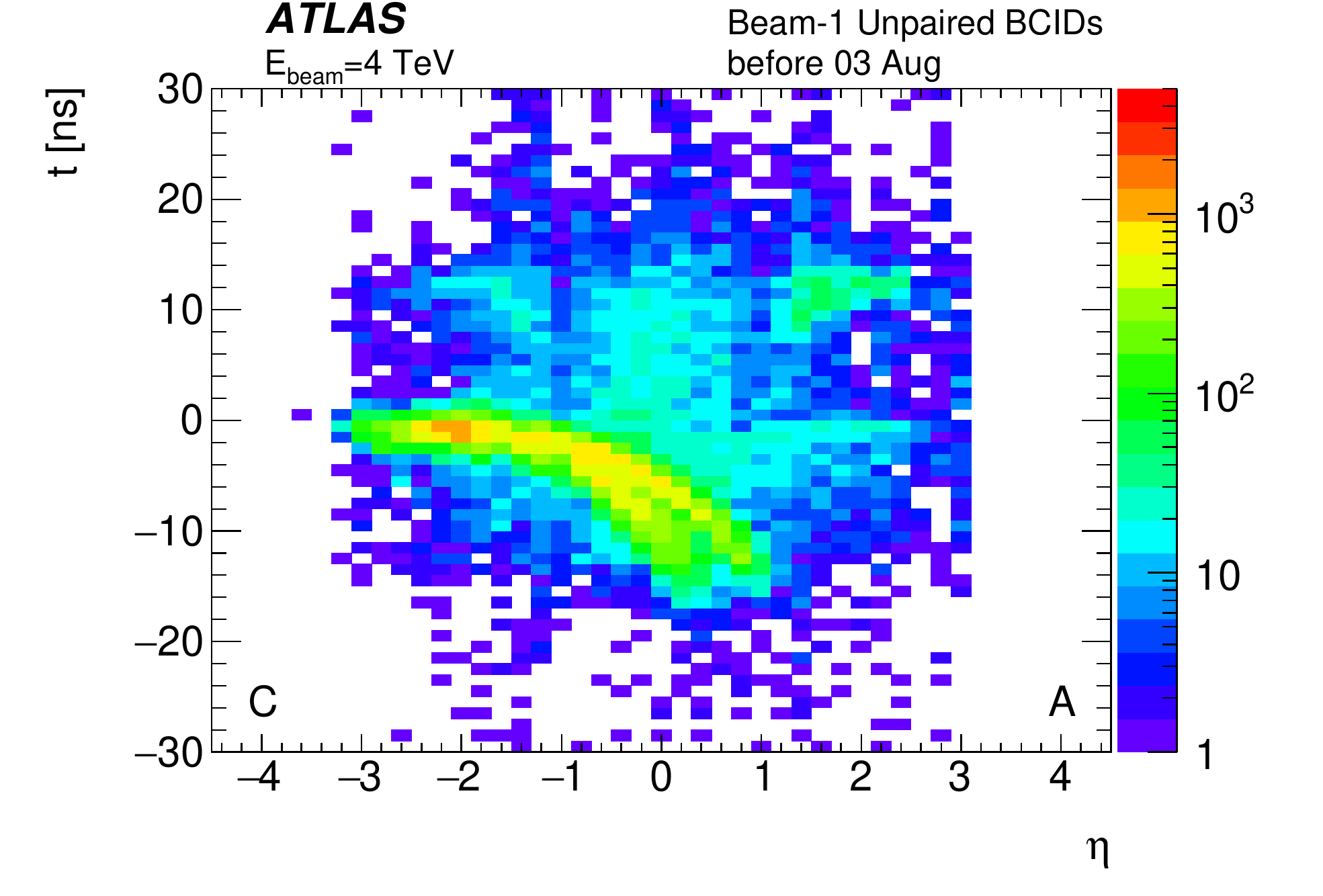}
\label{fig:ghostJetTiminga}
}
\subfigure[]{
\includegraphics[width=0.49\textwidth]{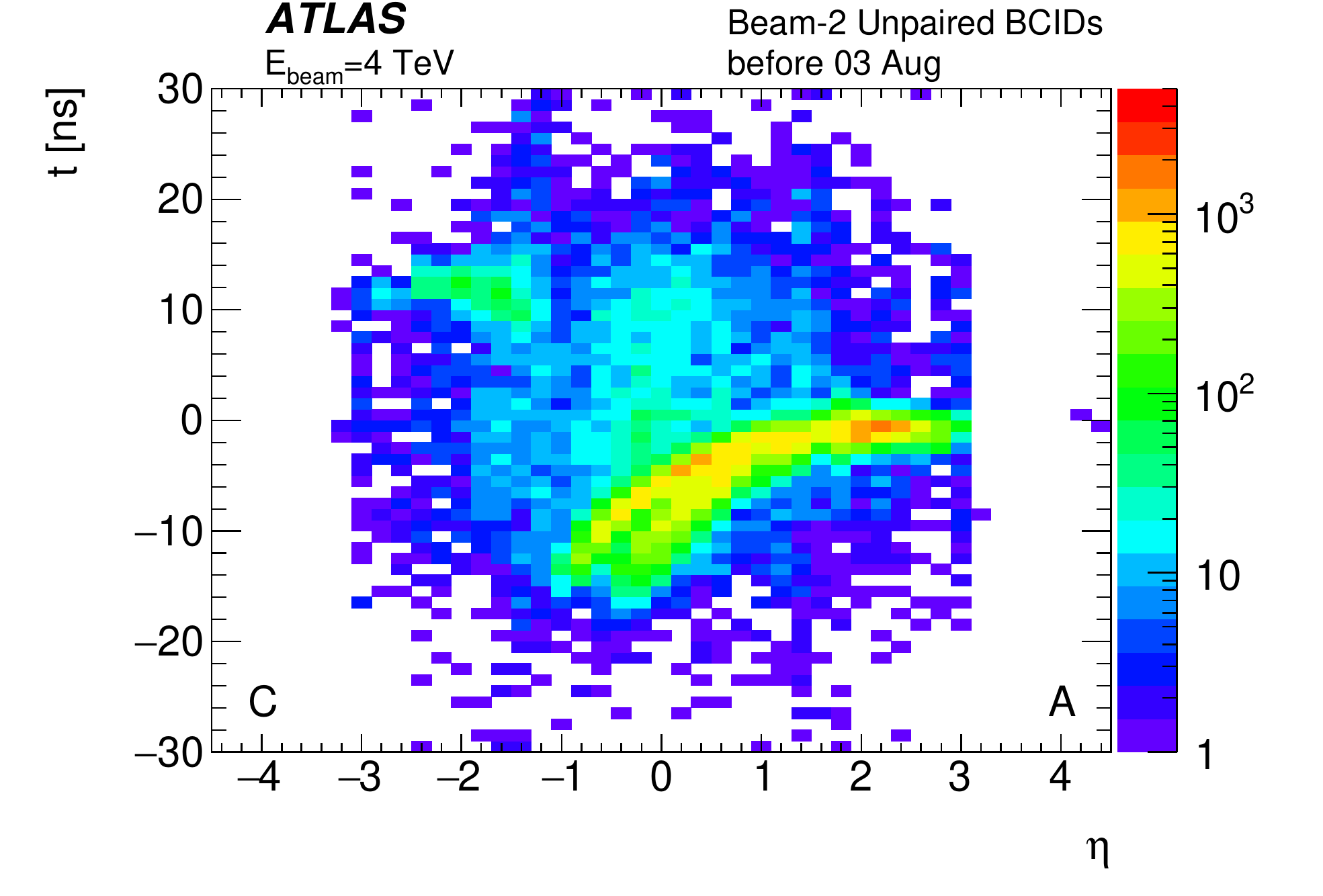} 
\label{fig:ghostJetTimingb}
}
} \\
\mbox{
\subfigure[]{
\includegraphics[width=0.49\textwidth]{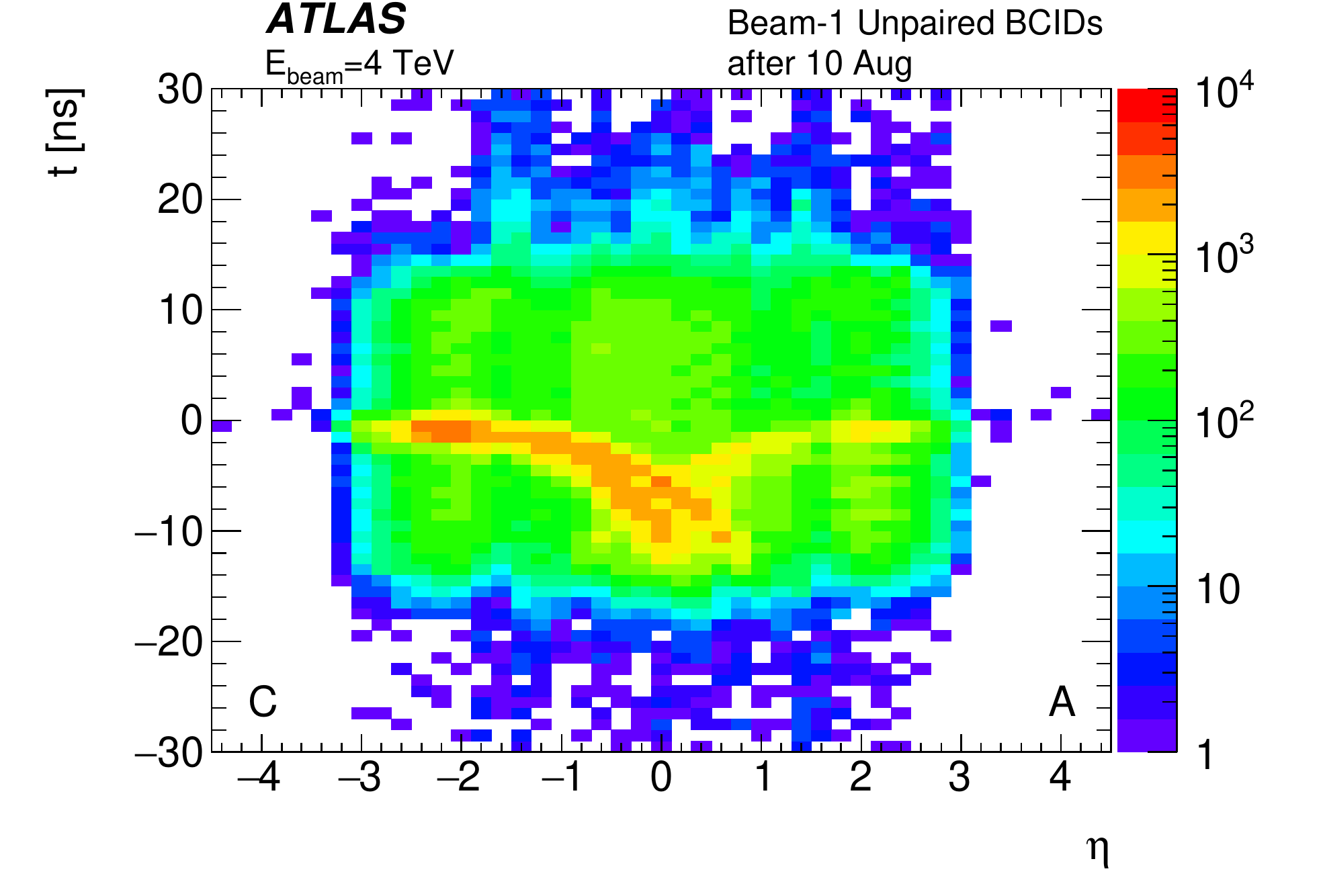}
\label{fig:ghostJetTimingc}
}
\subfigure[]{
\includegraphics[width=0.49\textwidth]{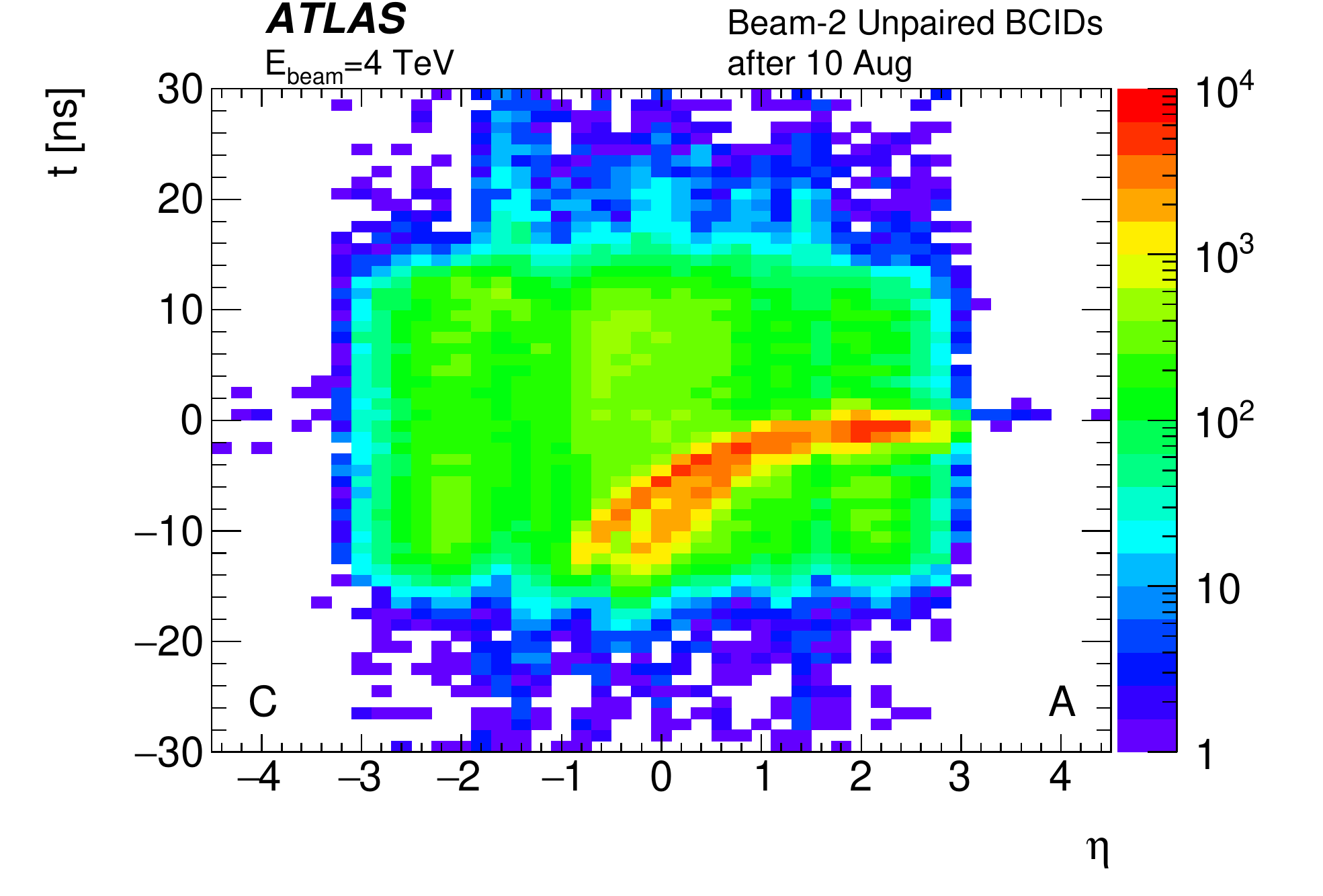}
\label{fig:ghostJetTimingd}
}
}
\caption{Jet times as a function of $\eta$ in unpaired BCID before 3 August (a,b) and after 10 August (c,d).
The (a,c) and (b,d) show unpaired BCIDs for beam-1 and beam-2, respectively.
Only fills with 1368 colliding bunches are considered. 
Negative $\eta$ corresponds to side C, i.e. outgoing beam-1.
}
\label{fig:ghostJetTiming}
\end{figure}

Figure\,\ref{fig:ghostJetTiming} shows the characteristic ``banana''-shape signature 
in the $\eta$--$t$ plane of the distribution of fake jets due to BIB muons that traverse the detector from one side to the other. 
The curvature of the ``banana'' depends on the radial position in the calorimeter\,\cite{backgroundpaper2011}. 
The larger curvature corresponds to the Tile calorimeter, the other tail is due to jets in the LAr calorimeter.
The jet time reconstruction is adjusted such that ideally $t=0$ is found for all jets originating in a $pp$ collision. 
At higher $|\eta|$ on the downstream side a small residual offset of approximately $-1\,$ns is observed because
the BIB muons travel parallel to the beam and therefore reach any given position in the calorimeters earlier than 
the collision products.
The time-distributions are cut off around $\pm 12.5\,$ns due to the 25\,ns acceptance of the trigger within a BCID.

The data in figure\,\ref{fig:ghostJetTiming} are shown separately for periods before early August (a, b) and after mid-August (c, d). 
The motivation for this separation is related to the chromaticity changes of the LHC, 
which cause significant differences to the fake jet distributions in the $\eta$--$t$ plane:
\begin{itemize}
\item In data before early August the ``banana'' shapes are clearly distinguished against a low background. In addition a concentration of jets 
at $|\eta|\sim 2$ and $t\sim 10$\,ns can be seen.
These are caused by upstream fake jets in the following bunch which arrives 50\,ns later, i.e. for the proper bunch crossing they
would appear at $t\sim -40$\,ns, but such an early arrival means that they fall into the trigger window two BCIDs earlier. 
\item After mid-August, most structures are masked by an almost
uniform pedestal covering the entire area in the $\eta$--$t$ plane, where jets are reconstructed.
An analysis of the azimuthal distribution of this pedestal revealed that it exhibits the characteristic $\phi$-structure
of beam backgrounds\,\cite{backgroundpaper2011}, which suggests that it is not random noise or \agpp{}, but real background. This suggests 
significantly higher levels of background from de-bunched ghost charge after the chromaticity changes, which will be further discussed in section\,\ref{sect:ghostBG}.
\end{itemize}

\begin{figure}
\centering 
\mbox{
\subfigure[]{
\includegraphics[width=0.49\textwidth]{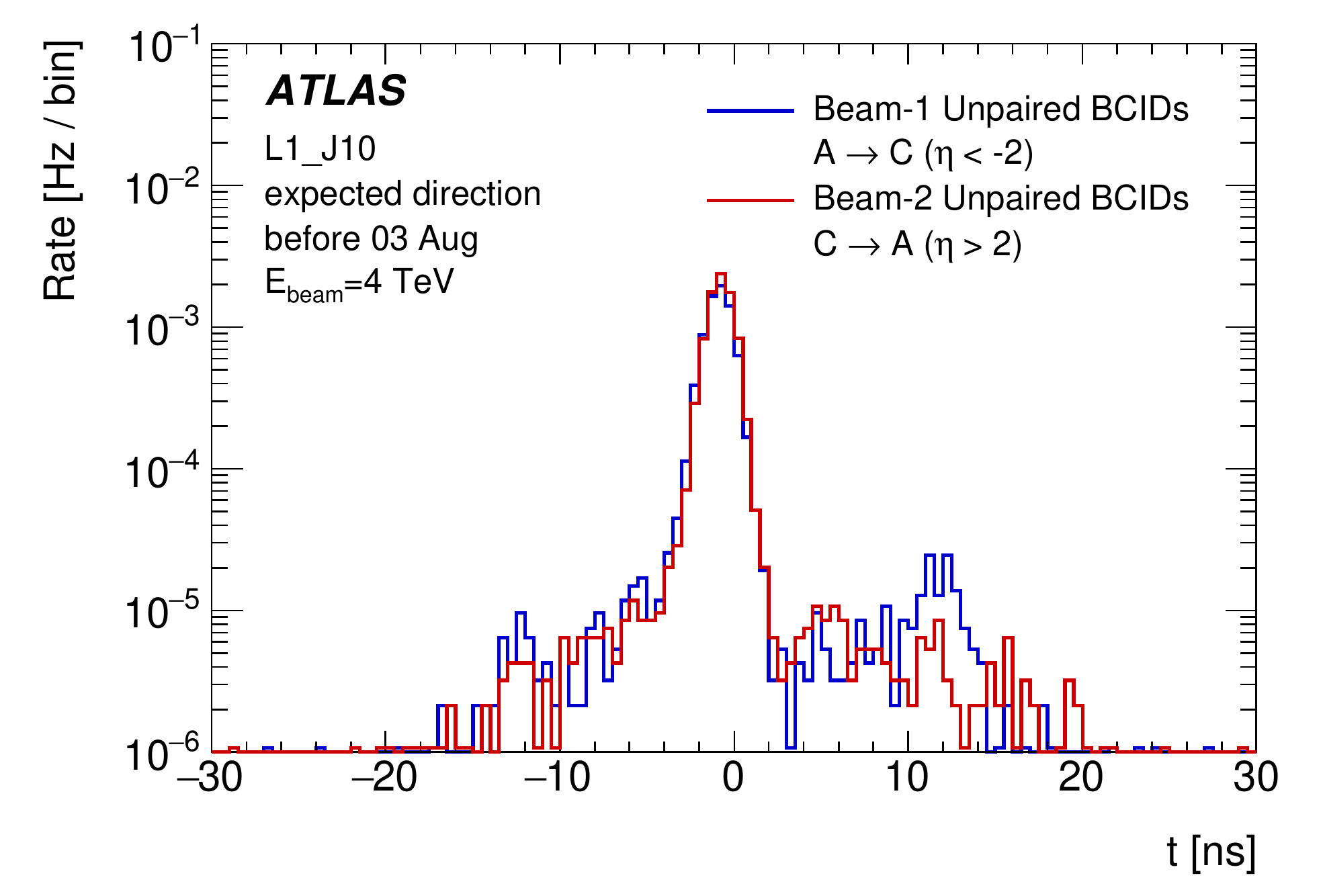}
\label{fig:ghostJetEta2a}
}
\subfigure[]{
\includegraphics[width=0.49\textwidth]{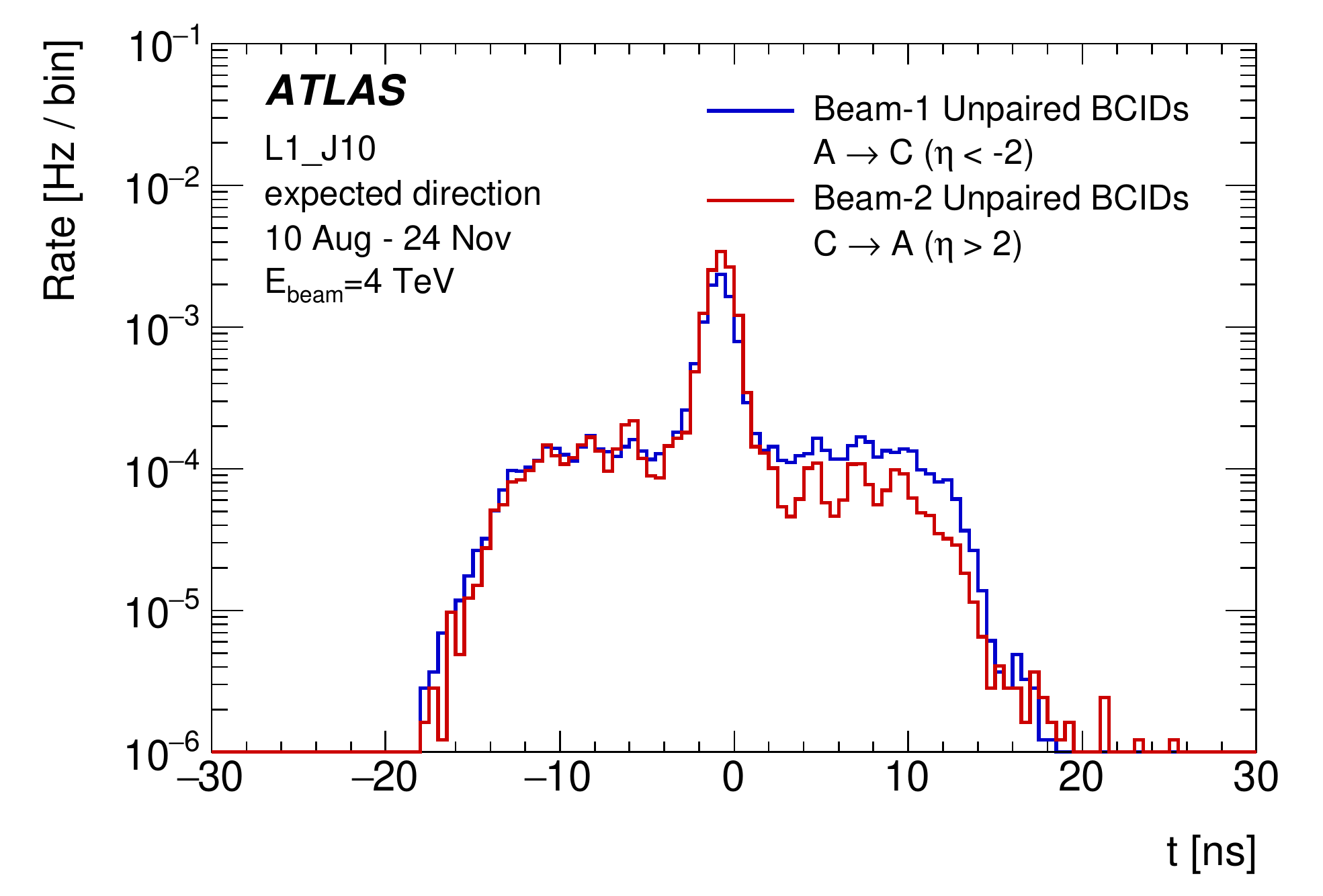}
\label{fig:ghostJetEta2b}
}
}
\caption{Jet times for fake jets at $|\eta|>2$ without a reconstructed vertex for the period before 3 August (a) and
between 10 August and the swap of the unpaired trains (b). Only fills with 1368 colliding bunches are considered. 
The plots are normalised by the live-time, the rate is averaged over the six unpaired BCIDs per beam and the bin width is 0.5\,ns.
}
\label{fig:ghostJetEta2}
\end{figure}

The explanation for an almost uniform jet distribution in the $\eta$--$t$ plane comes from the fact that the de-bunched ghost charge 
is equally distributed among all RF buckets and not just the nominal one.
In practice one expects a ``banana'' signature from the ghost bunch  every 2.5\,ns. The pedestal appears uniform because the
jet times are smeared out at central $\eta$. But focusing on $|\eta|>2$ the timing signal of the outgoing beam background is sharp enough to
resolve a detailed time structure of RF buckets, as shown in figure\,\ref{fig:ghostJetEta2}. A comparison of figures\,\ref{fig:ghostJetEta2a} and \ref{fig:ghostJetEta2b}
illustrates the appearance of the uniform pedestal and the RF bucket structure which is characteristic of de-bunched ghost charge.

Another striking feature in figure\,\ref{fig:ghostJetTimingc}, compared to figure\,\ref{fig:ghostJetTiminga}, is the appearance of a pronounced opposite 
direction ``banana'' in the nominal RF bucket occupied by unpaired bunches of beam-1. This is due to background associated with beam-2 ghost bunches (ghost-BIB) and the reasons 
for the appearance after the chromaticity changes will be discussed in section\,\ref{sect:ghostBG}.
No such ``banana'' in beam-1 direction is observed in figure\,\ref{fig:ghostJetTimingb}, nor in figure \ref{fig:ghostJetTimingd}.

\begin{figure}
\centering 
\mbox{
\subfigure[]{
\includegraphics[width=\textwidth]{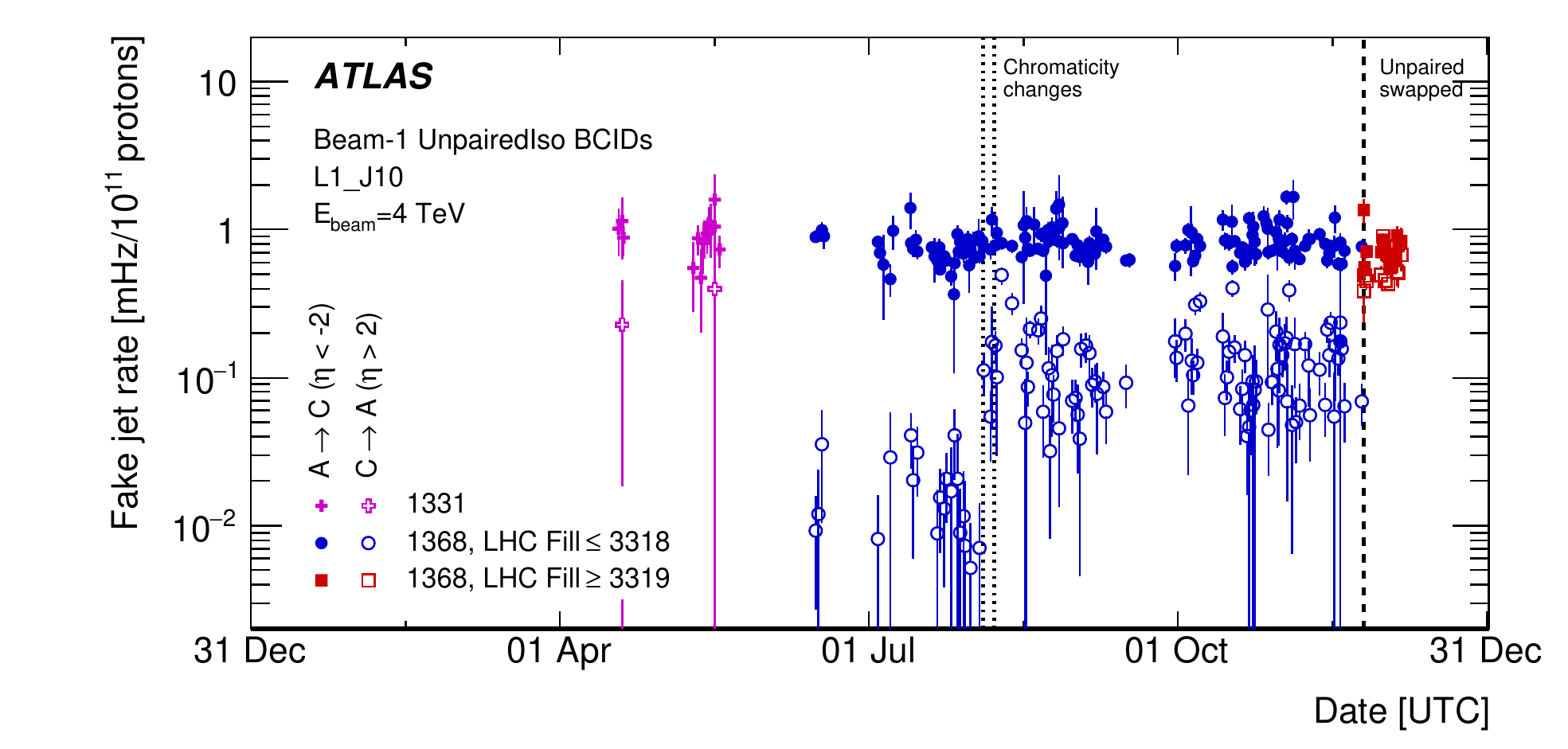}
\label{fig:ghostFakeJetYeara}
}
}\\
\mbox{
\subfigure[]{
\includegraphics[width=\textwidth]{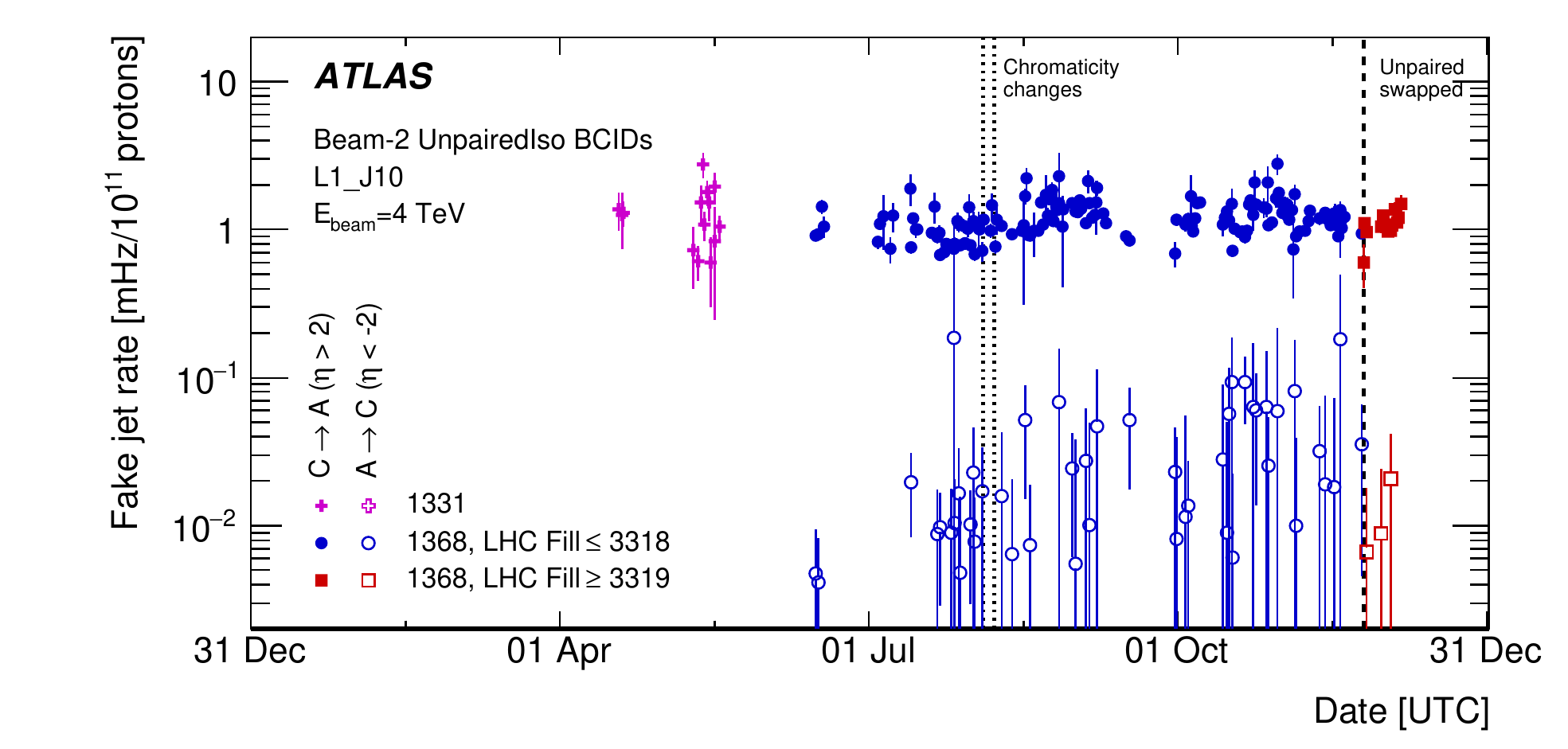}
\label{fig:ghostFakeJetYearb}
}
}
\caption{Rate of fake jets in the nominal RF buckets ($-2\,$ns $<t<0\,$ns) associated with 
beam-1 ($\eta<-2$) and beam-2 ($\eta>2$)
are shown for the unpaired isolated bunches defined for beam-1 (a) and beam-2 (b). 
Each point corresponds to one LHC fill. Solid symbols represent jets with a timing 
consistent with the beam direction, while open symbols show jets with timing in the
opposite direction.
Events with jets at $4\,$ns $<t<6\,$ns are subtracted in order to correct for the pedestal.
A vertex veto is applied to remove ghost collision contributions. The vertical lines
indicate important changes in the beam conditions, as detailed in table\,\protect\ref{breakpoints}.
}
\label{fig:ghostFakeJetYear}
\end{figure}

Figure\,\ref{fig:ghostFakeJetYear} shows the rate of fake jets at $|\eta|>2$ as a function of time, where contributions from beam-1 (beam-2) are expected at negative (positive) $\eta$.
The contribution from the nominal RF bucket is enhanced by restricting the jet time window to $-2\,\textrm{ns}<t<0\,\textrm{ns}$ and subtracting the pedestal contribution, 
estimated from $4\,\textrm{ns}<t<6\,\textrm{ns}$ (see figure\,\ref{fig:ghostJetEta2}).
Figure\,\ref{fig:ghostFakeJetYear} also shows the rate of jets in the direction opposite to the unpaired bunch, which 
can be attributed to ghost-BIB. In figure\,\ref{fig:ghostFakeJetYeara} these ghost-BIB rates show  a sudden jump in 
early August, when the LHC chromaticity change took place, consistent with the appearance of the opposite direction ``banana'' 
in figure\,\ref{fig:ghostJetTimingc}. 
The opposite direction background in beam-1 unpaired bunch positions, shown in
figure\,\ref{fig:ghostFakeJetYearb} remains at low level throughout the year with
no evident jump in mid-August. A detailed discussion of these observations is deferred to section\,\ref{sect:wrongDirJets}.

\subsection{Non-collision backgrounds in colliding bunches}
\label{sec:filled}

Fake jet rates can also be measured in the colliding bunches where they lead to
events with \MET{} balancing the transverse momentum of the fake jet, overlaid on top of minimum-bias interactions.
Such events are 
recorded by \MET{} triggers. In the following, the lowest 
unprescaled 
\MET{} trigger available
throughout 2012 with an $80\,\GeV$ threshold at L1 will be used together with
an offline selection of $\MET>160\,\GeV$ and a requirement of a jet with $\pt>120\,\GeV$ and $|\eta|<2.5$.

The missing transverse momentum is calculated from the vector
sum of the measured muon momenta and 
reconstructed calorimeter-based objects (electrons, photons, taus, and jets),
as well as calorimeter energy clusters within $|\eta|<4.9$\,\cite{met} that are not associated to any of these objects.
Energy deposits reconstructed as tau leptons are calibrated at the jet energy scale.

Noise spikes are mostly suppressed by standard quality requirements at the data reconstruction stage, described in section\,\ref{sec:calojets}.
The remaining significant noise contributions are identified in the $\eta$--$\phi$ distribution of jets
and the corresponding regions are masked in the offline analysis.

Multi-jet processes where a jet is mismeasured 
are efficiently suppressed by requiring a minimum azimuthal separation 
of $\Delta\phi(\textrm{jet},\MET)>0.5\,$rad between the missing transverse momentum direction and all jets with $\pt>30\,\GeV$ and $|\eta| <4.5$.

\begin{figure}[ht]
\centering 
\mbox{
\subfigure[]{
\includegraphics[width=0.49\textwidth]{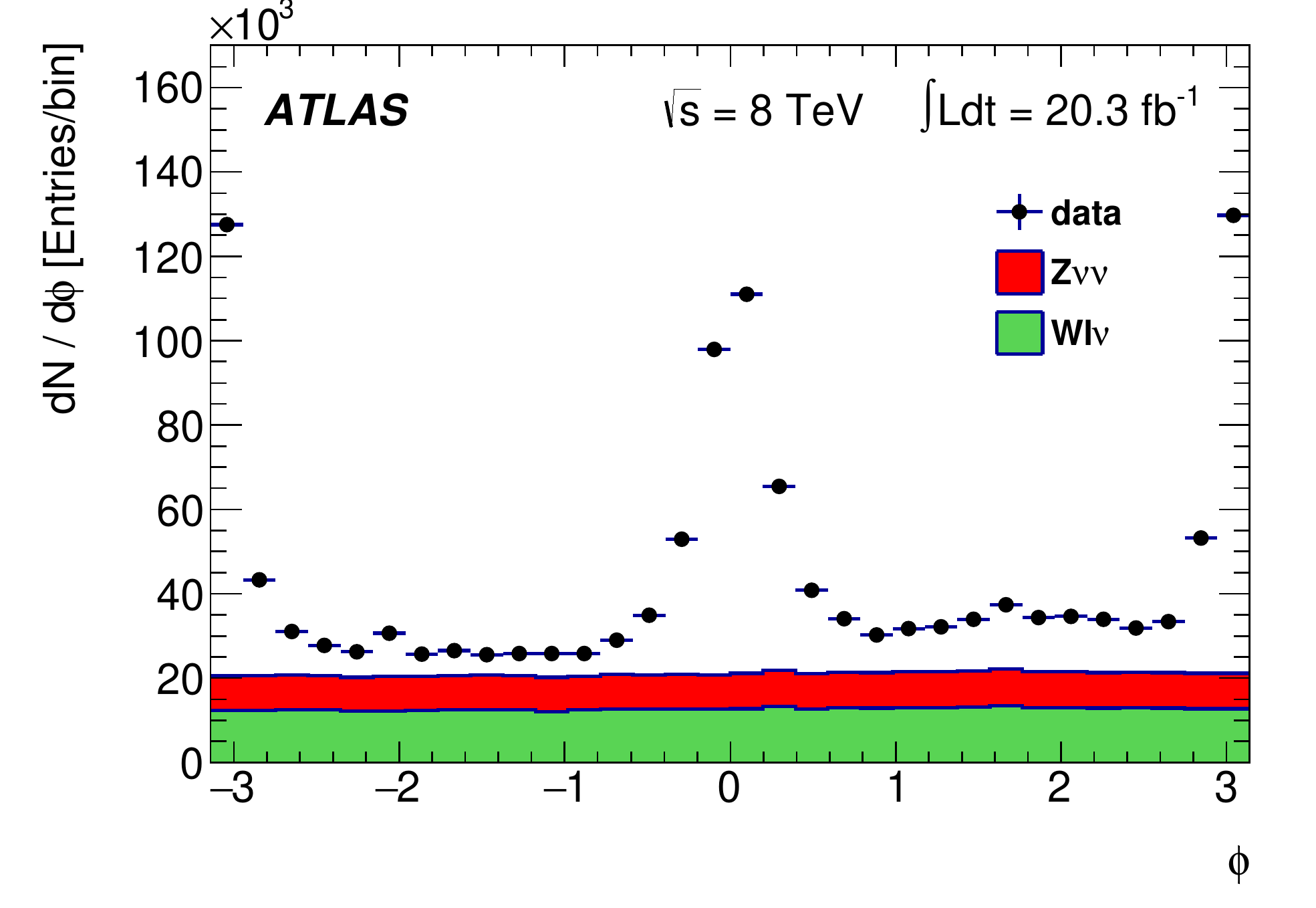}
\label{fig:bibjet_filled_alla}
}
\subfigure[]{
\includegraphics[width=0.49\textwidth]{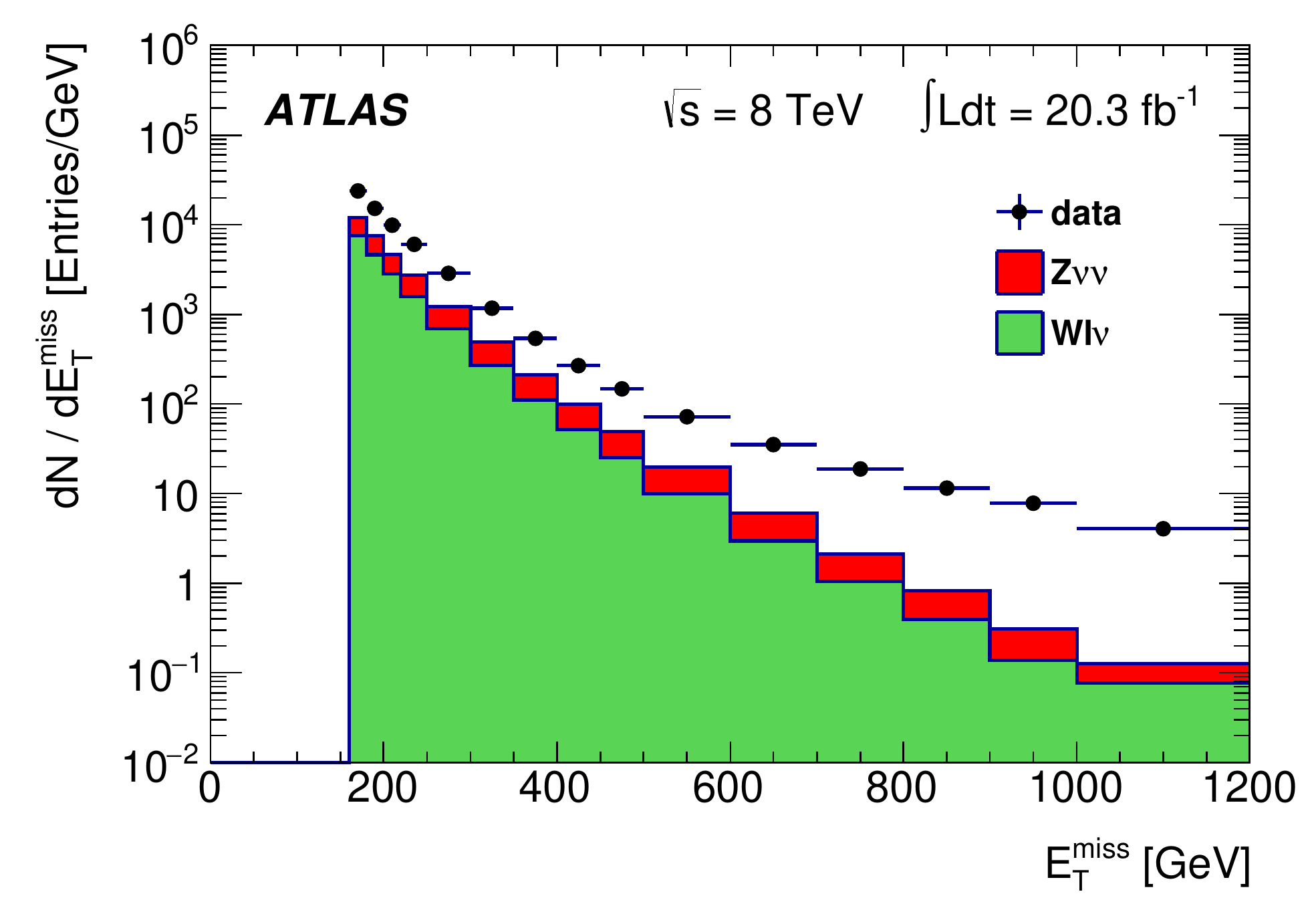}
\label{fig:bibjet_filled_allb}
}
}
\caption{
Distributions of the leading jet $\phi$ (a) and \MET{} (b) in the events from colliding bunches triggered by the \MET{} trigger with an offline requirement of $\MET>160\,\GeV$.
Data is compared to the Standard Model expectation from the $Z\rightarrow\nu\nu$+jets and $W\rightarrow l\nu$+jets electroweak processes. Other $pp$ collision products, such as top and dibosons, that contribute to the total expectation by less than approximately 10\% over the whole $\MET$ spectrum are not shown.
}
\label{fig:bibjet_filled_all}
\end{figure}

The dominant Standard Model processes passing this selection are
$Z\rightarrow\nu\nu$+jets and $W\rightarrow l\nu$+jets.
Further processes involving top quarks and dibosons, that contribute to the total Standard Model expectation by less than approximately 10\% over the whole $\MET$ spectrum, are neglected in this study.
Figure~\ref{fig:bibjet_filled_all} compares Monte Carlo (MC) simulations and recorded events in 2012 data, passing the selection described above.
In the MC simulations the $Z\rightarrow\nu\nu$+jets and $W\rightarrow l\nu$+jets electroweak processes are
generated using Sherpa~1.4.1~\cite{Gleisberg:2008ta}, including leading-order matrix elements for up to five partons in the final state and assuming massive $b$/$c$-quarks, with the CT10~\cite{Lai:2010vv} parton distribution functions.
The Monte Carlo expectations are normalised to next-to-next-to-leading-order (NNLO) perturbative QCD predictions using
DYNNLO~\cite{wcrosssection1, wcrosssection} and  MSTW2008 NNLO parton distribution function set~\cite{mstw}.
The generated events are interfaced with the GEANT4\,\cite{Agostinelli:2002hh} detector simulation.

Approximately 1.4 million data events are selected from the full 2012 dataset,
out of which more than half correspond to non-collision backgrounds.
The azimuthal distribution of the leading jet, shown in figure\,\ref{fig:bibjet_filled_alla}, exhibits clear 
spikes at $\phi=0$ and $\pi$, that are characteristic for beam-induced backgrounds\,\cite{backgroundpaper2011}, 
on top of the uniform distribution from the $pp$ collision products described by the Monte Carlo simulation.
Figure\,\ref{fig:bibjet_filled_allb} shows that the spectrum of fake jets in the \MET{} distribution is harder than the one 
expected from the electroweak processes.

The plots in figure\,\ref{fig:bibjet_filled_all} reveal that with a simple selection, based on jets and \MET{} only, almost twice as many events were 
selected than predicted by Monte Carlo simulations.
In the context of $pp$ collision data analyses, this illustrates how crucial it is to design an efficient jet cleaning strategy.
In order to reduce this large NCB contribution to a sub-percent level, a suppression power of approximately $10^3$ is needed (see reference\,\cite{backgroundpaper2011} for further discussion).

Fake jets due to noise or BIB muons have the common feature that there are usually no tracks
connecting the measured calorimeter signal with the primary vertex. This is also true for fake jets induced by CRB,
provided the path of the CRB-muon is not passing close to the IP. Furthermore, both noise and calorimeter deposits due to BIB muons
often are contained in a single layer of the barrel calorimeter.
These characteristics of NCB motivate use of the following quantities for the jet cleaning:
\begin{itemize}
\item Charged particle fraction, $\fch$, which is the fraction of the total transverse momentum of a jet coming from tracks with $\pt > 500\,\MeV$.
\item Maximum energy fraction in any calorimeter layer, $\fmax$.
\end{itemize}

\begin{figure}[ht]
\centering 
\mbox{
\subfigure[]{
\includegraphics[width=0.49\textwidth]{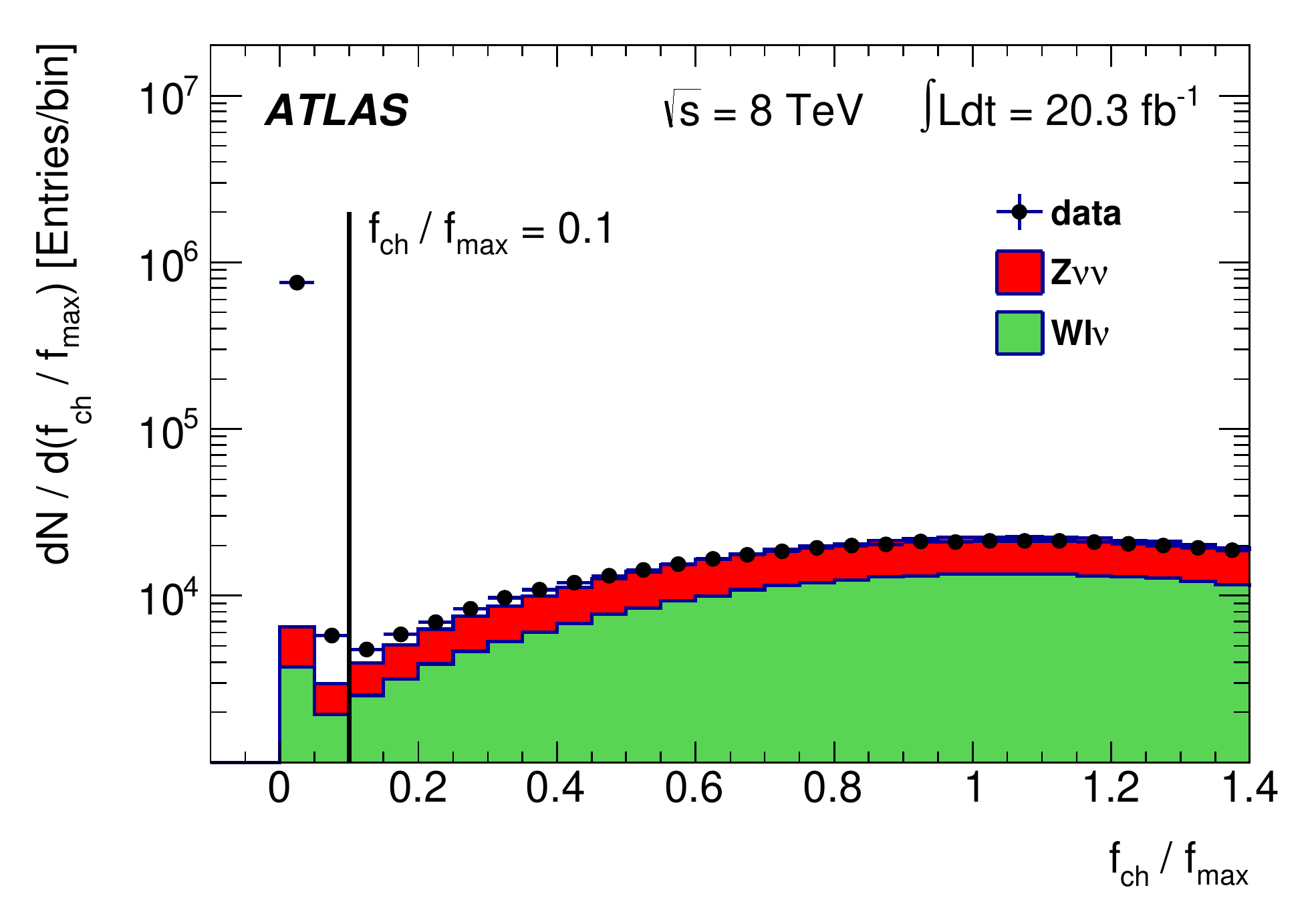}
\label{fig:bibjet_filled_ncba}
}
\subfigure[]{
\includegraphics[width=0.49\textwidth]{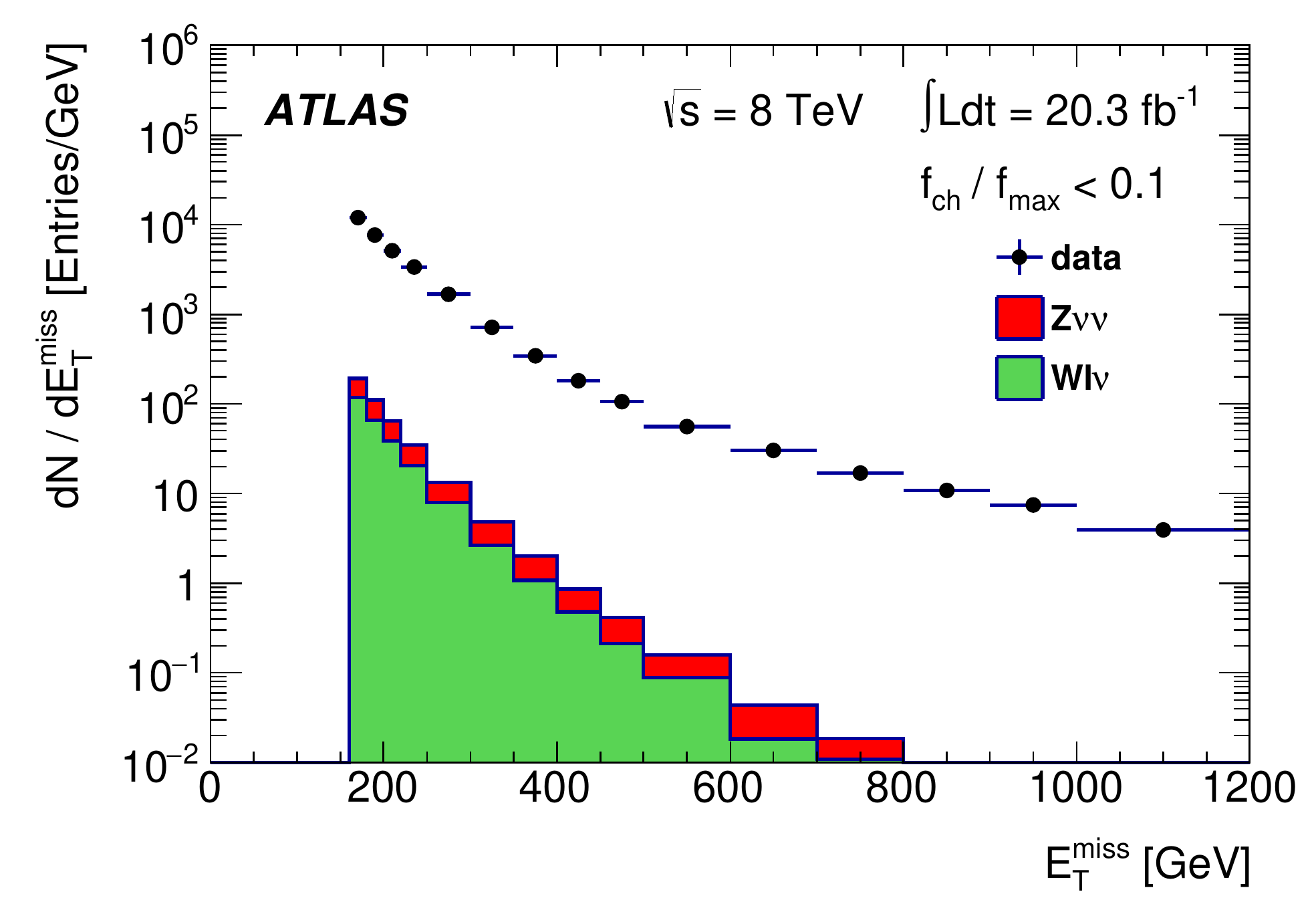}
\label{fig:bibjet_filled_ncbb}
}
}
\caption{
The distribution of the ratio $\fch/\fmax$ for leading jets in the events from colliding bunches triggered by the \MET{} trigger with an 
offline requirement of $\MET>160\,\GeV$ (a). The value $\fch/\fmax=0.1$, used to separate NCB from $pp$ collision processes, is indicated by the black line. 
The \MET{} distribution for the NCB events determined by the $\fch/\fmax<0.1$ selection (b).
Data is compared to the Standard Model expectation from the $Z\rightarrow\nu\nu$+jets and $W\rightarrow l\nu$+jets electroweak processes. Other $pp$ collision products, such as top and dibosons, that contribute to the total expectation by less than approximately 10\% over the whole $\MET$ spectrum are not shown.
}
\label{fig:bibjet_filled_ncb}
\end{figure}

The distribution of the ratio $\fch/\fmax$ for leading jets, shown in figure~\ref{fig:bibjet_filled_ncba}, suggests that $\fch/\fmax>0.1$ is
an efficient cleaning selection.
While for higher values the data are well described by the Monte Carlo simulation, NCB cause an excess at low values, which amounts to about two orders of magnitude with respect to the simulation.
By requiring $\fch/\fmax<0.1$, a sample of fake jets can be extracted for which the Monte Carlo simulation predicts a 1\% contamination from $pp$ collision processes.
Figure\,\ref{fig:bibjet_filled_ncbb} shows that the purity of BIB events becomes even higher with increasing \MET{}.

\begin{figure}[h!]
\centering 
\includegraphics[width=\textwidth]{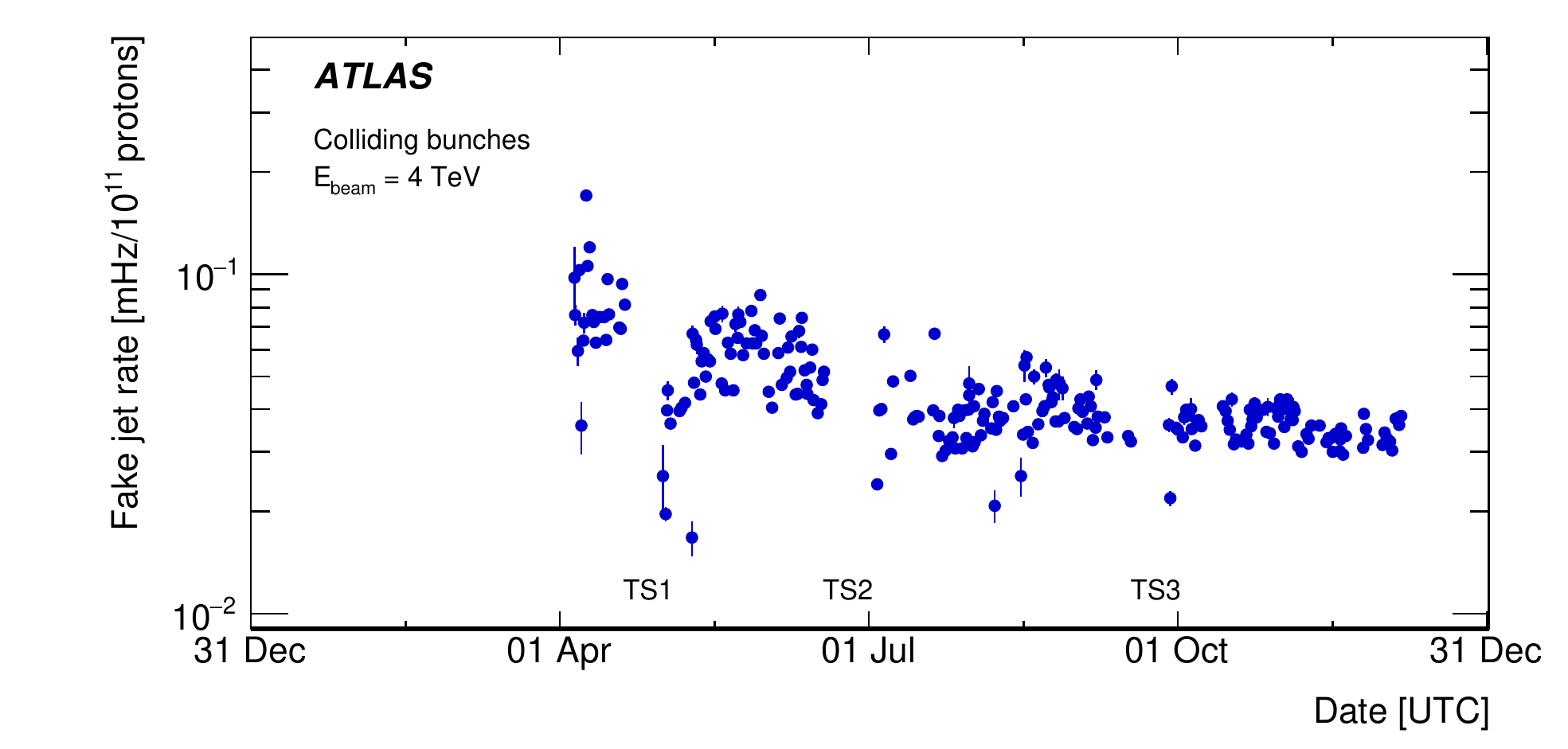}
\caption{The rate of NCB in colliding bunches triggered by the \MET{} trigger with an offline selection of $\MET>160\,\GeV$.
The $\sim1\%$ contamination of $pp$ collision products is not subtracted.
Each point corresponds to one LHC fill. The technical stops are indicated in the plot.
}
\label{fig:bibjet_rate_paired}
\end{figure}

\begin{figure}[h!]
\centering 
\includegraphics[width=\textwidth]{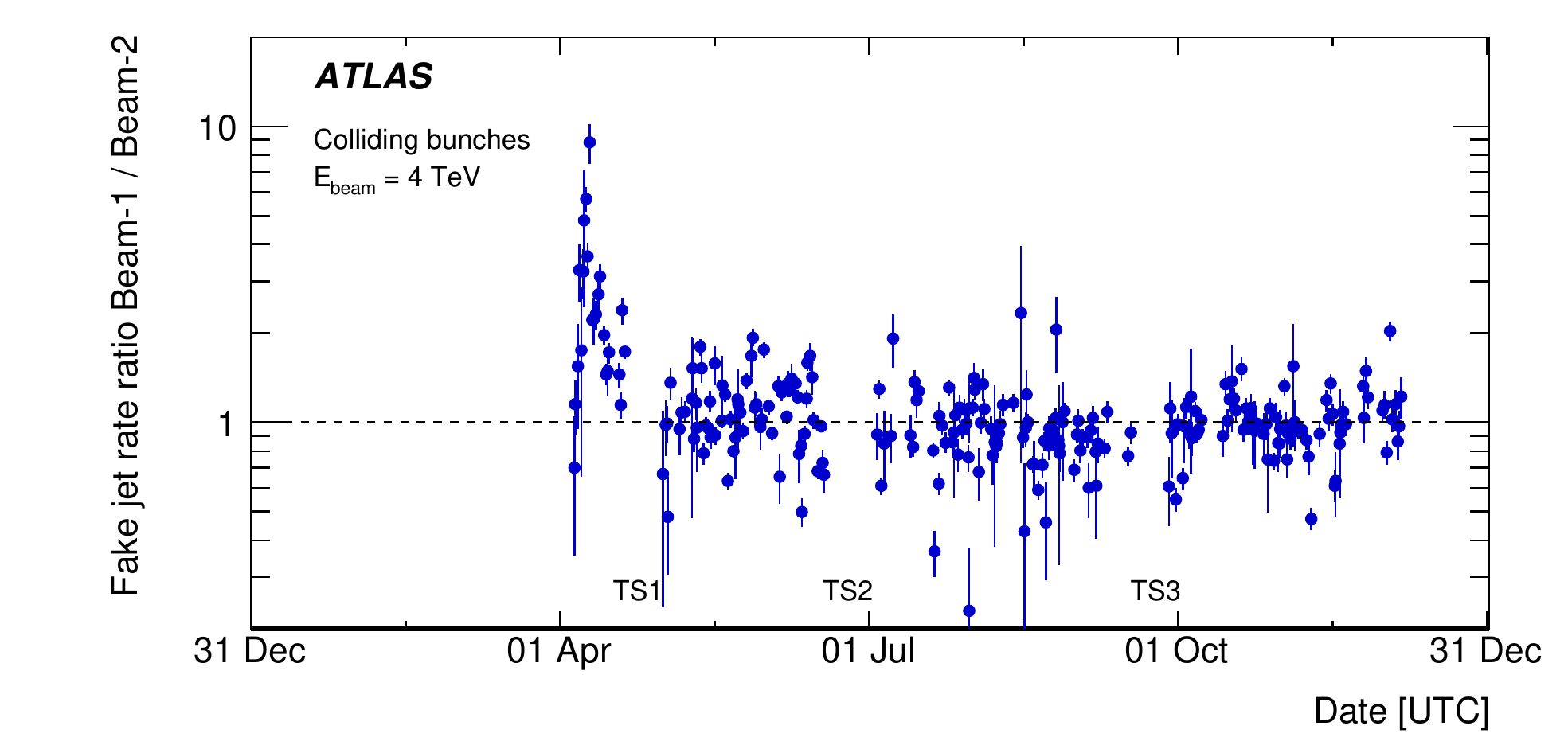}
\caption{Evolution of the ratio of the NCB levels associated to beam-1 and beam-2 in the events triggered by the \MET{} trigger with an offline selection of $\MET>160\,\GeV$ in colliding bunches. The beams are separated according to the 
leading jet $\eta$, where the beam-1 (beam-2) contribution is taken from $\eta<-1.2$ ($\eta>1.2$).
Each point corresponds to one LHC fill. The technical stops are indicated in the plot.
}
\label{fig:bibjet_ratio_paired}
\end{figure}

Figure\,\ref{fig:bibjet_rate_paired} shows the evolution of the NCB rate observed in the colliding bunches using this inverted cleaning selection.
A gradual decrease of fake rates is observed in the early part of the year, possibly related to continued vacuum conditioning by beam scrubbing. 
The rates become stable after the June technical stop (TS2) where the fill pattern with 1368 colliding bunches was mostly used.
The early drop is not observed in figure\,\ref{fig:ghostFakeJetYear}, but it should be noted that figure\,\ref{fig:bibjet_rate_paired} requires 
much more energetic jets. The differences, therefore, could point at a different origin of the BIB generating the jets.
As demonstrated in figure\,\ref{fig:ghostJetTiming}, the contribution from individual beams can be isolated at high $\eta$.
Since the contribution from BIB in the selected fake jet sample dominates over CRB at low $\MET$ (see section\,\ref{sec:cosmic}),
this allows studying the difference of BIB rates associated with the two beams. 
In figure\,\ref{fig:bibjet_ratio_paired}
similar rates are observed in both beams after the April technical stop (TS1) while the beam-1 rate is significantly higher at the beginning of 
2012 data-taking. 

\subsection{Fake jets in cosmic-ray events}
\label{sec:cosmic}

In analogy to BIB, CRB muons entering ATLAS can generate fake jets by radiative processes in the calorimeters.
The rate and properties of such jets have been studied using two dedicated data-samples taken in November 2012 while the LHC machine was in cryogenics 
recovery, i.e. during which beams were not present in the LHC.
The CRB data were recorded with all ATLAS detector sub-systems operational and a total live-time of 30.4 hours.

The fake jet properties are investigated in the data selected by a single-jet L1 trigger with a jet $\pt$ threshold of 30\,\GeV.
An offline jet transverse momentum requirement of $\pt>40\,\GeV$ ensured full trigger efficiency. 
The trigger was active in 3473 out of 3564 BCIDs and data rates quoted below are corrected for this 2.5\% live-time inefficiency.

The rate of CRB muons decreases rapidly with energy \cite{Gaisser:1990vg}. Over most of the energy range bremsstrahlung
is the most important radiative process that can result in a large local energy loss. The cross sections of
all radiative processes increase slowly with muon energy. The bremsstrahlung spectrum follows roughly 
a $1/\nu$-dependence, where $\nu$ is the fraction of the muon energy transferred to the photon\,\cite{Kelner:1997cy}.
Most of the energy loss of muons is due to continuous processes and since they are effectively minimum ionising particles, they are expected to lose on average $\sim$40\,\GeV{} 
when passing through the overburden from the surface to the ATLAS cavern. However the presence of the access shafts 
allows lower energy muons to reach ATLAS. Several samples of Monte Carlo simulated CRB muon events at the surface are therefore generated in different energy ranges, 
covering $50\,\GeV<E_{\mu}<100$\,\TeV,
and  simulated with 
GEANT4.
The samples are normalised to the differential CRB muon flux parametrised in reference~\cite{Dar:1983pt}.

\begin{figure}[h]
\centering 
\includegraphics[width=0.7\textwidth]{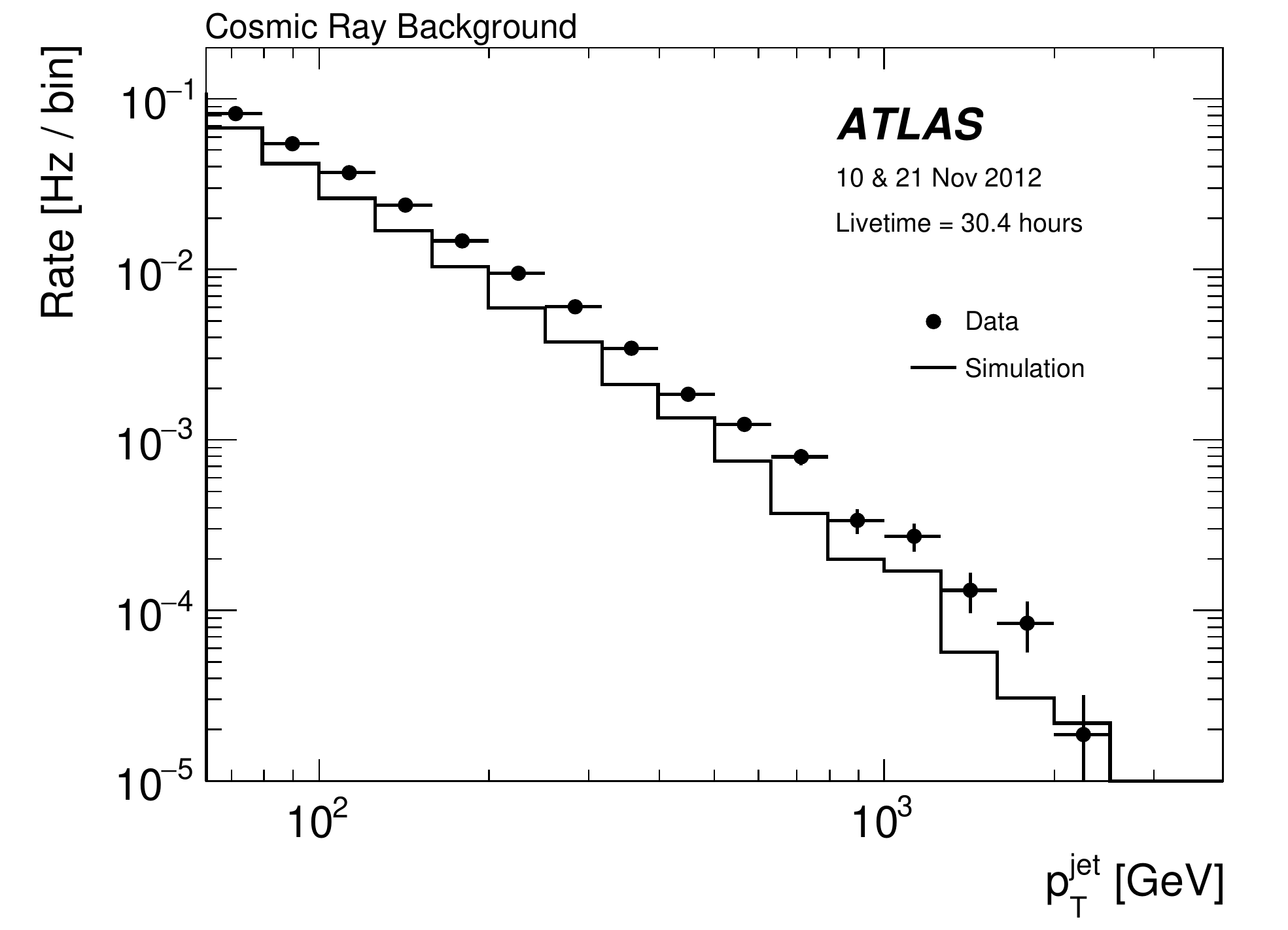}
\caption{Distributions of the transverse momentum of the leading reconstructed jet in the selected events from the dedicated CRB data-taking.
Data are compared to the CRB-muon Monte Carlo simulation.
}
\label{fig:cr_jetlog10pt}
\end{figure}

The distributions of the apparent transverse momentum, calculated with respect to the nominal beamline, are shown in figure\,\ref{fig:cr_jetlog10pt} for the leading reconstructed jets in selected events.
The total rates in data and Monte Carlo agree to within $\sim$50\%. While such a discrepancy would be considered large for simulation of $pp$-collisions, it is reasonably good
agreement for this kind of simulation. The uncertainties in the shape and normalisation of the parametrisation of the muon flux at the surface are about 10\%. More
significant uncertainties are related to the muon transport through the overburden, i.e. the exact density and composition of the $\sim$60\,m thick soil above the experiment.
The principal message of figure\,\ref{fig:cr_jetlog10pt} is, that the simulations describe well the spectrum of the CRB related fake jets up to the highest apparent $\pt$-values.

\begin{figure}[h]
\centering 
\includegraphics[width=0.7\textwidth]{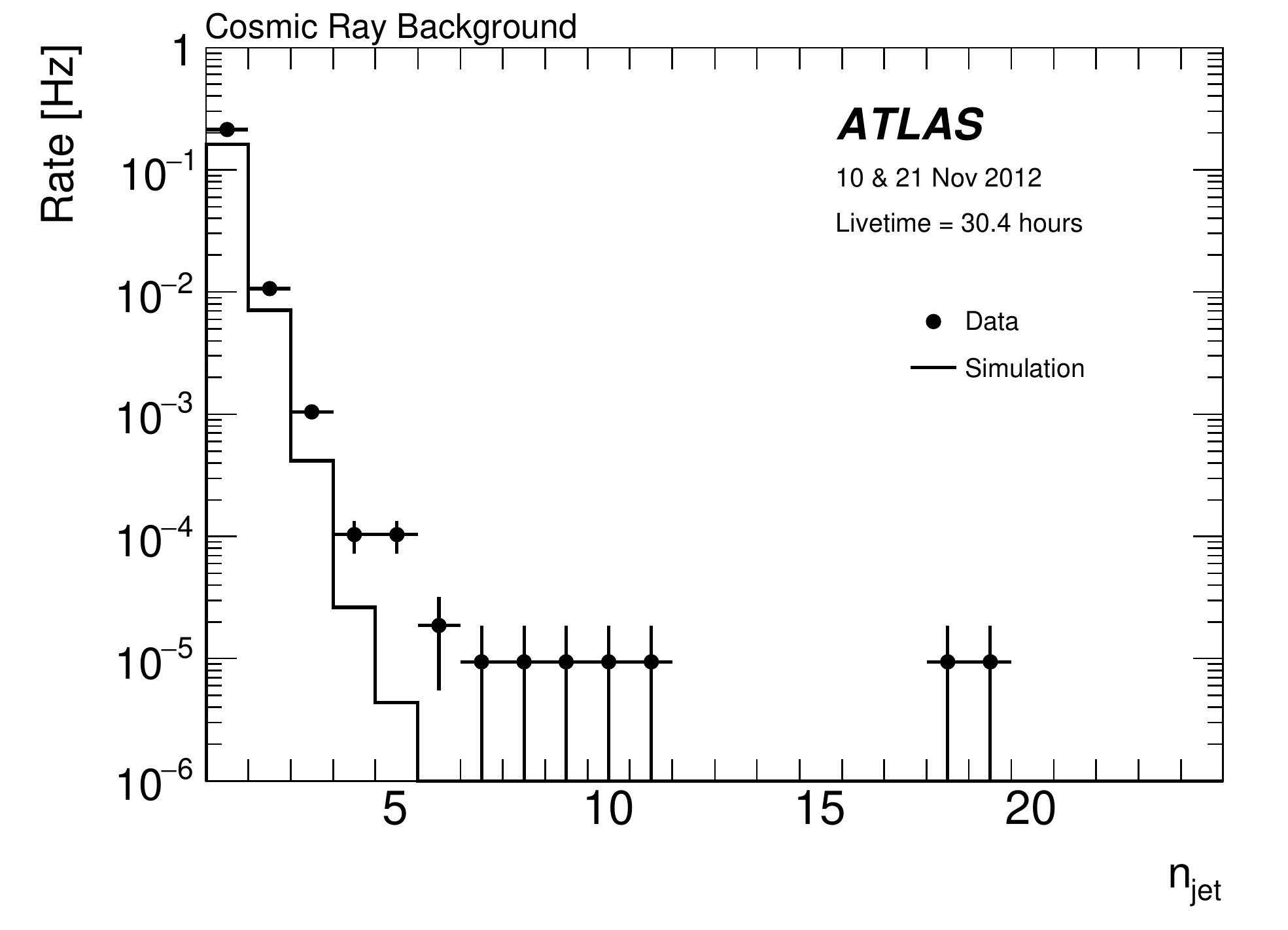}
\caption{Distribution of the multiplicity of jets with $\pt>10\,\GeV$ and $|\eta|<4.5$ in events containing at least one jet with $\pt>40\,\GeV$,
using the dedicated CRB data, compared to the CRB-muon Monte Carlo simulation.}
\label{fig:cr_njet}
\end{figure}

Figure\,\ref{fig:cr_njet} shows the distribution of jet multiplicities ($n_{\rm jet}$) for selected events containing at least one jet with $\pt>40\,\GeV$.  
The data and Monte Carlo distributions agree to within $\sim$\,30\% for $n_{\rm jet}<3$, indicating that the Monte Carlo simulation of single CRB muons provides a 
reasonable representation of the data at low jet multiplicities. 
The deviation of the absolute normalisation is related to that observed in figure\,\ref{fig:cr_jetlog10pt}. 
For jet multiplicities above two, however, the number of selected data events significantly exceeds 
the Monte Carlo expectations. This is to be expected due to the presence of multiple-muon events in the data generated by extensive air 
showers~\cite{Achard:2004ws,Avati:2000mn,Abdallah:2007fk}, which are not modelled by the Monte Carlo generator.

\begin{figure}[t!]
\centering 
\includegraphics[width=0.7\textwidth]{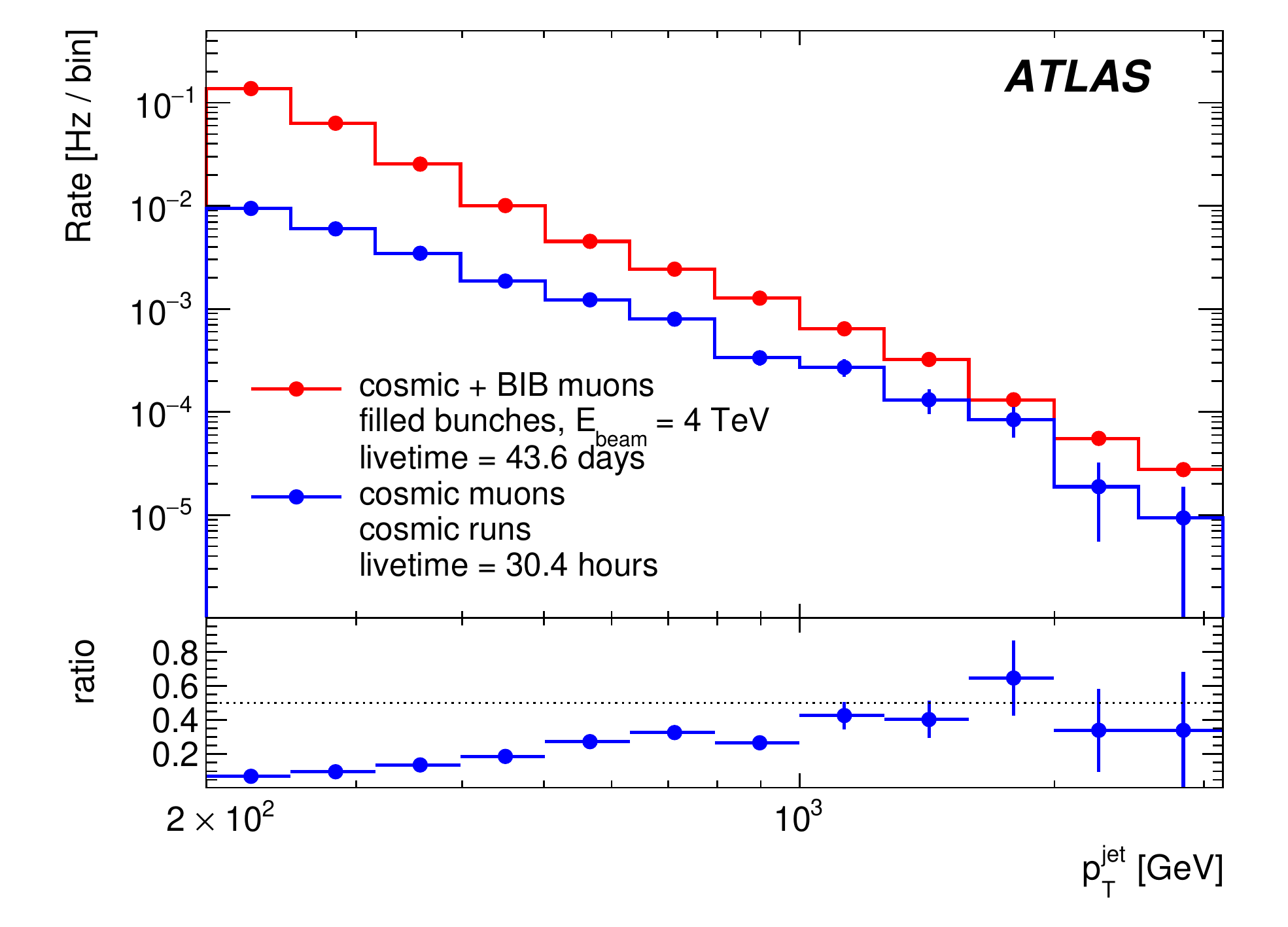}
\caption{Distributions of the leading jet transverse momentum in the fake jet samples obtained from the colliding bunches and the dedicated CRB data.
The ratio indicates the fraction of CRB in the NCB sample.
Only the fills with 1368 colliding bunches are taken in the former case.
}
\label{fig:fakerate}
\end{figure}

The fake jet rates from the dedicated CRB data are also directly compared to the rates of NCB obtained from the colliding bunches in section\,\ref{sec:filled}. 
For this, additional selection criteria are applied in the CRB data in order to ensure consistency of the fiducial volumes of the two samples. Therefore, the CRB data are 
further restricted to contain only leading jets in the tracker acceptance $|\eta|<2.5$ that pass the inverted cleaning selection $\fch/\fmax<0.1$ used to select the NCB events 
from the colliding bunches. 
Furthermore, the same kinematic and topological selection as in section\,\ref{sec:filled} is imposed.
A comparison of the rates is shown in figure\,\ref{fig:fakerate} for jet $\pt>200\,\GeV$. The higher jet $\pt$ threshold is chosen in order to avoid turn-on effects
caused by the $\MET$ selection in the CRB events where both the muon and the induced fake jet are reconstructed in such a way that the two objects 
weigh against each other in the $\MET$ calculation.
The NCB rate from the colliding bunches is corrected for the live-time of the 2012 data-taking with 1368 colliding bunches. As in the case 
of the rate from the CRB data, the rate is scaled up such that it corresponds to all 3564 BCIDs being filled with a nominal bunch.
As stated in section~\ref{sec:filled}, the noise contribution in the NCB rate is suppressed, i.e. the dominant components 
are fake jets from BIB and CRB.
The plot shows that the fake jet rate from BIB is 10 times higher than from CRB at $\pt\sim200\,\GeV$ 
but the difference gradually decreases with increasing $\pt$.
Beyond $\pt\sim 1\,\TeV$, the CRB fake jets contribute at a similar level as BIB jets, although the CRB dataset becomes statistically limited there, preventing a firm statement.

%% file: ghost.tex
\section{BIB from ghost charge}
\label{sect:ghostBG}

Although the intensity of ghost bunches is very low, they still can produce BIB, just like nominal bunches. 
When other contributions are sufficiently suppressed by suitable selection, those small BIB signals can be seen both by the
BCM and as fake jets. This section will describe several observations made on those ghost-BIB signals.

\subsection{BCM background from ghost charge}
\label{sect:bcmGhostBG}

In section\,\ref{sect:beamgas} it was shown that the L1\_BCM\_AC\_CA background for unpaired bunches is dominated by beam-gas. If this were
true also for ghost bunches, then the BIB rate from them should scale with intensity.

It was already alluded that the small early peaks in downstream modules, seen in figure\,\ref{fig:bcmresponse}, can be attributed to BIB from
ghost bunches in the opposite beam. However, the relative heights of those peaks with respect to the peaks from the unpaired bunches appear too
high with respect to typical injected ghost bunch intensities, which according to figure\,\ref{fig:ldmplot} are at least a factor $10^{3}$ lower
than those of unpaired bunches.

\begin{figure}
\centering 
\mbox{
\subfigure[]{
\includegraphics[width=0.49\textwidth]{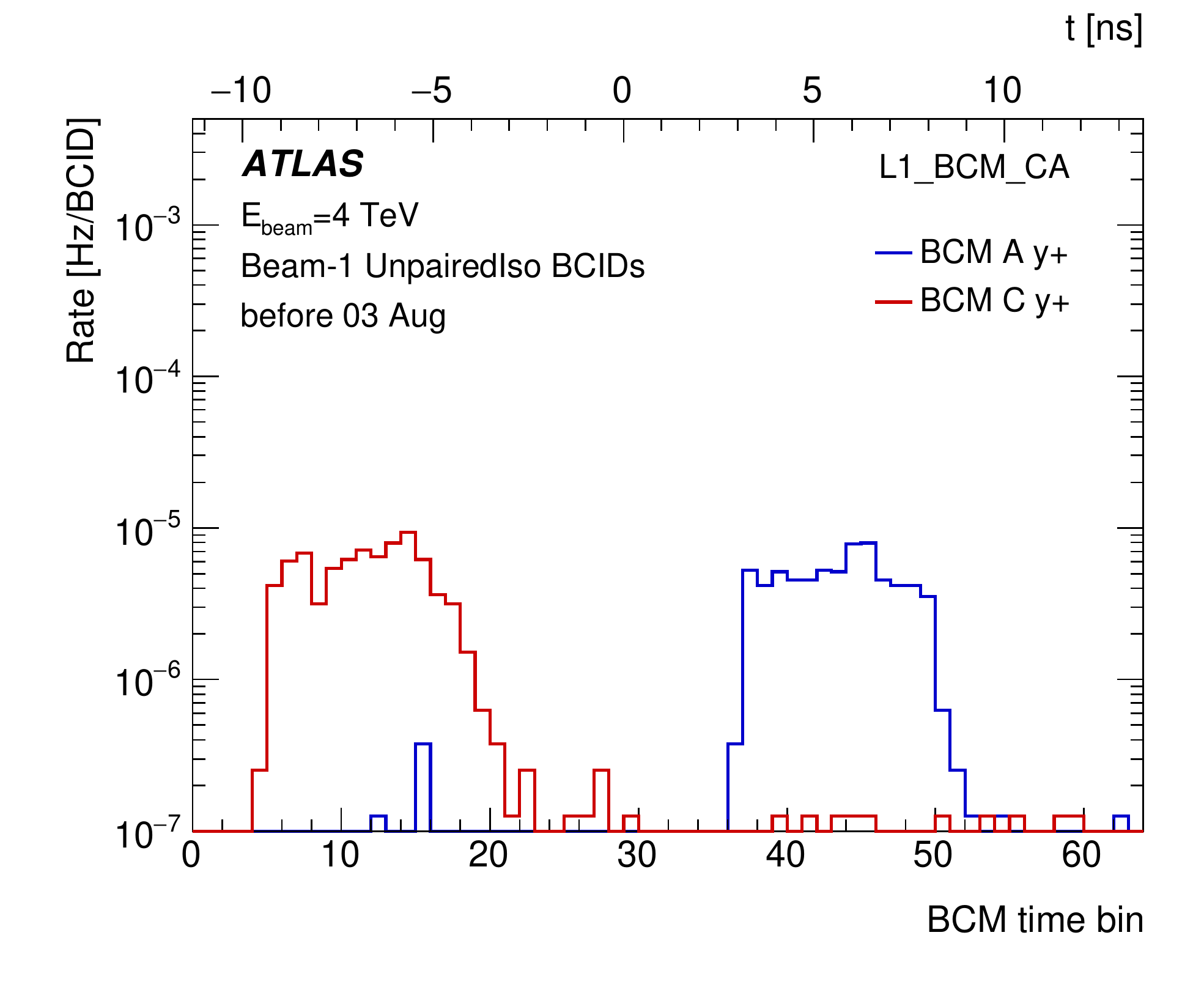}
\label{fig:bcmbeamgas-dir2-4a}
}
\subfigure[]{
\includegraphics[width=0.49\textwidth]{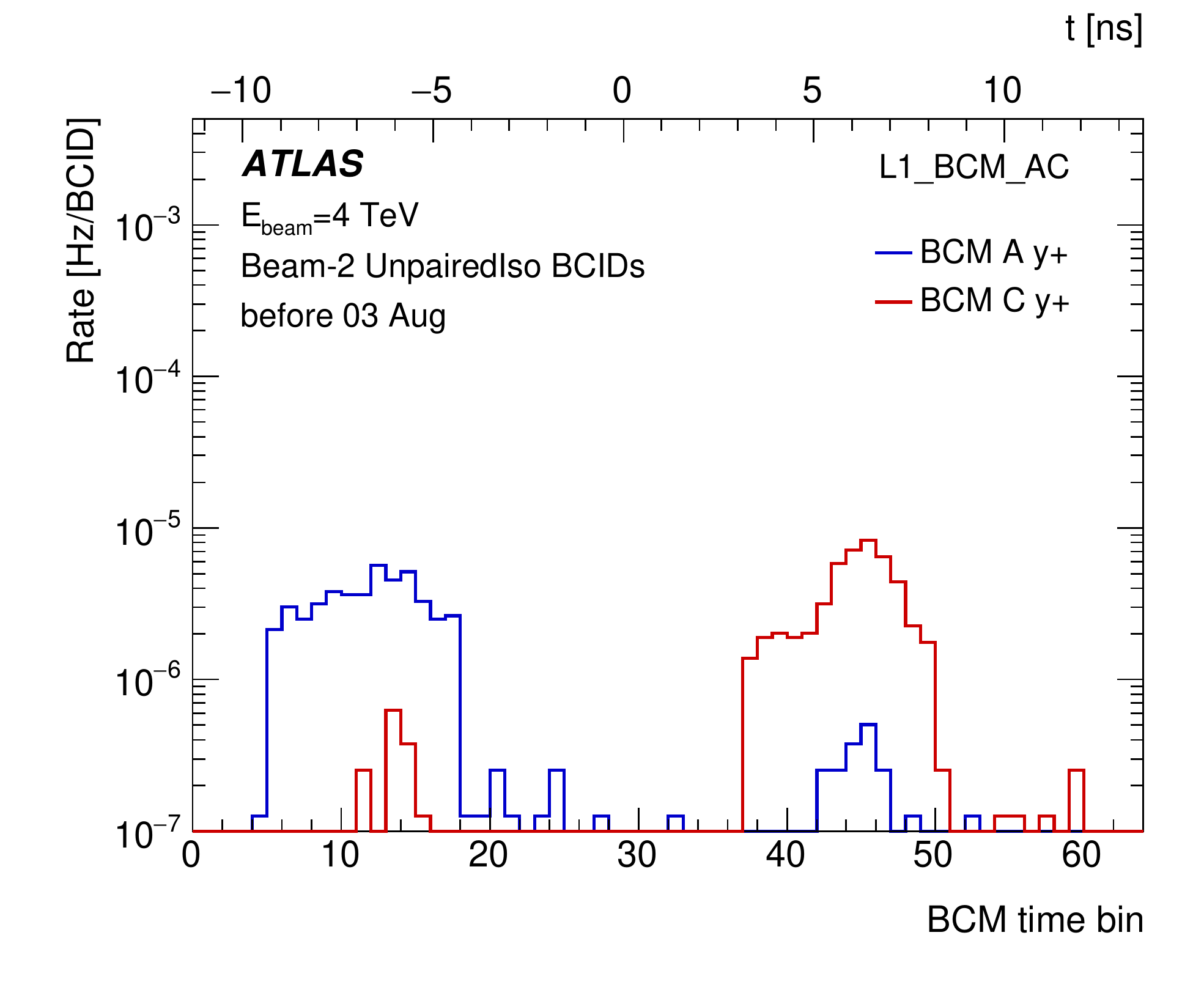} 
\label{fig:bcmbeamgas-dir2-4b}
}
} \\
\mbox{
\subfigure[]{
\includegraphics[width=0.49\textwidth]{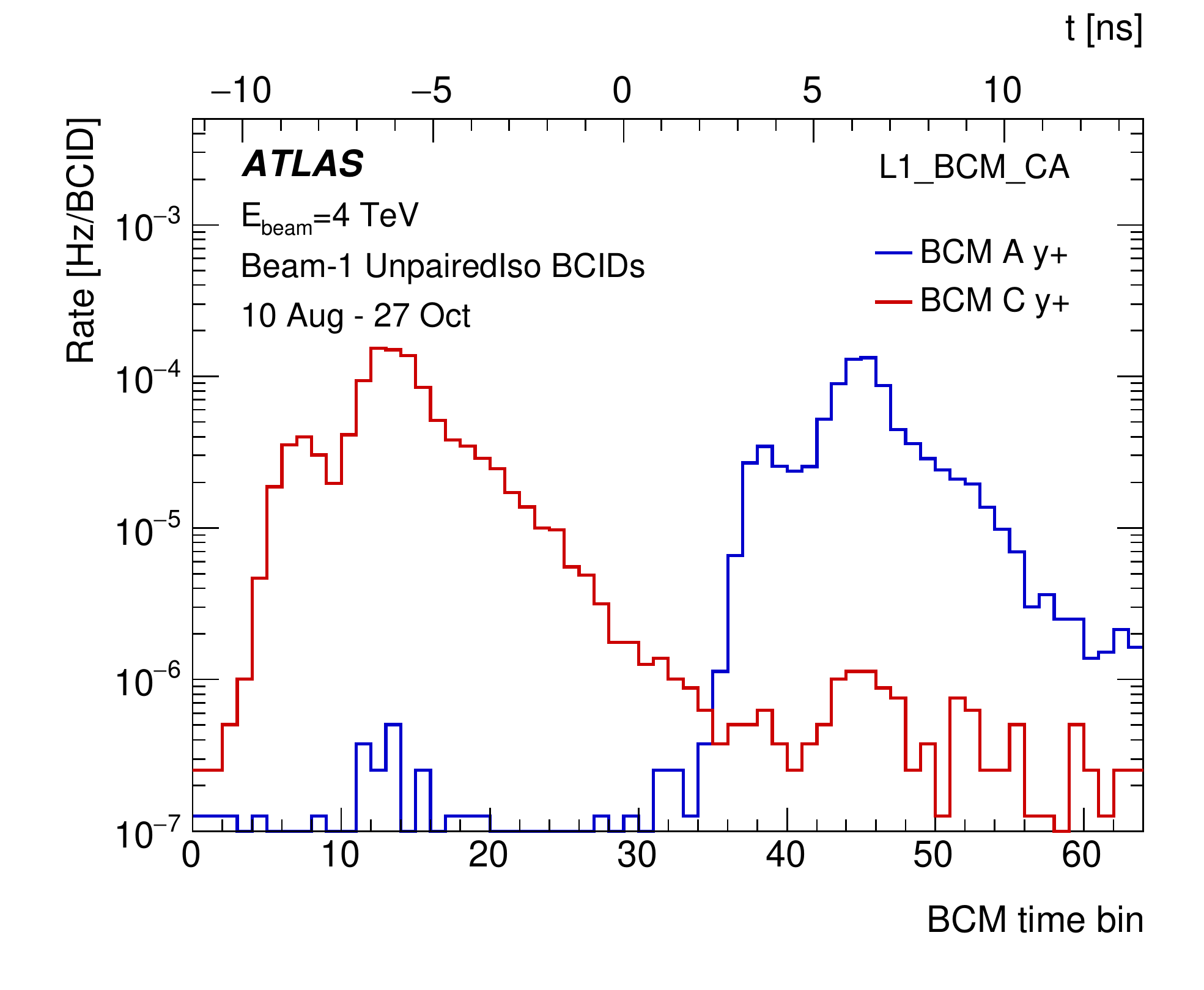}
\label{fig:bcmbeamgas-dir2-4c}
}
\subfigure[]{
\includegraphics[width=0.49\textwidth]{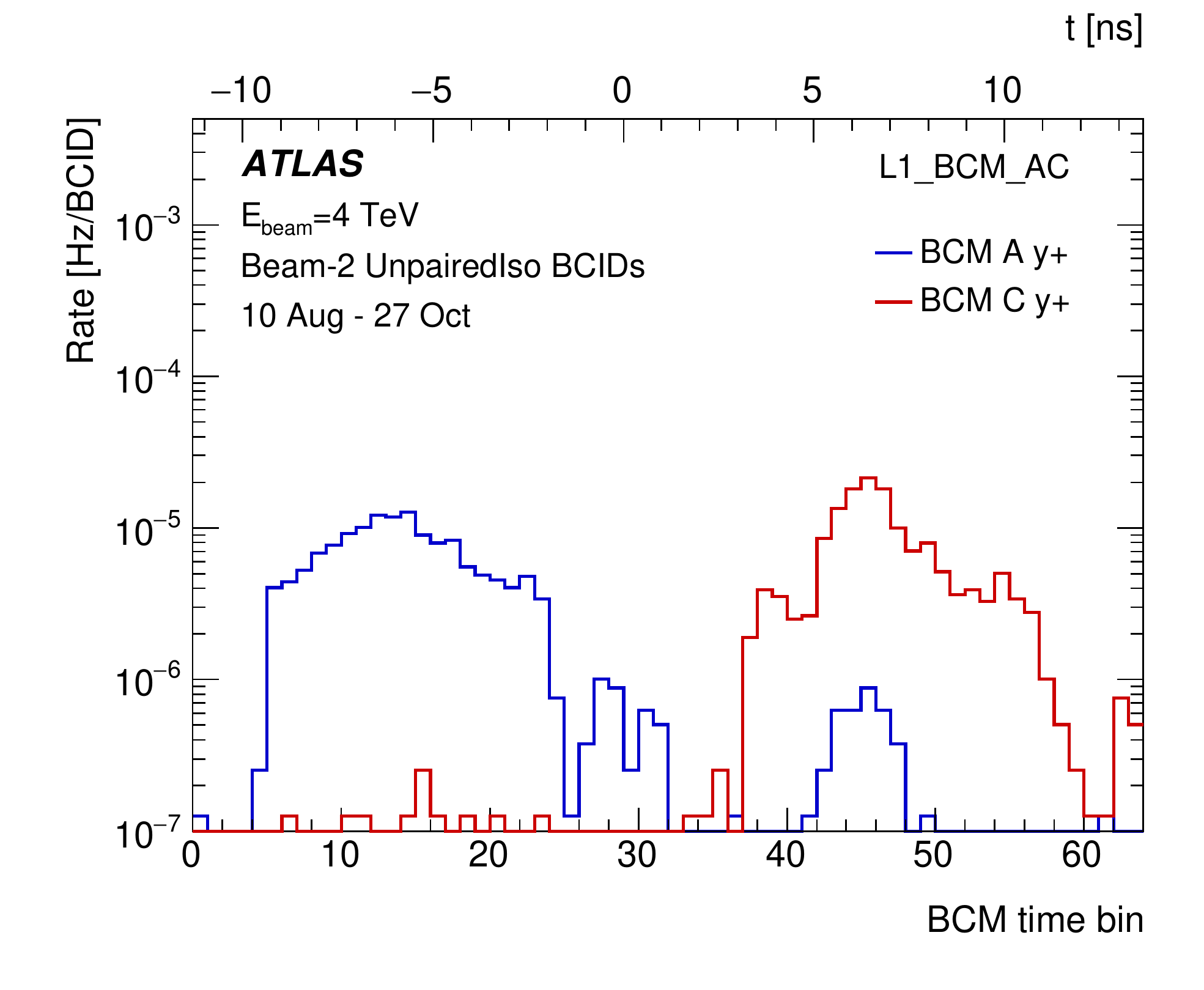}
\label{fig:bcmbeamgas-dir2-4d}
}
}
\caption{Hits in the individual BCM stations in the events triggered by L1\_BCM\_CA\_UNPAIRED\_ISO for beam-1 BCIDs (a, c)
and L1\_BCM\_AC\_UNPAIRED\_ISO for beam-2 BCIDs (b, d), i.e. in all cases opposite to the unpaired bunch direction.
The plots (a) and (b) show data before 3 August while (c) and (d) show data from 10 August until the onset of the BCM noise.
Only fills with 1368 colliding bunches are considered.}
\label{fig:bcmbeamgas-dir2-4}
\end{figure}

The ghost-BIB signals can be extracted from figure\,\ref{fig:bcmresponse} by considering only the events which have a BCM background trigger in the direction
opposite to the unpaired bunch. The rates after this selection are shown in figure\,\ref{fig:bcmbeamgas-dir2-4} for both beams.
The data are divided into periods before and after the LHC chromaticity changes. 

In the plots corresponding to fills before the chromaticity changes the opposite direction signals are distributed almost uniformly in the trigger window, 
with only a slight hint of a peak structure. This suggests that most of the rate is due to random coincidences of the pedestal.
However, from 10 August onwards, clear peaks are seen, especially in events triggered by L1\_BCM\_CA, i.e. in beam-2 direction.
The early and in-time peaks have very similar size and shape, indicating that the peaks are due to genuine
background tracks and not accidental coincidences of noise or \agpp{}. Furthermore, sub-peaks can be distinguished, 2.5\,ns (6 bins) 
before the main peak. This corresponds to the spacing of the RF buckets in the LHC, and further confirms that these peaks are related to 
ghost bunches. 

After mid-August the ghost-BIB is much larger for beam 2, i.e. in beam 1 unpaired positions. This seems to be in conflict with 
figure\,\ref{fig:bcmGhostColl}, which indicates that the intensity of injected ghost bunches (colliding with unpaired bunches of 
beam 2) is higher in beam 1.

The main peak seen in figure\,\ref{fig:bcmbeamgas-dir2-4} after 10 August is located in the bucket paired with the unpaired bunch in 
beam-1, i.e. it should be probed by the collision rate of figure\,\ref{fig:bcmGhostColl}.
The BCID-dependence of the rate, obtained by this collision probing, is consistent with the product of the bunch intensity data from the 
LDM for fill 3005 (26 August), shown in figure\,\ref{fig:ldmplot}(b). Thus both the collision probing and LDM show more ghost charge in
beam-1, while the BIB rates in the direction opposite to the unpaired bunches indicate more background from beam-2 ghost bunches.

This apparent paradox can be resolved by postulating that the ghost-BIB is dominated by cleaning debris and not by beam-gas interactions. 
Under this assumption the asymmetry of the LHC, with respect to beam cleaning insertions, plays an important 
role in explaining the observations. The momentum cleaning, located in IR3, is only two octants away from ATLAS along beam-2, 
but 6 octants along beam-1. In addition beam-1 has to pass both IR5 (CMS) and IR7 with its very tight betatron collimation.
Thus, debris from momentum cleaning in IR3 is much more likely to make it to ATLAS in beam-2 than in beam-1.

Since the IR3 collimators are at very relaxed settings with respect to the normal momentum spread\,\cite{roderick-collimation}, 
protons contained in a normal bunch will not be intercepted in IR3. Protons de-bunched in the LHC itself, i.e.
having escaped their original buckets and drifting in the ring can develop large enough 
a momentum offset to reach IR3 collimator jaws. 
As explained in section\,\ref{sec:ncb} this de-bunched ghost charge is rather uniformly distributed in the entire ring and therefore
the contribution from IR3 debris to the BIB is roughly the same for all RF buckets, including those filled with a nominal bunch. 
A nominal bunch has at least 3 orders of magnitude higher intensity than a ghost bunch and since the beam-gas rate is proportional to intensity, 
the role of IR3 debris for nominal bunches is negligible compared to close-by beam-gas scattering. 
For the low-intensity ghost bunches, however, the small component of IR3 debris can constitute the dominant background.

\begin{figure}[t]
\centering 
\includegraphics[width=\textwidth]{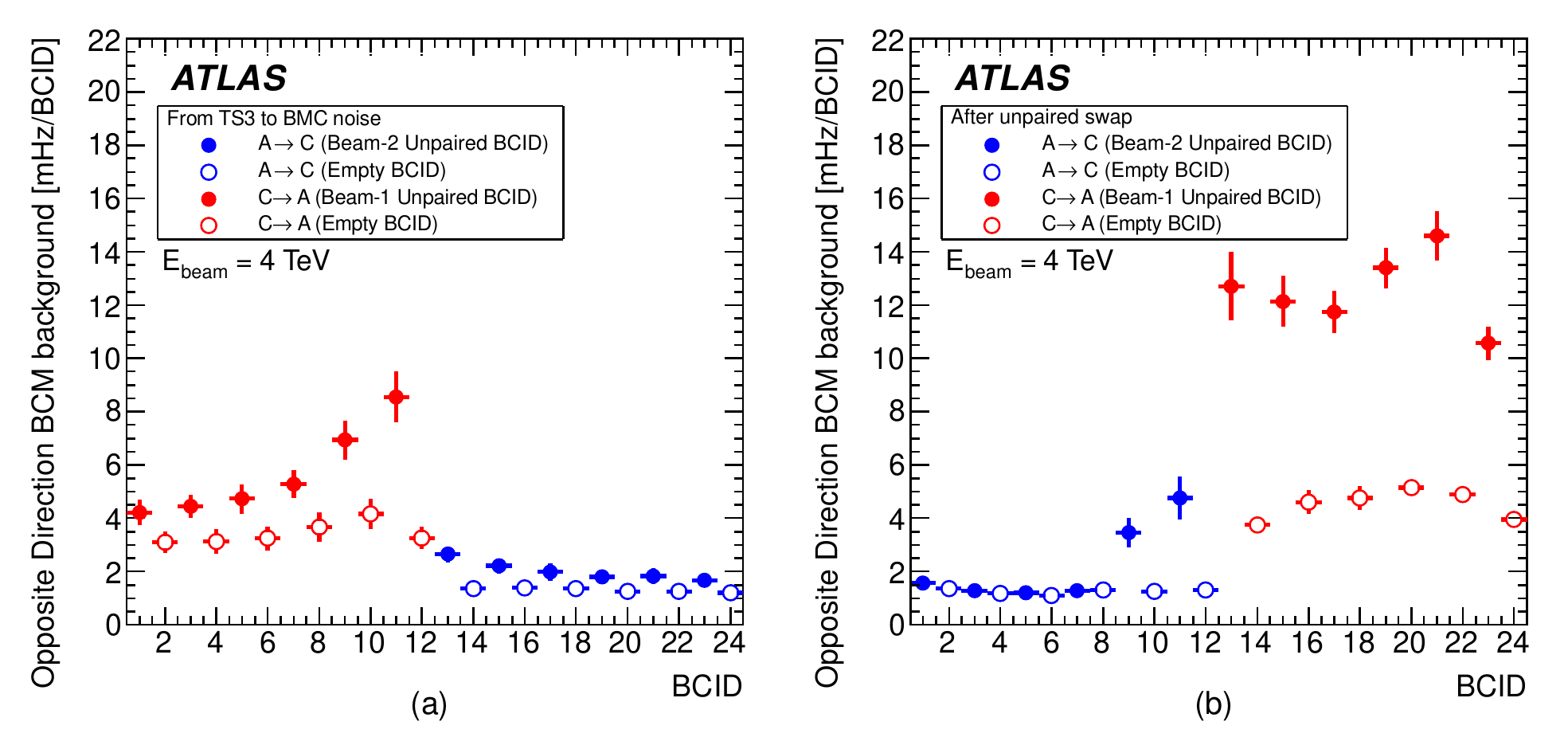}
\caption{L1\_BCM\_AC (L1\_BCM\_CA) trigger rates per BCID from ghost-BIB
before and after the swap of the unpaired trains. The BCM noise period has
been excluded. From each fill the first five hours of stable beams have been
considered. Fills shorter than that are not included. The mean values correspond 
to the average of several fills and the error bars indicate the RMS of the mean.
}
\label{fig:wdploterr}
\end{figure}

The swapping of the unpaired trains at the end of 2012 $pp$-operation provides a good means to test this 
hypothesis. The systematic tendency for higher intensity in trailing than heading injected ghost bunches implies that after the swap 
the intensity of injected beam-2 ghost bunches (L1\_BCM\_CA triggers) in beam-1 unpaired locations should be higher than before the swap. 
A comparison of per-BCID rates between mid-August and the onset of the BCM noise, and after the swap of unpaired trains
is shown in figure\,\ref{fig:wdploterr}. The data clearly indicate that even when the beam-1 train is first, 
there is more BIB from injected beam-2 ghost bunches than from beam-1, consistent with figure\,\ref{fig:bcmbeamgas-dir2-4}. 
After the swap of the unpaired trains the BIB from injected beam-1 ghost bunches is very low, while there is a large 
increase for beam-2, as expected.

\begin{figure}[t]
\centering 
\includegraphics[width=\textwidth]{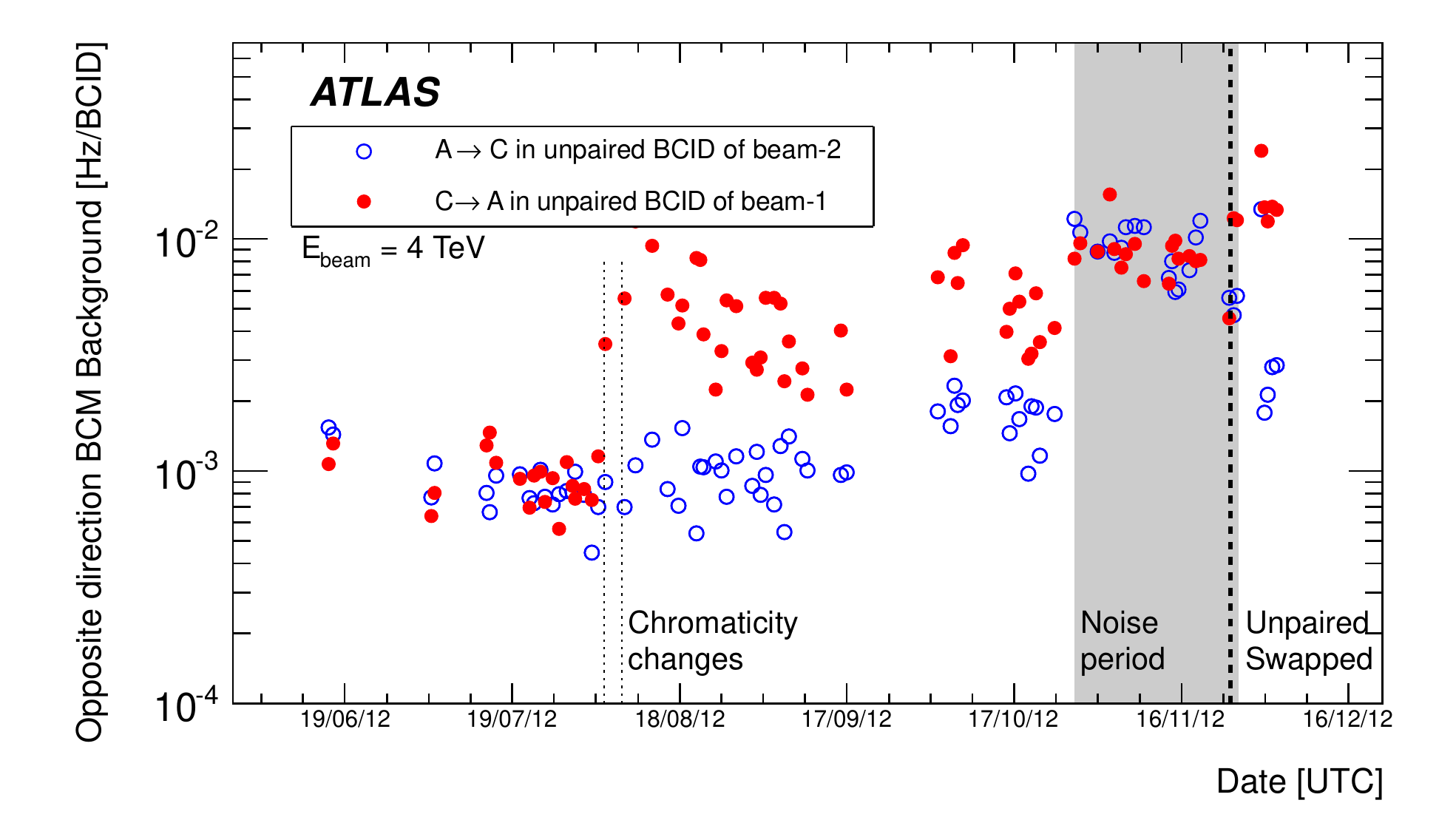}
\caption{Evolution of the opposite direction BCM background for both beams throughout 2012.
Only ghost bunches in BCIDs with an unpaired bunch in the other beam are considered. The
plot is restricted to fills with the 1368 colliding bunches pattern, which all have the 
same configuration of 6+6 unpaired bunches. 
Only the first five hours of stable beams
are considered and shorter fills are ignored. Data in the period between 27 October 
and 26 November suffer from the BCM noise and should be ignored.
}
\label{fig:beamgas_wrong}
\end{figure}

Figure\,\ref{fig:beamgas_wrong} shows the BCM background rates due to ghost bunches for all 2012 fills with 
1368 colliding bunches. A sudden increase of background from beam-2 ghosts is seen, coincident with 
the chromaticity changes in early August\footnote{Coincident with the chromaticity changes the LHC also switched octupole polarities on 7 August, which is another possible
candidate for having caused the jump. The data indicate, however, that the background started to rise already 
three days earlier, when the chromaticity optimisation started.}\,\cite{chromaticity}, while no rise is seen in the beam-1 direction. 
At the same time, the fill-to-fill scatter increases significantly.
Since chromaticity settings might affect how tight the momentum cleaning in IR3 is, this observation further supports 
the interpretation that the excess background, attributed to ghost bunches, originates from IR3. 

\subsection{Fake jets from ghost bunches}
\label{sect:wrongDirJets}

The rates in the C$\rightarrow$A direction in figure\,\ref{fig:beamgas_wrong} have a clear similarity to the corresponding 
fake jet rates due to BIB muons in the C$\rightarrow$A direction, observed in figure\,\ref{fig:ghostFakeJetYeara}. 
This strongly suggests that the jump in early August, which manifests itself in both background observables, has the same origin. 

The fake jet rates due to the ghost bunch in the colliding bucket can be extracted from the events without any vertex by selecting jets with time $-2\,\textrm{ns}<t<0\,\textrm{ns}$ at $\eta<-2$ and $\eta>2$ for beam-1 and beam-2, respectively.

\begin{figure}[t]
\centering 
\mbox{
\subfigure[]{
\includegraphics[width=0.49\textwidth]{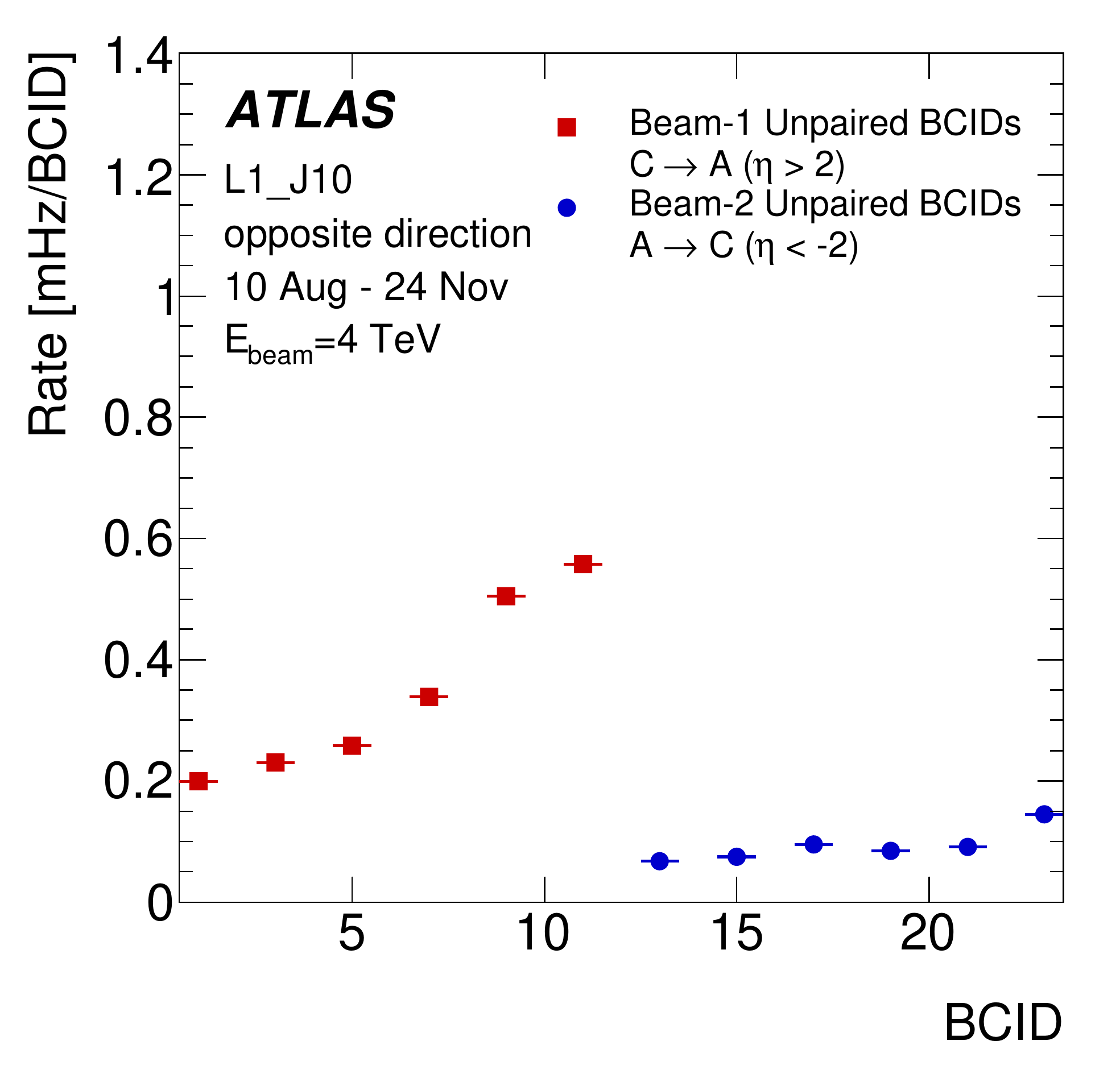}
\label{fig:wdplotJ10a}
}
\subfigure[]{
\includegraphics[width=0.49\textwidth]{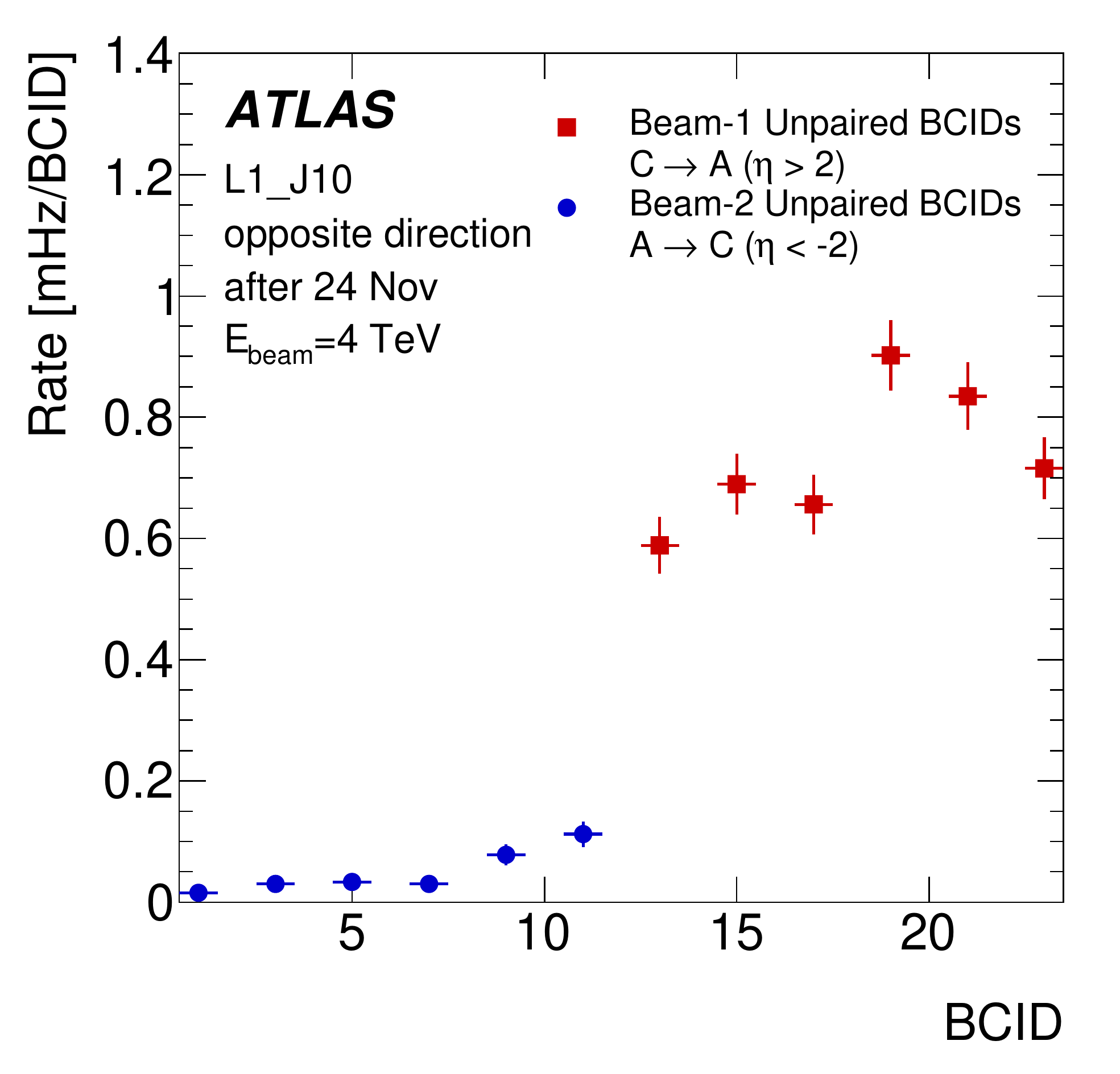}
\label{fig:wdplotJ10b}
}
}
\caption{Rate of fake jets associated with ghost-BIB from the opposite beam after mid-August (a) and after the swap of the unpaired trains (b).}
\label{fig:wdplotJ10}
\end{figure}

Figure\,\ref{fig:wdplotJ10} shows the rates obtained with this selection and averaged over fills before and
after swapping of the unpaired train. The similarity with figure\,\ref{fig:wdploterr} provides further evidence 
that the sources of the background are the same or at least strongly correlated.

The timing distributions for fake jets in the direction of the unpaired bunches at $|\eta|>2$ are shown in figure\,\ref{fig:ghostJetEta2} for data 
before early August and those after mid-August. Two peaks can be observed in figure\,\ref{fig:ghostJetEta2a} with the larger one at $t\sim0$, corresponding 
to BIB from the unpaired bunch. 
The large peak has equal size for both beams, which follows naturally from the very similar pressure distributions
on both sides, if the fake jets are entirely due to beam-gas events.
The beam cleaning, however, is asymmetric with respect to the two beams, so figure\,\ref{fig:ghostJetEta2} 
implies that either the beam-halo contribution to fake jets from unpaired bunches is negligible or halo losses of 
unpaired bunches are also very similar on the two sides. 

In  figure\,\ref{fig:ghostJetEta2b} as well  a prominent peak from the unpaired bunches is observed at $t\sim0$.
Its size is the same as in figure\,\ref{fig:ghostJetEta2a}, which implies that there is no change in the BIB from unpaired bunches.
Unlike in figure\,\ref{fig:ghostJetEta2a}, several smaller peaks are observed in figure\,\ref{fig:ghostJetEta2b}, especially for beam-2. 
These peaks reflect the 2.5\,ns time-structure 
of the LHC RF, which is consistent with them being due to ghost charge produced by de-bunching in the LHC. 
The level of these peaks, however, is only a factor 20--50 below the main peak, which is 
much less than the intensity difference (at least 3 orders of magnitude)  between nominal and ghost bunches. This
observation, that ghost bunches produce more background per unit intensity than nominal ones, 
provides further evidence that their losses are dominated by beam cleaning and not beam-gas interactions.

\begin{figure}
\centering 
\mbox{
\subfigure[]{
\includegraphics[width=0.49\textwidth]{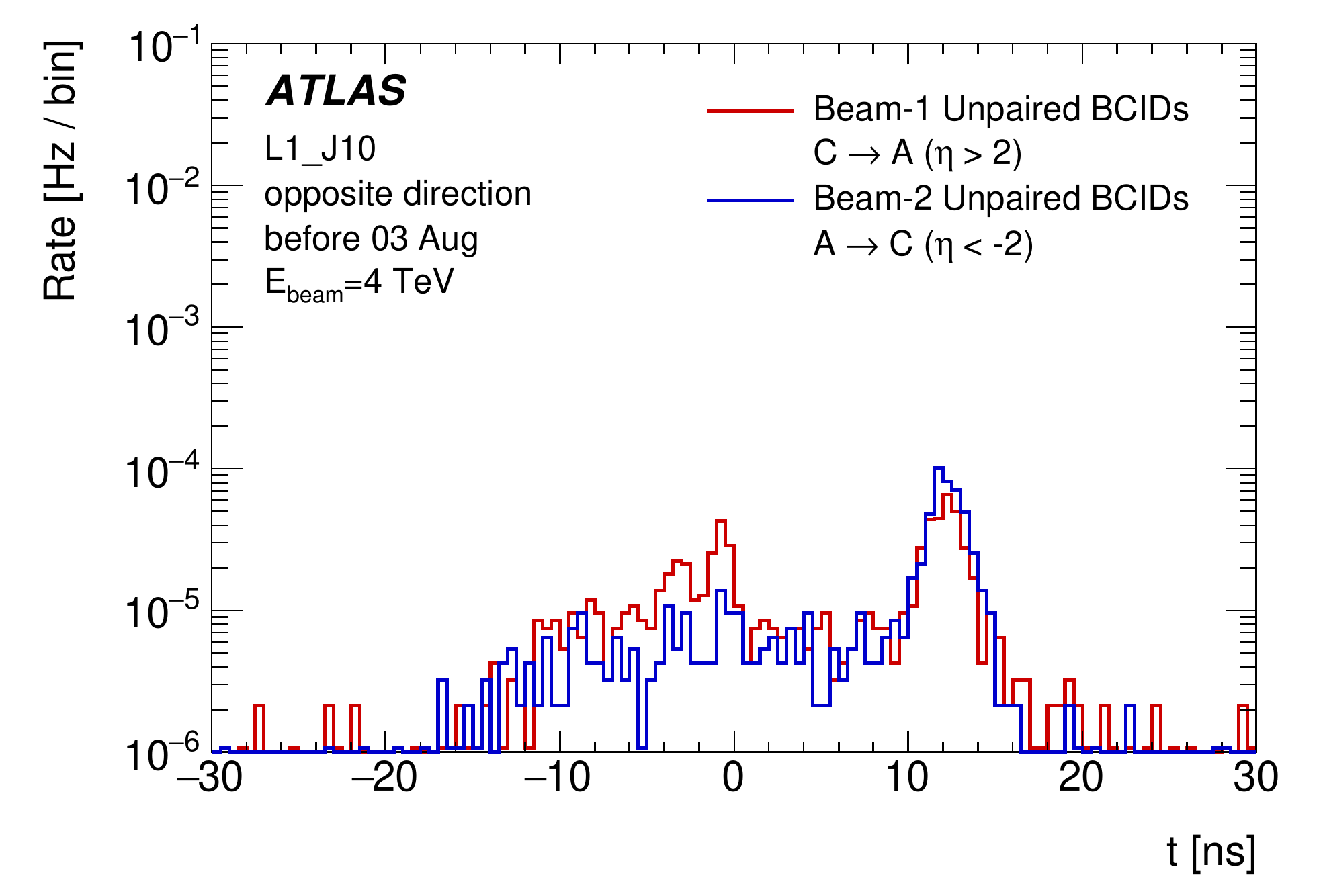}
\label{fig:ghostJetEta2wda}
}
\subfigure[]{
\includegraphics[width=0.49\textwidth]{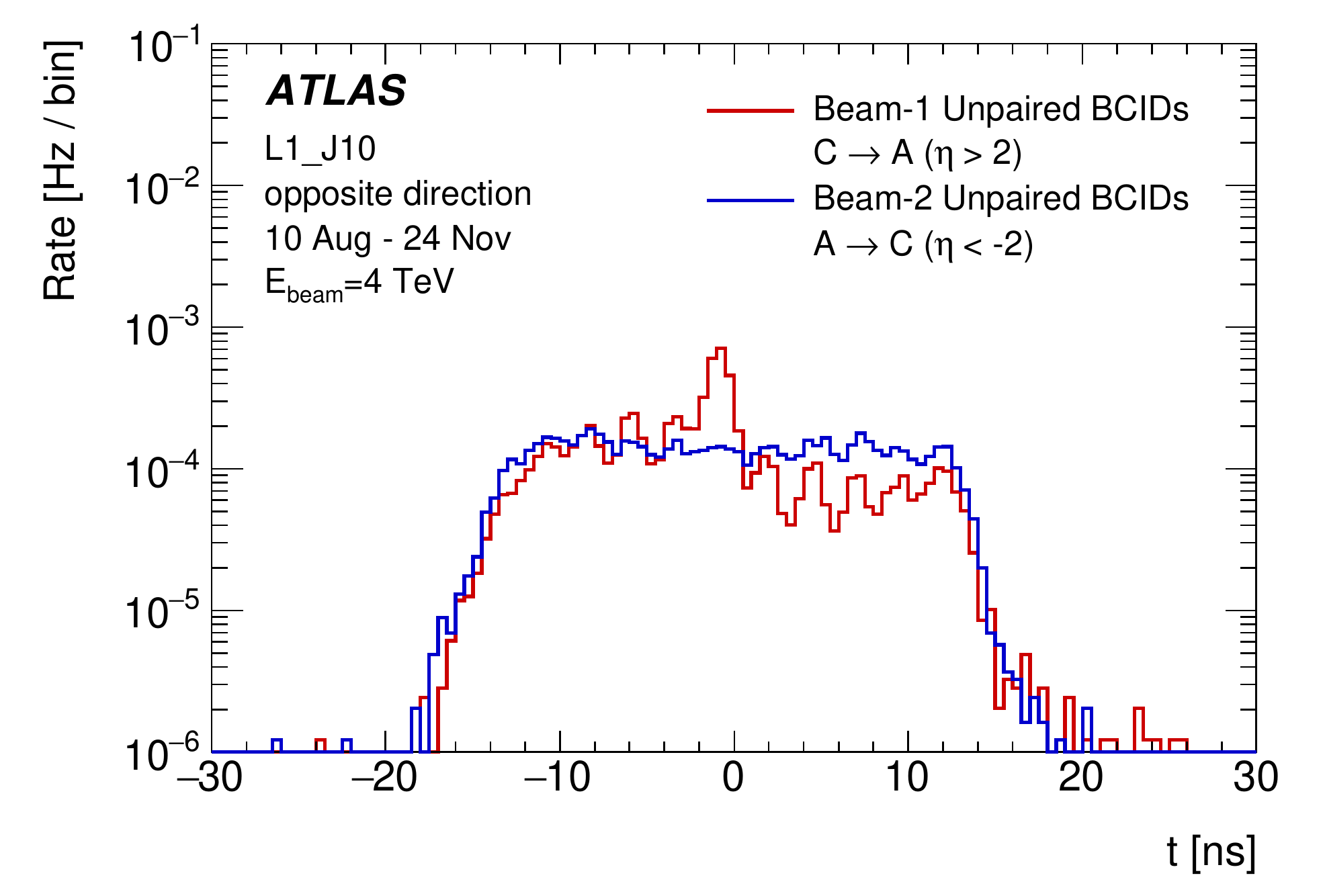}
\label{fig:ghostJetEta2wdb}
}
}
\caption{Jet times in the direction opposite to the unpaired bunch for fake jets 
at $|\eta|>2$ without a reconstructed vertex for the period before 3 August (a) and between 10 August and the 
swap of the unpaired bunches (b). Only fills with 1368 colliding bunches are considered.
The plots are normalised by the live-time, the rate is averaged over the six unpaired BCIDs per beam and the bin width is 0.5\,ns. 
}
\label{fig:ghostJetEta2wd}
\end{figure}

Figure\,\ref{fig:ghostJetEta2wd} shows the timing distribution of fake jets in the direction
opposite to the unpaired bunch, i.e. corresponding to ghost-BIB from the other beam.
Like in figure\,\ref{fig:ghostJetEta2}, smaller peaks reflect the RF structure and indicate that
the ghost charge is distributed among all buckets. It should be noted that there is no
central peak in the $A\rightarrow C$ direction, just an almost uniform pedestal. This is consistent
with the assumption that there is negligible fake jet production due to beam-1 ghost bunches and the observed rate 
is due to the smeared-out incoming beam-2 contribution where individual buckets cannot be resolved.\footnote{This is best understood 
from figure\,\protect\ref{fig:ghostJetTiming} where it is seen that downstream fake jets have a narrow time distribution, while upstream ones are smeared out.}
In the $C\rightarrow A$ direction, the side-peaks are clearly 
visible, but the absolute level is similar to the other direction. The reason is that in both directions
the rate is dominated by beam-2 ghost-BIB, but only in the $C\rightarrow A$ direction, i.e. downstream, is the time-spread small enough to 
resolve individual buckets (see figure\,\ref{fig:ghostJetTiming}). A comparison of the pedestals, i.e. the rates for $|t|>4$\,ns, reveals
that their levels are almost identical in figures\,\ref{fig:ghostJetEta2b} and \ref{fig:ghostJetEta2wdb}. This is consistent with the
pedestal being dominated by losses of the debunched ghost charge, which mostly are smeared out in time.

In figure\,\ref{fig:ghostJetEta2wdb} the peak in the nominal position, at $t\sim0$, is larger than the
other peaks. While it is established that the injectors generate more intense injected ghost bunches in the nominal RF buckets, it is also 
assumed that those bunches would not be sufficiently off-momentum to reach IR3 cleaning. It is possible, however, that those
low-intensity injected ghost bunches develop larger off-momentum tails than nominal ones.
Another observation that can be made in figure\,\ref{fig:ghostJetEta2wd} is that the side-peaks in the beam-2 (C$\rightarrow$A) 
direction are higher for earlier arrival, $t<0$, than for $t>0$. This might be related to a preferred direction of de-bunching from
the injected ghost bunch in the $t=0$ bucket: de-bunching is predominantly due to energy loss, as a result of which the particle moves to
a slightly smaller radius and arrives earlier on subsequent turns.

These observations of fake jets caused by de-bunched ghost charge, distributed all around the ring, imply that
some specific physics analyses, in particular searches for long-lived heavy particles, must
consider the presence of fake jet background due to BIB in all RF buckets. An 
example of an analysis, properly accounting for the possible presence of such backgrounds, can 
be found in reference\,\cite{R-hadron-paper}.

\subsection{Ghost bunches in the abort gap}

It was argued before that the de-bunched ghost charge, the source of beam-2 ghost-BIB, is spread 
uniformly over the ring. This implies that the effect should be visible also in BCIDs, which are not
paired with an unpaired bunch in the other beam. An observation of the beam-2 excess
in such BCIDs would also exclude any accidental coincidences involving the unpaired bunch.
Such an excess in the beam-2 (C$\rightarrow$A) direction can, indeed, be seen in the rates shown for empty bunches 
in figure\,\ref{fig:wdploterr}, but the cleanest possible sample of empty bunches can be found 
in the abort gap.

\begin{figure}[t]
\centering 
\includegraphics[width=\textwidth]{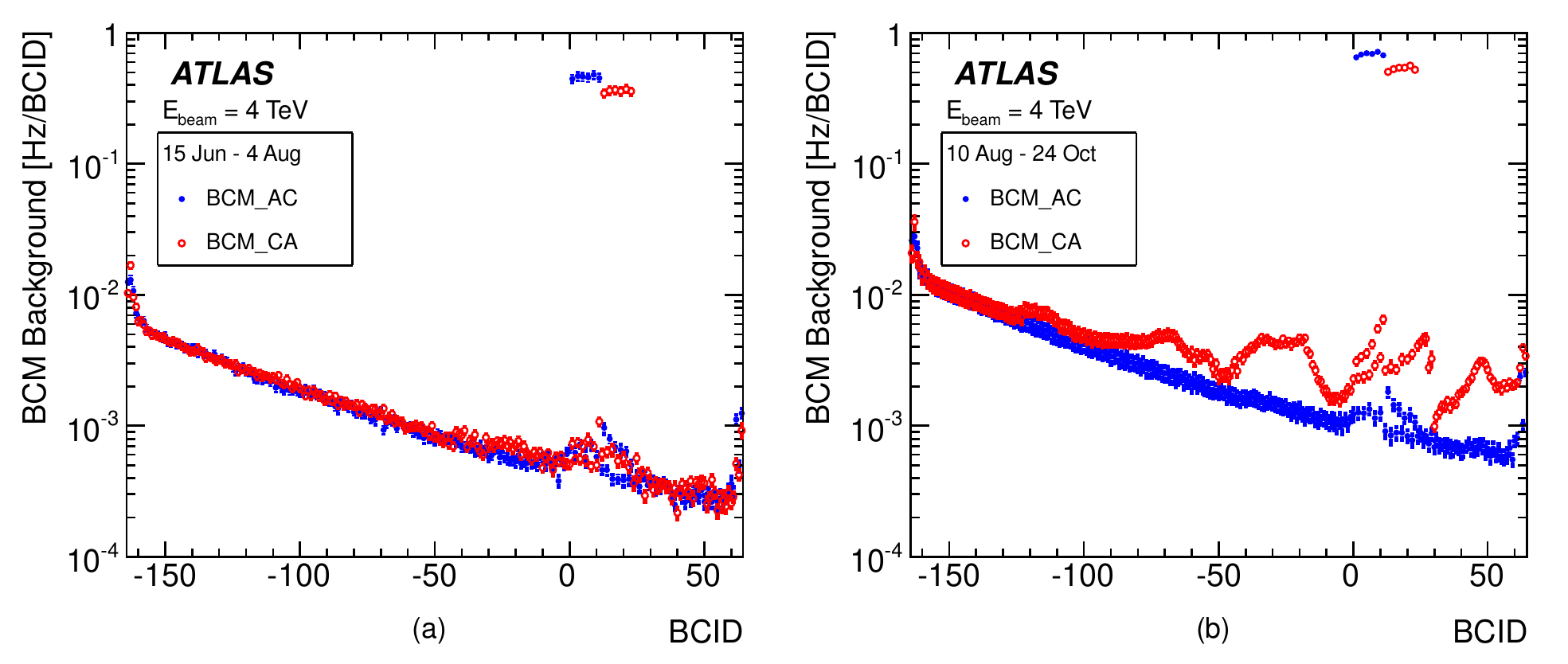}
\caption{L1\_BCM\_AC (beam-1) and L1\_BCM\_CA (beam-2) background rates in the abort gap region.
The points correspond to the average over all fills with the 1368 colliding bunch pattern, using the 
first five hours of stable beams and the error bars show the RMS of this average, thus reflecting 
any fill-to-fill variation. Fills shorter than five hours are not considered. Data shown are averages over
periods before (a) and after (b) the LHC chromaticity changes.
For plotting, the BCIDs in the abort gap have been re-mapped to negative values by BCID\,=\,BCID$_{\rm true}$\,-\,3564.
The high rates just above BCID=0 correspond to the unpaired bunches.
}
\label{fig:abortgapghost2}
\end{figure}

Figure\,\ref{fig:abortgapghost2} shows the background rates due to ghost-BIB  
as a function of BCID. It is evident that prior to the chromaticity changes
both beams produce very similar levels of BIB, dominated by accidental 
background-like coincidences from \agpp{} or noise. The gently falling slope is explained by the
decreasing \agpp{} following colliding trains before the abort gap.
The situation is entirely different after mid-August: while the same smooth slope is 
observed for beam-1, the BIB from beam-2 reveals a periodic structure of humps. 
As indicated by the small error bars -- which include the effect of fill-to-fill differences -- the pattern of the
humps is the same for all fills.

The presence of the humps in the fake jet rates cannot be verified since the fake jet analysis is 
based on recorded data and, due to the lack of a dedicated jet trigger in the abort gap, suitable data 
were not recorded in 2012.

An obvious question is if, and how, the humps evolve during a fill.
Since the rate is very low, there is not enough statistics to study the time-dependence 
for a single BCID in a given fill. Only an integration over longer time and many fills provides enough data
to clearly see the structures. 

An additional complication is that the random background from \agpp{} masks
the humps, especially at the start of the abort gap and early in the fill when the luminosity is high. 
This contribution, however, can be subtracted if it is assumed that the rate is fully random
and caused by noise and \agpp{} only. For this, the luminosity data have been used taking the
product of the BCM-TORx rates of both sides as an estimate for the random pedestal
coincidences.\footnote{This is valid only because the pedestal is uniformly distributed so that
early and in-time hits are equally likely.} Window-width and efficiency corrections are not
performed, instead the data are matched in the region of BCID 3400 to 3425 (shown as $-164$ to $-139$ in
figure\,\ref{fig:abortgapghost2}). The method reproduces very well the hump-free slope for beam-1 which
gives confidence that it is a valid estimate of the random background.

\begin{figure}[t]
\centering 
\includegraphics[width=\textwidth]{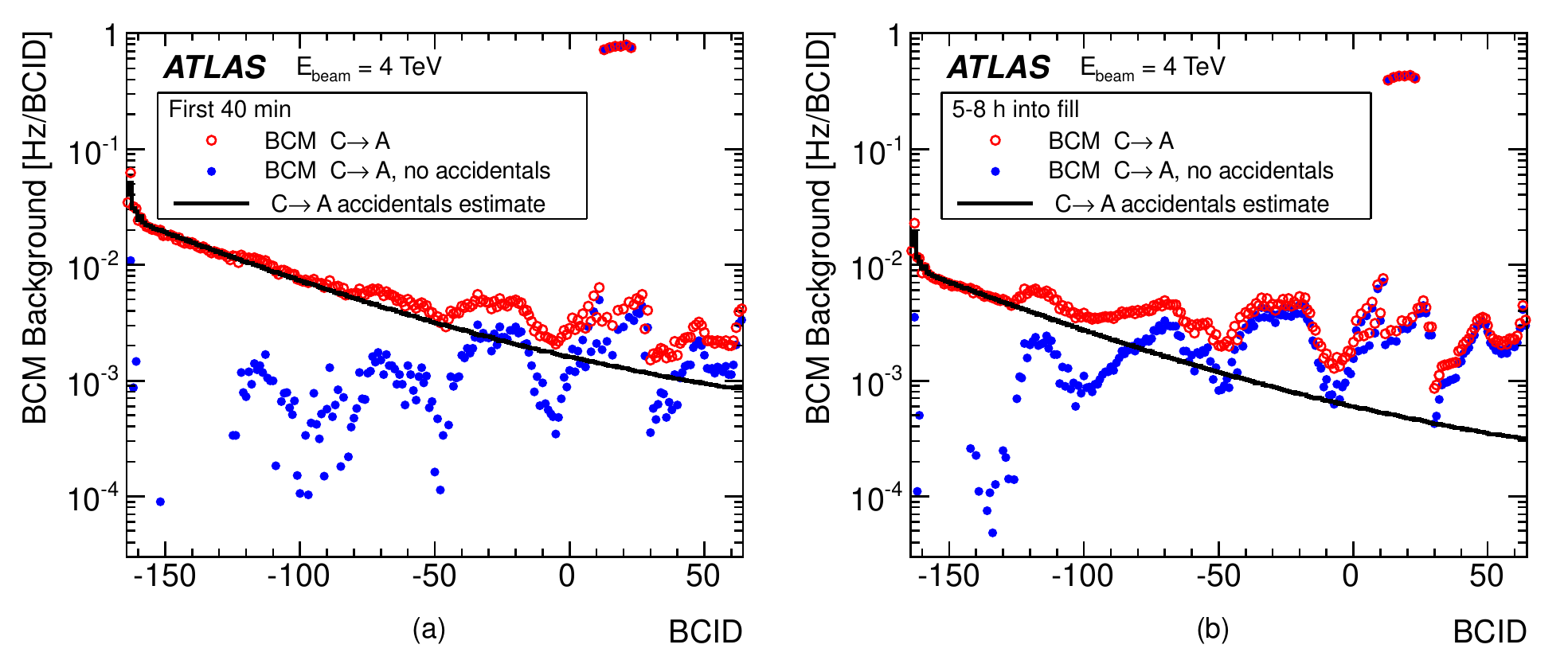}
\caption{L1\_BCM\_CA (beam-2) background rates in the abort gap region for two time-ranges during
stable beams. The open circles show the total data. The line represents an estimate of the
random coincidence rate, as explained in the text. The solid circles show the data with the 
random rate subtracted. The fills used for plot (b) are a subset of those used for plot (a).
Only fills between mid-August and the onset of the BCM noise have been considered.
For plotting, the BCIDs in the abort gap have been re-mapped to negative values by BCID\,=\,BCID$_{\rm true}$\,-\,3564.
}
\label{fig:humpsvstime}
\end{figure}

Figure\,\ref{fig:humpsvstime} shows a comparison of beam-2 ghost-BIB for the first 40\,minutes of STABLE BEAMS
and for the time between 5--8\,h into STABLE BEAMS mode, using all available fills (of which there were more
for the 40\,minutes plot). 
The rate after random background subtraction reveals that the humps remain in the same position and their magnitude 
shows only a slight sign of increase with time. The decrease of the random background in the course of the fill 
causes the humps to show up more clearly.

\begin{figure}[t]
\centering 
\begin{subfigure}[]{
\includegraphics[width=\textwidth]{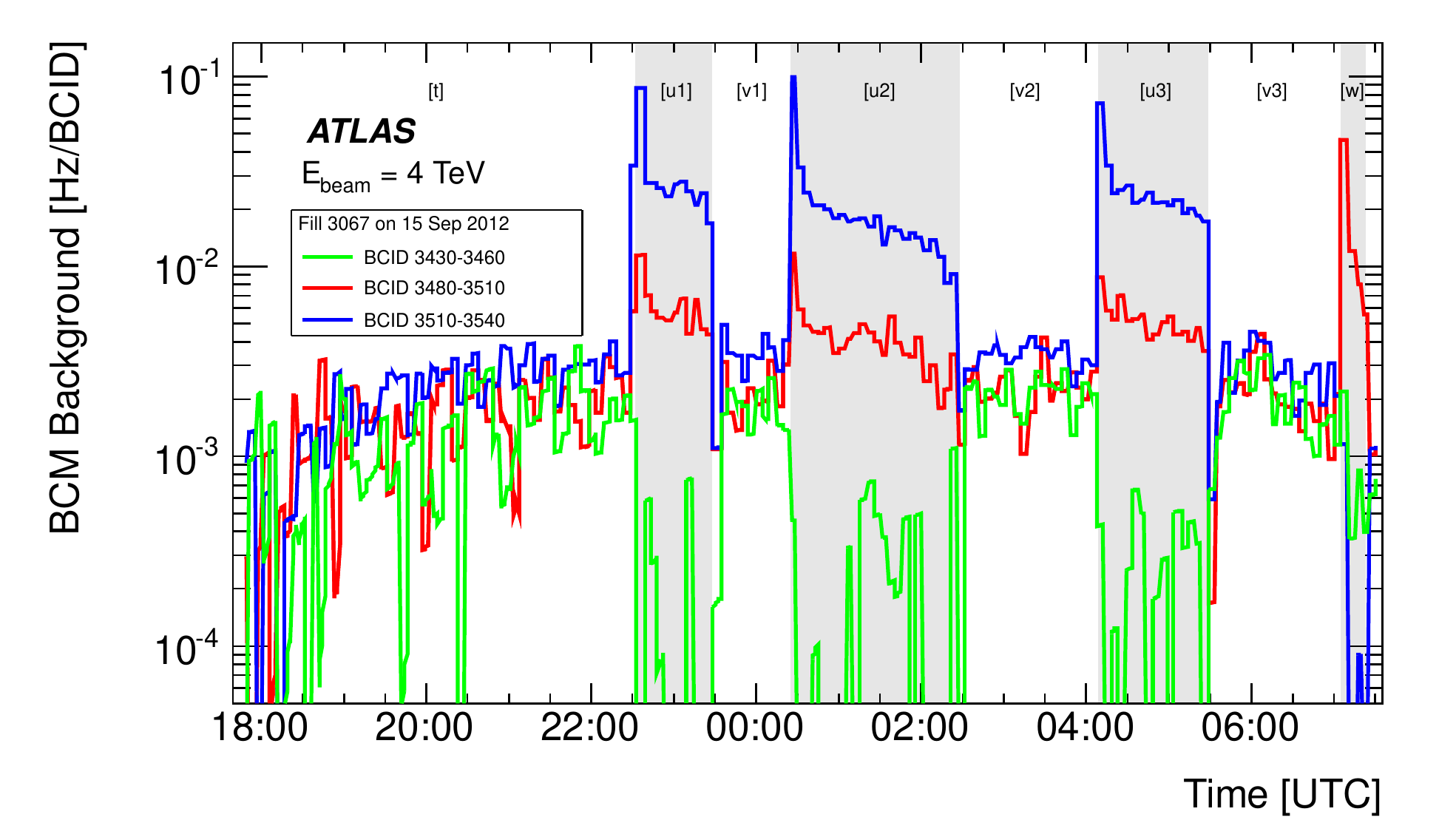}
\label{fig:humpsCleaning}}
\end{subfigure} 
\begin{subfigure}[]{
\includegraphics[width=\textwidth]{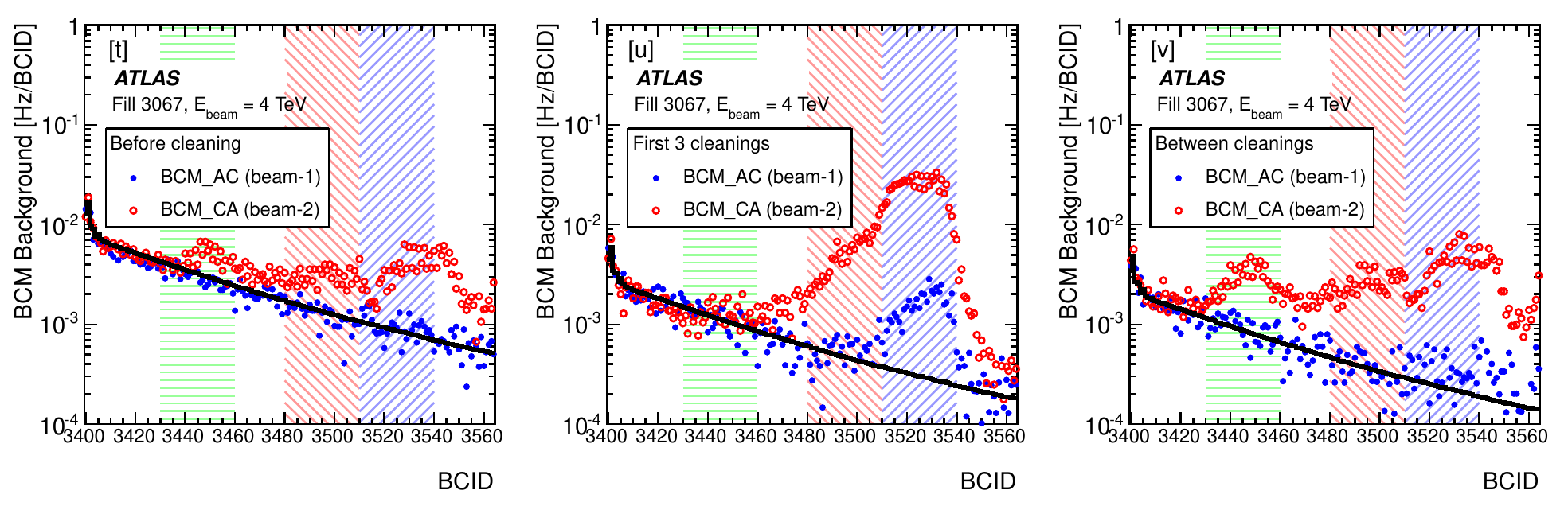}
\label{fig:acplot-f3067}}
\end{subfigure} 
\caption{Time development (a) of the beam-2 hump sizes in the abort gap in a fill where abort gap cleaning 
was repeatedly active (shaded areas). The random coincidence background has been subtracted. The BCID-ranges 
corresponding to the three histograms are indicated in figure (b) where the background in the abort gap for 
both beams is shown. The data labelled [u] and [v] have been averaged
over the periods indicated in (a), i.e. [u] shows an average over [u1]-[u3] of figure (a), 
and correspondingly for [v]. The hatched areas in figure (b) correspond to the three histograms, i.e. BCID-ranges, shown in figure (a).
The curve indicates the estimated random coincidence background.}
\end{figure}

In the second half of 2012 $pp$ operation, an abort-gap cleaning method\,\cite{gapcleaningnote} was 
exercised in several fills. The principle of abort gap cleaning is that excitations are introduced by 
the transverse damper system of the LHC.  Figure\,\ref{fig:humpsCleaning} shows how BIB rates in 
different BCID-ranges of the abort-gap react to the cleaning in fill 3067. The cleaning was switched on and 
off several times during the fill.
The BCID ranges are shown in figure\,\ref{fig:acplot-f3067}, which
gives per-BCID rates averaged over the periods indicated in figure\,\ref{fig:humpsCleaning}.
The solid black curve in figure\,\ref{fig:acplot-f3067} shows the estimated contribution from random coincidences.
This has been subtracted in figure\,\ref{fig:humpsCleaning}.

Figures\,\ref{fig:humpsCleaning} and \ref{fig:acplot-f3067} show that as soon as the cleaning is switched on a 
large hump develops in the latter part of the abort gap while the hump in the early part is strongly suppressed.
This suggest that the gap is filling up from the ``right'' 
(in figure\,\ref{fig:acplot-f3067}), i.e. off-momentum particles tend to move faster and arrive earlier. 
When the cleaning is switched off, a reduction of BIB rates is seen for all BCID ranges shown in figure\,\ref{fig:humpsCleaning}, 
but only in the first bin (1 luminosity block, i.e. typically 60\,s) following the cleaning. Thereafter the BIB rates 
return to their pre-cleaning levels.
This indicates  that the time-scale to repopulate the abort gap is of the order of one minute, which
is in good agreement with the estimate presented in reference\,\cite{Elena-Epac2004}.

The last cleaning period seen in figure\,\ref{fig:humpsCleaning} shows a different behaviour. It appears that the
excitation affects a different BCID range. Unfortunately the period is too short to obtain enough statistics 
in order to include it in figure\,\ref{fig:acplot-f3067}. 

The cleaning was active for both beams simultaneously. In figure\,\ref{fig:acplot-f3067} also the BIB rates  from 
beam-1 are shown and it can be seen that, when the cleaning is active, losses are significant enough for a visible hump to develop
also for beam-1. 

Finally, it should be emphasised that figures\,\ref{fig:abortgapghost2} -- \ref{fig:acplot-f3067} show a quantity
that depends on the loss rate of protons in the abort gap, not the abort gap population itself. This is also the
reason why the humps increase when the cleaning is switched on.

%% file: special.tex
\section{Backgrounds in special fills}

During 2012 operation three fills with special conditions exhibited backgrounds 
which motivated detailed studies. The first was a normal physics fill with
exceptionally high background in beam-1. Studies of this fill provide
insights into the reconditioning of the vacuum following an accidental
local gas release. 

The second case is a dedicated high-$\beta^*$ fill with very intense periodic
beam cleaning. The low beam intensity and luminosity of this fill resulted in 
strongly reduced beam-gas rates and afterglow. At the same time the losses in
the beam cleaning insertions were temporarily very high. In these particular 
conditions the correlation between beam cleaning and backgrounds in ATLAS could 
be studied in detail. The results obtained provide interesting insights and 
suggest similar studies for the normal optics in LHC Run-2.

The third fill to be discussed is a test with a 25\,ns bunch spacing, performed 
at the end of 2012 proton operation. The backgrounds observed during this
short test establish a reference point for LHC Run-2, in which
a 25\,ns bunch spacing is the baseline.

%% file: run213816.tex
\subsection{High background fill}
\label{sec_highbackgroundrun}
\label{sect:run213816}

In  LHC fill 3252 abnormally high BIB levels from 
beam-1 were observed in ATLAS for the first 90 minutes of stable beams. During
this time the background gradually decreased from about a 40-fold excess
back to almost the normal level. At the same time no excess was seen in beam-2
backgrounds.

\begin{figure}[t]
\centering 
\includegraphics[width=\textwidth]{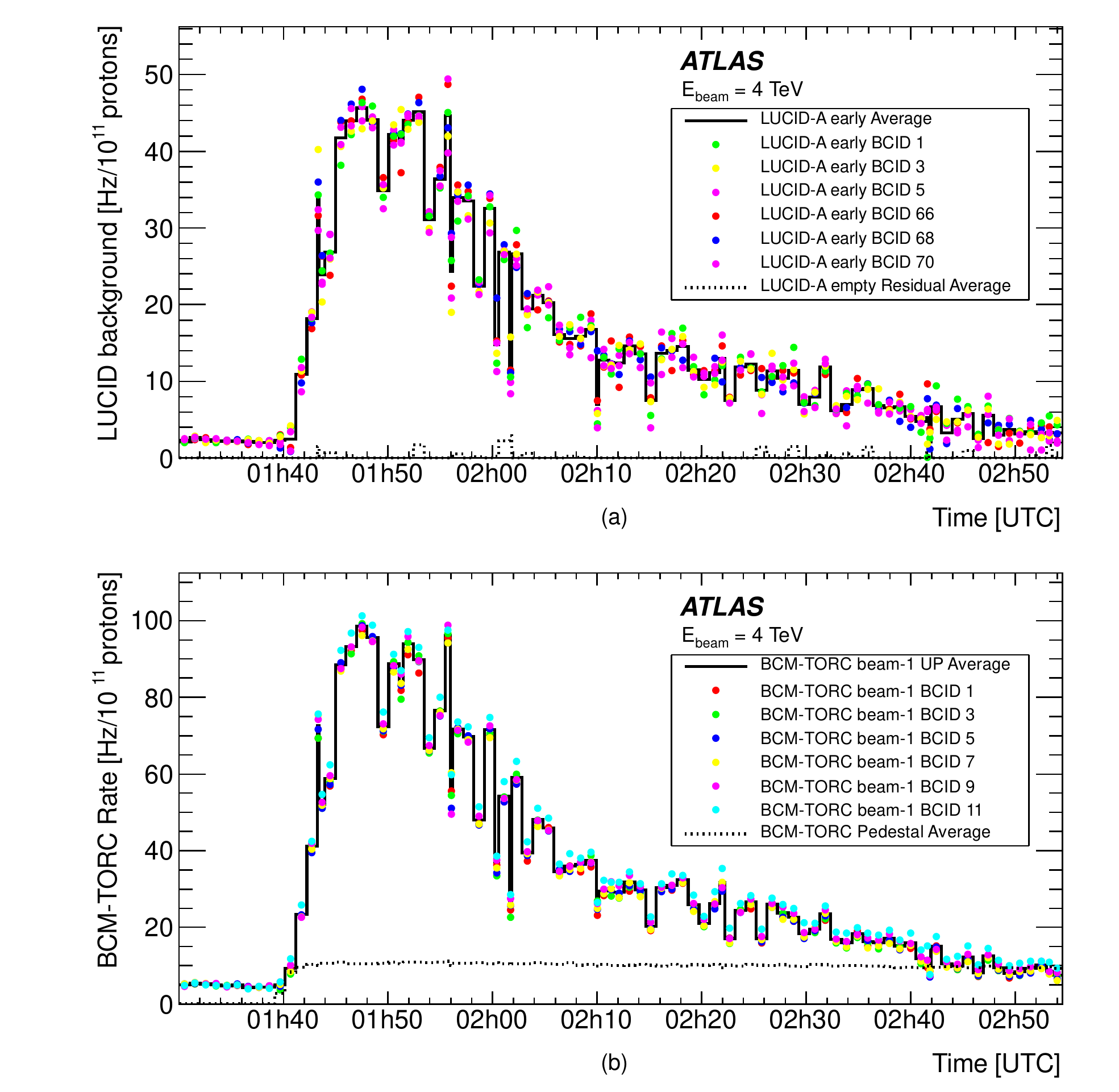}
\caption{Beam-1 background as a function of time at the start of
stable beams in LHC fill 3252, as seen by LUCID (a) and BCM (b). 
The lines show the averages over BCIDs, while the symbols show individual BCIDs. 
The bins correspond to ATLAS
luminosity blocks, the widths of which varied from 10\,s to the nominal 60\,s.
The dotted histograms are normalised per BCID, since there is no applicable 
bunch intensity.}
\label{fig:bcm_213816}
\end{figure}

The large background was seen by all inner detectors which were capable
of resolving it from the debris from $pp$ collisions. 
Due to their $\sim$40\,$\upmu{\rm s}$ integration time, the BLMs close to ATLAS 
were all dominated by $pp$ collision products and unable to resolve an 
excess of this size.
Monitors in the cleaning insertions showed no structure with similar magnitude 
and temporal behaviour as the observed background, neither did vacuum
gauges in IR1.

Figure\,\ref{fig:bcm_213816} shows the detailed time-structure of the background,
extracted from the single-sided luminosity data of BCM and using the 
incoming bunch signal of LUCID, as described in section\,\ref{sect:lucid}. 
The LUCID method allows for monitoring of the background associated with the first three 
unpaired bunches of beam-1, and the first three paired bunches of the 
first train, i.e. bunches in BCIDs 66, 68 and 70. For all other bunches the LUCID 
signal overlaps with the overwhelming luminosity signal from bunch-bunch or 
bunch-ghost collisions.

At the A-side of LUCID the background signal from the first colliding bunches appears in BCIDs 61, 63 and 65.
The pedestal level for these was estimated as an average over the empty BCIDs 57, 58 and 59.
For the first unpaired bunches, the background signal
appears in BCIDs 3559, 3561 and 3563, for which the pedestal is evaluated from BCIDs 3556--3558. 
In both cases the pedestal was found to be about three times larger than the background signal. 
However, since the pedestal is dominated by afterglow 
from trains several $\upmu{\rm s}$ earlier, it is very constant and could be subtracted in order to extract the signal.
Thereafter the remaining signal is normalised by the bunch intensity. 
In order to estimate the uncertainty introduced by this method, the subtraction was applied also on
BCIDs 60, 62 and 64, where no signal is expected. The result is shown by the dotted line 
in the upper plot of figure\,\ref{fig:bcm_213816}. The level, consistent with 0, indicates that contributions 
other than the beam background are efficiently removed by the pedestal subtraction.

With the BCM-TORx rates the unpaired bunches can be monitored but the
rates include a contribution from ghost collisions.\footnote{The 
dedicated L1\_BCM\_AC\_CA background rates would have less contamination but
their per bunch rates are available only with 300\,s time resolution, which is too 
coarse for this analysis.} As shown in figure\,\ref{fig:ghostAGCorr}, the
unpaired bunch signals are on top of a smoothly dropping background, which is
removed by subtracting the rate in the preceding BCID. This pedestal estimate
is indicated by the dotted line in the lower plot of figure\,\ref{fig:bcm_213816}.
The remaining beam background signal is normalised by the bunch intensity.

The data from the two independent detectors exhibit a similar spiky time structure, which is 
clearly beyond statistical fluctuation.\footnote{With a count rate of about 0.003 per bunch, 
the statistical error for a 60\,s LB is around 2.3\%.} All bunches follow the same structure. 

The origin of the high background could be attributed to an incident a few hours earlier: 
a quench of the inner triplet on side A of ATLAS
had warmed up the beam-pipe and desorbed some gas from the cold bore. This event resulted in
a pressure surge at vacuum gauges on the IP-side of the triplet ($p22$) and a slower increase in the gauge 
on the other side ($p58$), indicating that gas was distributed throughout the entire triplet 
length and certainly part of it was absorbed on the beam screen.

The inner triplet and its beam-screen are exposed to a significant heat-load from $pp$ collision debris, thus,
at the beginning of the next fill, the beam-screen was scrubbed by the 
secondaries created in the collisions. This led to a pressure increase within 
the triplet, but since the beam pipe was now cold again, there was sufficient
cryo-pumping that the gas did not diffuse to the gauges. Therefore, normal
pressure levels were reported, while the actual pressure within the triplet
must have been significantly above normal.

\begin{figure}[t]
\centering 
\includegraphics[width=\textwidth]{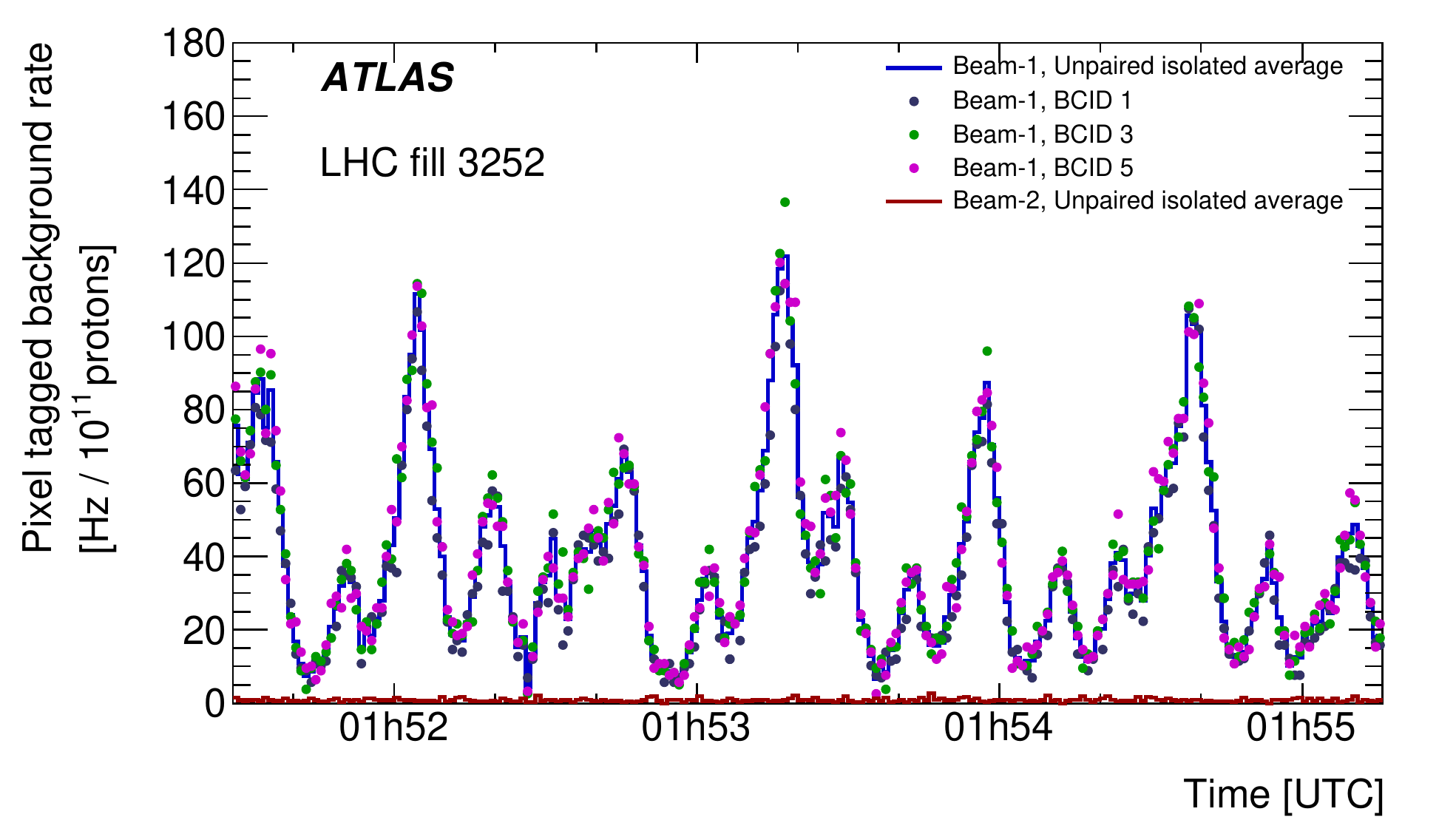}
\caption{Background rate measured with the Pixel tagging method over 4 minutes during
the period of highest rate. For beam-1 the average and the three unpaired isolated BCIDs are 
shown separately. For beam-2, with normal background, only the average is shown.}
\label{fig:pixel-run213816}
\end{figure}

A third method to observe the background is provided by the Pixel
detector using the background tagging tool developed in reference\,\cite{backgroundpaper2011} and described in
section\,\ref{sec:tagger}.
Unlike the luminosity data of BCM and LUCID, used for figure\,\ref{fig:bcm_213816},
the Pixel-tagged events are not sensitive to afterglow. 
The overall shape of beam-1 background seen by this method is perfectly consistent with figure\,\ref{fig:bcm_213816},
but since the tagged events have exact time-stamps, it is possible to significantly improve the time resolution.
Figure\,\ref{fig:pixel-run213816} shows the time-dependence of the background over four minutes close to the
maximum and a clear substructure is resolved. All unpaired isolated BCIDs exhibit 
a consistent periodic fluctuation with a frequency of the order of 0.1\,Hz and significant amplitude. 
The temperature of the beam-screen is actively controlled and, indeed, the frequency of the background oscillations 
is very similar to the temperature fluctuations measured at the inlet of the beam screen cooling loop, suggesting
that the rate of outgasing is strongly coupled to the beam screen temperature.

\begin{figure}[t]
\centering 
\includegraphics[width=\textwidth]{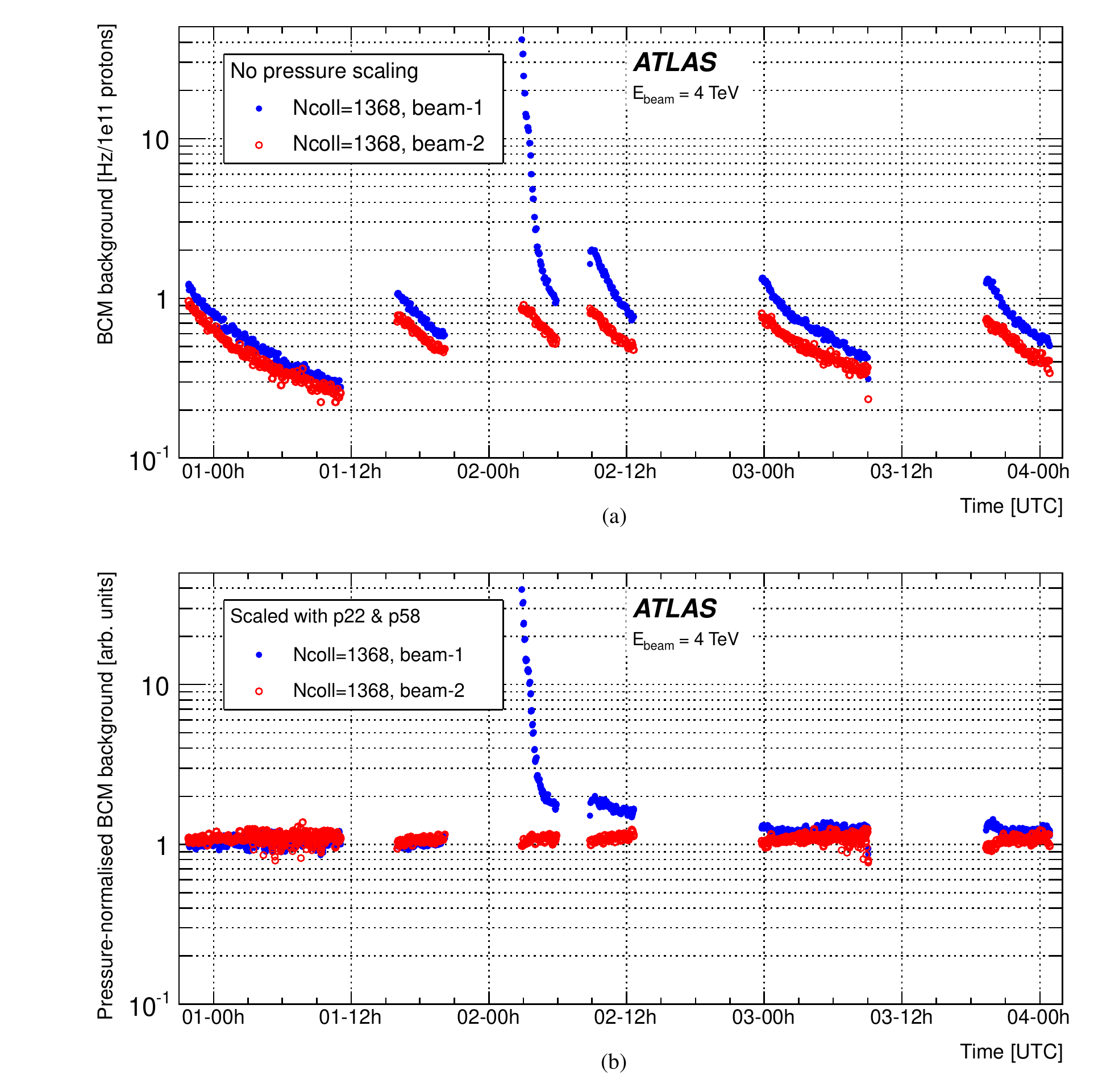}
\caption{BCM\_AC (beam-1) and BCM\_CA (beam-2) backgrounds in the high-background
fill and a few fills before and after. Unlike the luminosity data in figure\,\protect\ref{fig:bcm_213816},
these triggers have negligible background contamination, but suffer from a time-resolution of only 300\,s.
Rates are shown before (a) and after (b) normalisation with the residual pressure. The high-background fill, 3252, is 
seen as the large excess in the middle of the plots.}
\label{fig:bcmAC-run213816}
\end{figure}

In figure\,\ref{fig:bcmAC-run213816}(b) the pressure scaling, derived in section\,\ref{sect:vacuum}, 
is applied to the rates, but has no reducing effect on the excess. This indicates that the additional
background originates from a source that has no correlation with the pressure at 22\,m.
The scaling also reveals that even the next fills are affected, indicating that the scrubbing of
the contaminated beam screen continues into several subsequent fills.
In fact a close examination of figure\,\ref{fig:bcmAC-halo-all2012}(b) reveals that the scrubbing continues for 
about one week until the level of beam-1 background has dropped to the same level as of beam-2.
This observation, made in figure\,\ref{fig:bcmAC-halo-all2012}, demonstrates that the pressure scaling 
significantly increases the sensitivity to any backgrounds not correlated with the 22\,m pressure.
This enables the detection of much smaller additional effects than without the pressure scaling.

\begin{figure}
\centering 
\mbox{
\subfigure[]{
\includegraphics[width=0.49\textwidth]{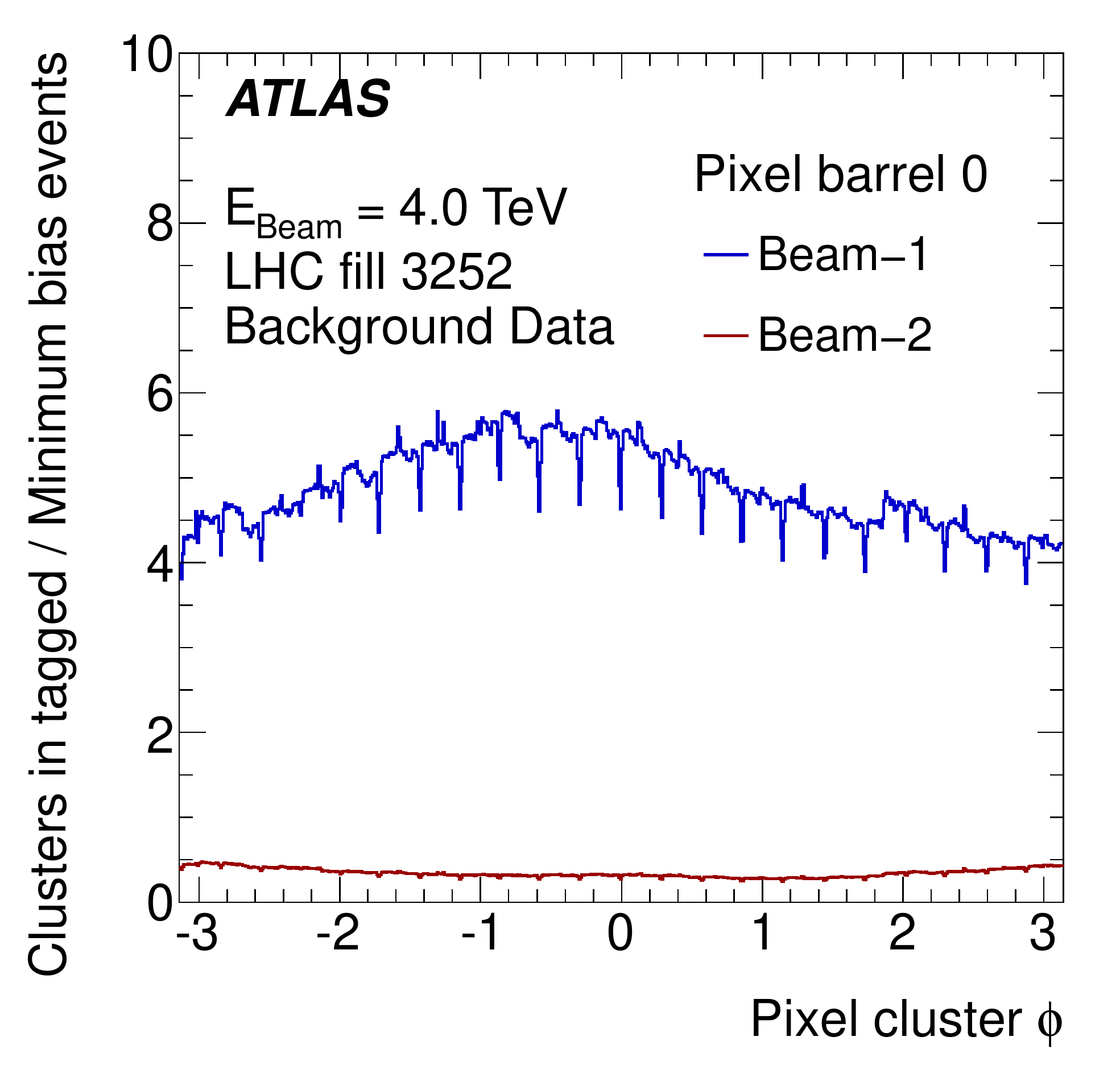} 
\label{fig:pixelPhia}
}
\subfigure[]{
\includegraphics[width=0.49\textwidth]{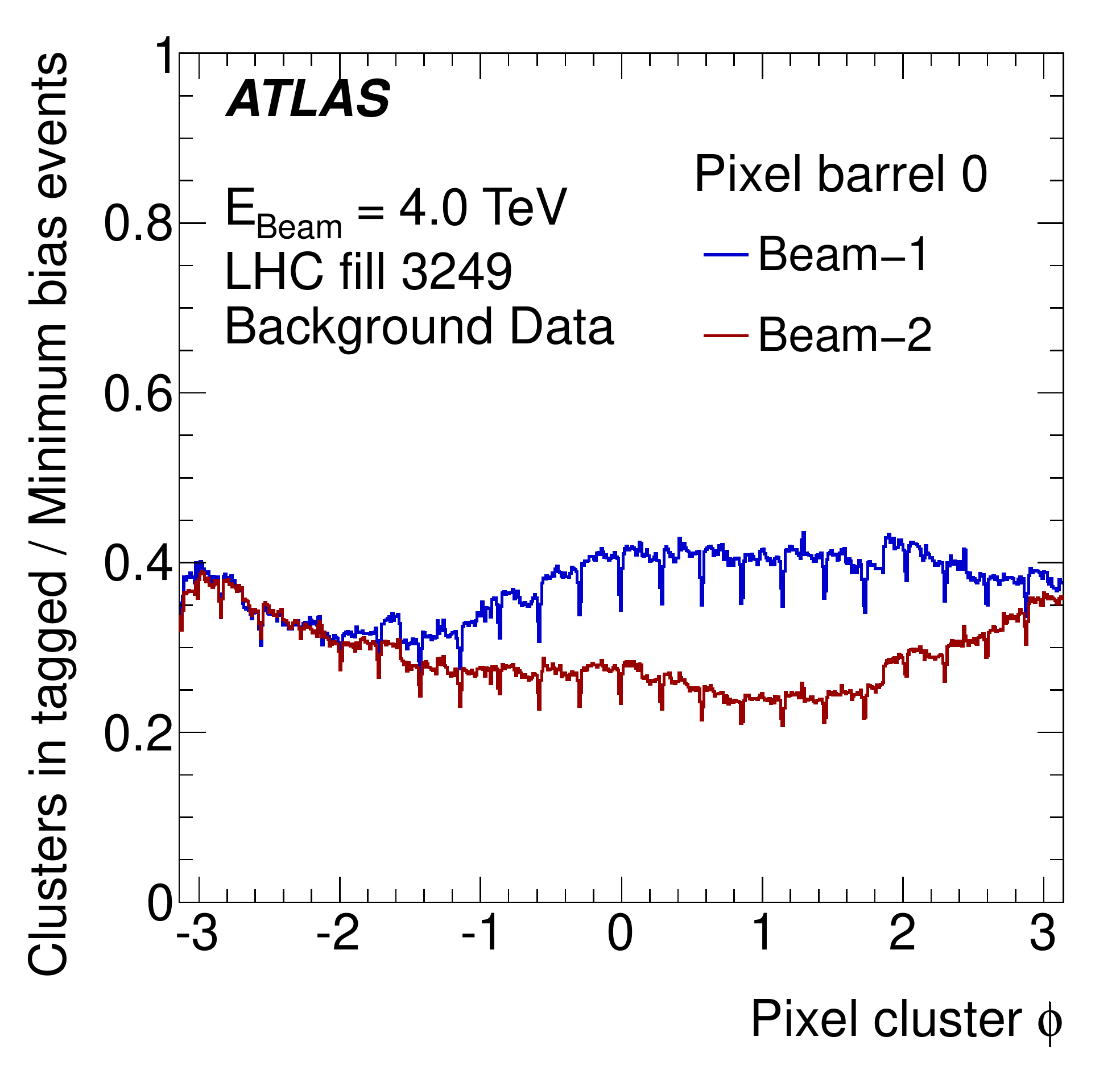}
\label{fig:pixelPhib}
}
}
\caption{Azimuthal distribution of clusters in events tagged as background during the
high-background phase in fill 3252 (a) and during the normal background fill 3249 (b). 
Rates have been normalised by minimum-bias events in the same fill to compensate 
for inefficiencies and overlaps of individual modules. 
Taking into account the very different vertical scale on the two plots -- the histograms for beam-2 are in fact almost identical 
in the two fills. The small dips reflect a residual effect of the module overlaps which have different effects for tracks from the IP and from upstream BIB events. 
}
\label{fig:pixelPhi}
\end{figure}

The Pixel-based background tagger can be used to extract a detailed azimuthal dependence of the background in order to
compare it with that of a normal fill. Efficiency differences of individual Pixel modules lead to a rather non-uniform azimuthal
distribution of cluster rates even for minimum-bias events, which produce secondaries uniformly in $\phi$. 
In order to compensate for this and extract the azimuthal asymmetry of the BIB, the rates of clusters in events tagged 
as BIB have been normalised by the cluster rates of minimum bias events in the same fill. 
The results for the innermost layer of the Pixel detector are shown
in figure\,\ref{fig:pixelPhi}. The very high rate of beam-1 background in fill 3252 is again illustrated
by the difference of the level of beam-1 and beam-2 backgrounds seen in figure\,\ref{fig:pixelPhia}. The 
more interesting aspect, however, is that the azimuthal distribution is different from that seen in 
figure\,\ref{fig:pixelPhib}, which represents a normal fill. This indicates that
the source distribution along the $z$-axis must be different in the two cases such that the secondaries 
traverse different magnet configurations. As argued before, in the high-background fill the beam-gas events
must be localised in the inner triplet. 
In section\,\protect\ref{sect:beamgas} it is established that the background seen by the BCM correlates with the 
pressure at 22\,m, i.e. the IP-end of the triplet, but this does not exclude the possibility of contributions 
from more distant locations, as long as they also correlate with the 22\,m pressure.
It is worth noting that the integral of the beam-1 histogram in figure\,\ref{fig:pixelPhib} is found to be
29\,\% larger than that of beam-2. This difference is consistent with the 28\,\% found in sections\,\ref{sect:afterglow} 
and \ref{sect:beamgas} for the beam-1/beam-2 ratio of the BCM backgrounds. This agreement, and the observation of a different 
azimuthal distribution, support the assumption formulated at the end of section\,\ref{sect:beamgas} that the background of
beam-1 has either an additional contribution or originates from a different pressure distribution than that of beam-2.

%% file: highbeta.tex
\subsection{Background in high-$\beta^*$ fills}
\label{sect:highbeta}

LHC fill 3216, with $\beta^*$=1000\,m, had a very low interaction rate, of the order of 0.005 $pp$ collisions per bunch crossing,
and a sparse bunch pattern with only three nominal intensity bunches per beam, two colliding in BCIDs 101 and 1886 and
one unpaired in BCID 901 and 992 for beam-1 and beam-2, respectively. 
Because the forward detectors, ALFA\,\cite{alfareference} and TOTEM\,\cite{totemreference,totemreference2}, approached the beam with their
roman pots to a distance of only 3.0 nominal $\sigma$, the collimators in IR7 were temporarily moved as close as 2.0 nominal $\sigma$ in 
order to clean the immediate surroundings of the beam core.
Around the same time IR3 collimators were closed to 5.9 $\sigma$ while TCTs in IR1 and IR5 were retracted to 17 $\sigma$ throughout the fill.
These and other collimator settings are summarised in table\,\ref{tab:apertures}.
This tight cleaning resulted in very large beam losses over short durations of time. Against the 
very low luminosity background, these loss-spikes provided a particularly clean environment to study the relationship between
proton losses at the primary collimators and background seen in ATLAS. It has to be emphasised,
however, that the LHC optics was very different from nominal so these studies provide interesting
insights, but do not allow conclusions for normal operating conditions.

Unfortunately, the time-resolution of 300\,s, at which the per-BCID L1\_BCM\_AC\_CA rates were recorded in 2012, is too 
coarse with respect to the duration of the loss peaks during cleaning.
The very low luminosity and sparse bunch pattern, however,
resulted in negligible afterglow, allowing the use of BCM luminosity data from one side of the detector 
only. For paired bunches this method suffers from the rate due to luminosity which, although very low, prevents from
seeing small background spikes. For the unpaired bunch the sensitivity to background is 
about an order of magnitude better.

An even cleaner background measurement was obtained from the LUCID
detector, using the signal of the incoming beam in the upstream detector, as described in section\,\ref{sect:lucid}.
As will be seen, a comparison between the two detectors is very useful, since in some cases it confirms
real background while in others it uncovers mechanisms that result in wrong association of spikes to bunches.

\begin{figure}[t]
\centering 
\includegraphics[width=\textwidth]{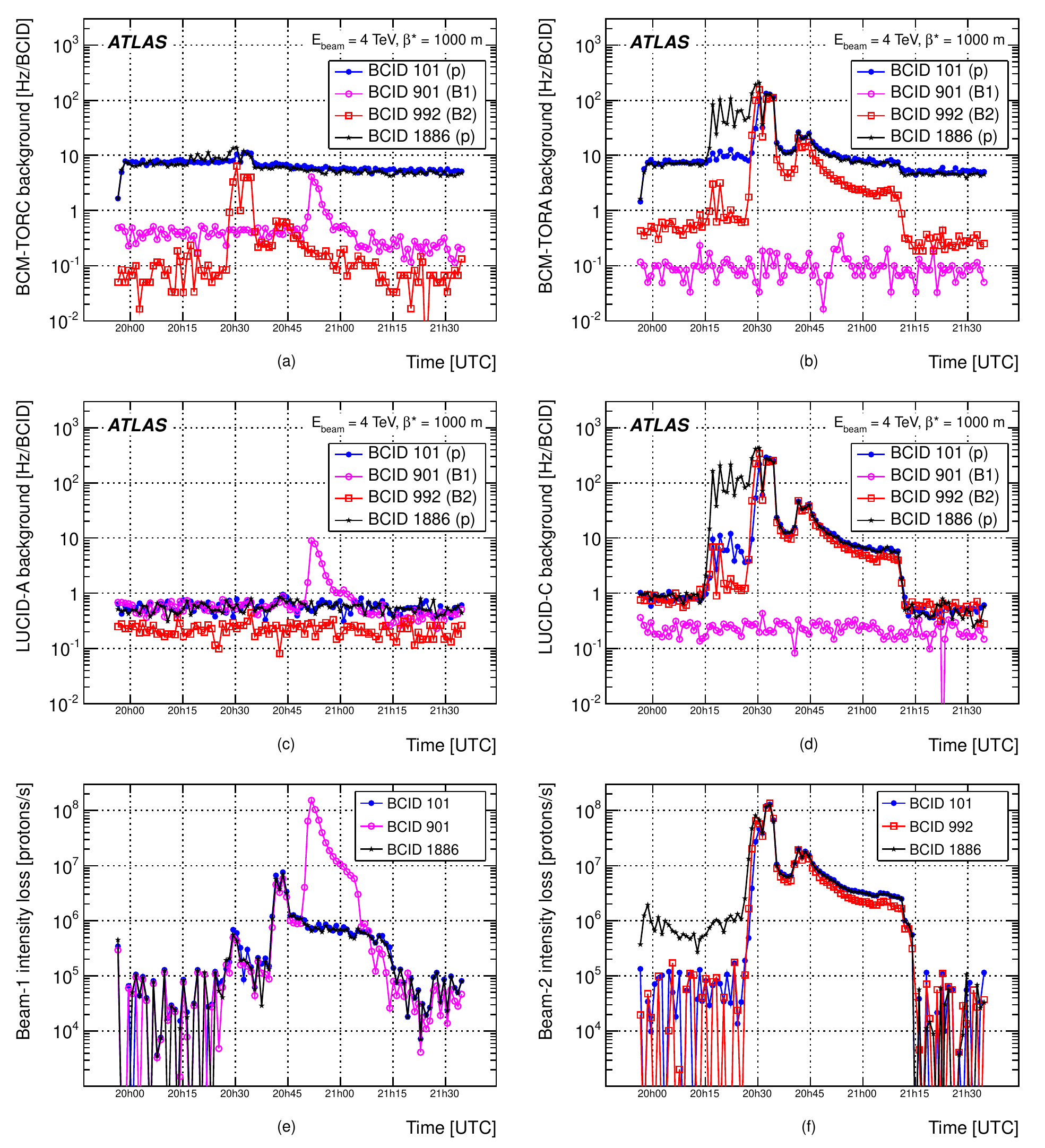}
\caption{Background seen by BCM (a, b) and LUCID (c, d) in the high-$\beta^*$ LHC fill 3216 in 
the paired (101, 1886) and unpaired (901, 992) bunches. Plots (a) and (c) show the hit rates from beam-1,
while plots (b) and (d) show those from beam-2. Plots (e) and (f) give the corresponding bunch intensity 
loss rates for beam-1 and beam-2, respectively.}
\label{fig:alfaPlot-all}
\end{figure}

\begin{figure}[t]
\centering 
\includegraphics[width=\textwidth]{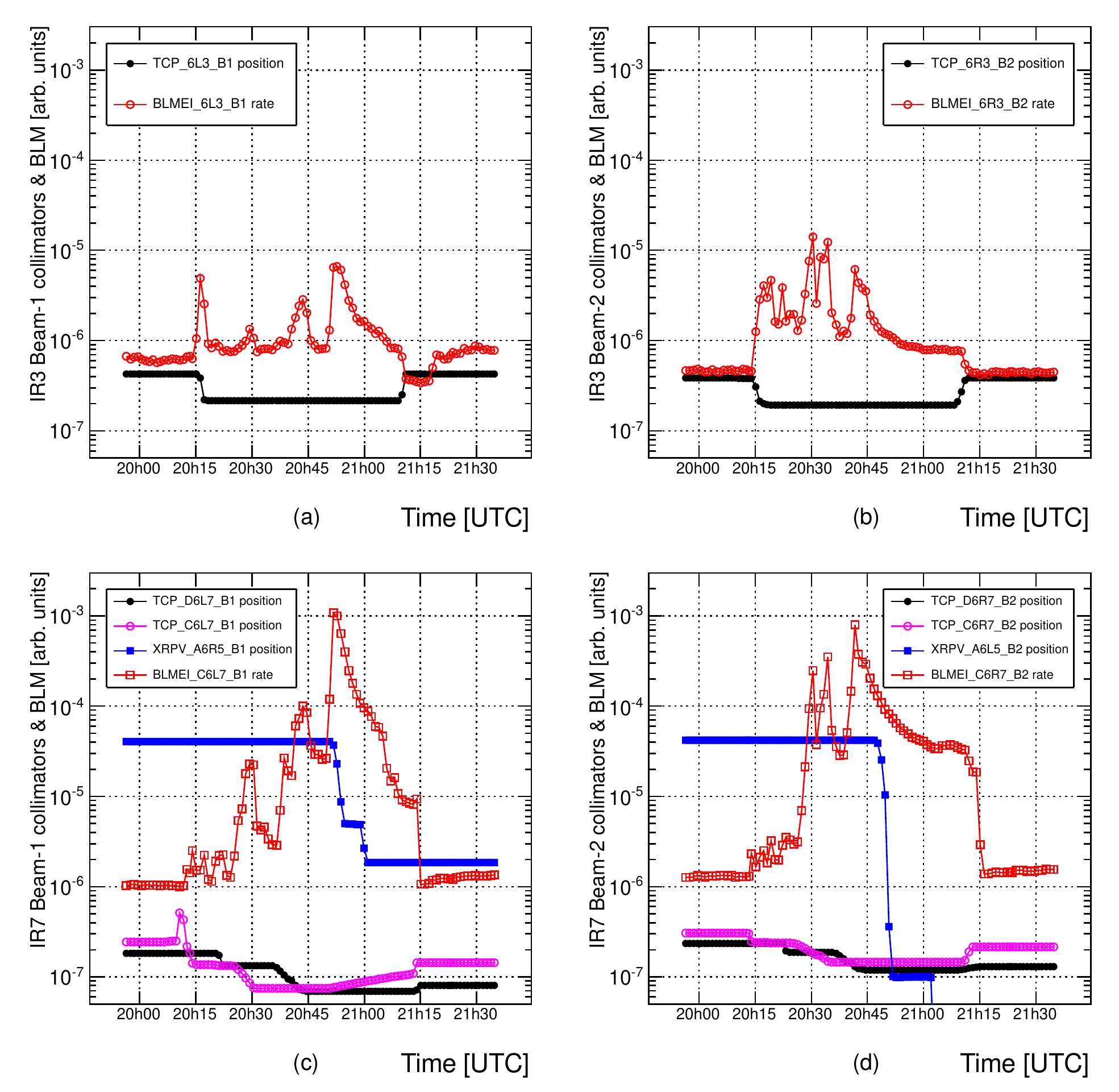}
\caption{Loss rates (BLMEI\_6x3, BLMEI\_C6x7) and collimator positions (TCP\_6x3, TCP\_C6x7 \& TCP\_D6x7) in IR3 (a, b) and in IR7 (c, d). 
Also shown in plots (c) and (d) are the positions of some roman pots (XRPV\_A6x5) in IR5. In the instrument codes the ``x'' stands for 
either ``R'' or ``L'', depending on the side of the experiment, as defined in section\,\protect\ref{sect:lhc}.
For all quantities the vertical unit is arbitrary. Only the relative loss rates are significant
and the collimator positions only indicate the times of change.
}
\label{fig:timberalfa}
\end{figure}

Figures\,\ref{fig:alfaPlot-all}(a--d) shows the observed beam-1 and beam-2 backgrounds as seen by BCM and LUCID while figures\,\ref{fig:alfaPlot-all}(e--f)
indicate the loss in beam intensities.
It can be seen that, especially in beam-2, large backgrounds are observed already at 20:15\,UTC, about 10
minutes before any appreciable beam intensity loss is seen. A detailed analysis of BLM 
readings in the cleaning insertions, summarised in figure\,\ref{fig:timberalfa} together with the relevant collimator positions, reveals 
that the observed intensity losses are caused by IR7 collimators closing between 20:30 and 20:40\,UTC (figure\,\ref{fig:timberalfa}(c--d)). 
Collimators in IR3 (figure\,\ref{fig:timberalfa}(a--b)), however, close 
already at 20:15\,UTC and while they cause no visible intensity loss in figure\,\ref{fig:alfaPlot-all}(e) or (f) their
movement and local losses seen in IR3 BLMs correlate perfectly with beam-2 background increase seen in ATLAS (figure\,\ref{fig:alfaPlot-all}(b) and (d)). 
This leads to the interpretation that ATLAS backgrounds from beam-2 are much more sensitive to losses in IR3 than those in IR7. This is
not too surprising given that, for beam-2, IR3 is only two octants away from ATLAS.

One interesting feature is that none of the BLMs between IR3 and IR1 record any excess loss during the background-spike.
Unlike in normal low-$\beta^*$ optics, in this high-$\beta^*$ configuration the TAS in the experimental area is a limiting 
aperture, reaching closer to the beam envelope than any other element between IR3 and IR1. Simulations confirm that 
the debris produced by protons scattering on IR3 collimators can make it all the way to the TAS in IR1\,\cite{bruce1000m}.

The fact that ATLAS is more sensitive to IR3 for beam-2 due to the relative vicinity might be taken to suggest that,
by symmetry, there should be more sensitivity to IR7 for beam-1. This, however, is not supported
by a comparison of figure\,\ref{fig:alfaPlot-all} with figure\,\ref{fig:timberalfa}.
In the figure\,\ref{fig:alfaPlot-all}(e) an intensity loss exceeding $10^8$\,p/s can be seen
for BCID 901 of beam-1 at 20:50\,UTC. A double spike of equal magnitude is seen in figure\,\ref{fig:alfaPlot-all}(f) at 
20:30 for all bunches of beam-2.
Despite the similar intensity loss, a comparison of figures\,\ref{fig:alfaPlot-all}(c) and \ref{fig:alfaPlot-all}(d) 
reveals that the observed background spike in beam-2 is a factor 30 larger than in beam-1.
Interestingly, figure\,\ref{fig:timberalfa}(b) indicates that the IR3 collimators on beam-2 do not move at 20:30\,UTC when the 
largest beam-2 intensity losses are observed. Figure\,\ref{fig:timberalfa}(d) confirms that the beam-2 losses at 20:30\,UTC are caused by
IR7 collimators closing -- and at the same time large losses are recorded by the BLMs in IR7. The reason that also in this case beam-2 gives in ATLAS larger 
backgrounds than beam-1 is that some protons scattered in IR7 get intercepted in IR3, which according to table\,\ref{tab:apertures}
is the second tightest aperture in the ring. This cross-talk from IR7 to IR3 is confirmed by a comparison of figures\,\ref{fig:timberalfa}(b) and (c)
where a clear correlation of loss rates seen by the BLMs is observed, especially for the double peak at 20:30--20:35\,UTC.

The observation that more background is seen from IR3 than IR7, even if the related intensity loss is much less, is explained by a 
higher efficiency of the secondary collimators and absorbers in IR7 compared to IR3. This has the effect that a smaller fraction of the 
protons scattered by the primary collimators actually escape IR7 than IR3\,\cite{bruce1000m}.

Figure\,\ref{fig:alfaPlot-all} reveals several other interesting features:
\begin{itemize}
\item All bunches seem to behave differently with respect to collimation. This is seen, especially,
in figure\,\ref{fig:alfaPlot-all}(b) (and also \ref{fig:alfaPlot-all}(d)) where backgrounds from individual beam-2 bunches, especially BCID 101 and 992, 
differ by up to two orders of magnitude between 20:15\,UTC and 20:30\,UTC. 
Since these backgrounds originate from IR3, it suggests that the off-momentum tails of the bunches differ significantly from each other.
\item Pronounced background (figure\,\ref{fig:alfaPlot-all}(c)) and intensity loss (figure\,\ref{fig:alfaPlot-all}(e)) peaks are seen 
at 20:50\,UTC for BCID 901 of beam-1, but a comparison with figure\,\ref{fig:timberalfa}(c) reveals that it is not related to collimators 
closing, but rather to a roman pot that was moving closer to the beam in IR5. It is not understood why only the unpaired bunch is affected.
\item Figure\,\ref{fig:alfaPlot-all}(a) shows that the BCM observes a significant double-peak in background at 20:30\,UTC for BCID 992 in 
the beam-1 direction, although BCID 992 actually is an unpaired bunch of beam-2. No trace of such a peak is seen by LUCID.
\end{itemize}

Concerning the last of these observations, a comparison with beam-2 backgrounds, shown in figure\,\ref{fig:alfaPlot-all}(b), reveals 
that the shape and time-alignment of the apparent background spikes in beam-1 match the larger spikes seen in beam-2. 
This observation is consistent with interpreting the beam-1 excess as afterglow from beam-2 halo impinging on the TAS, i.e. an effect 
similar to \abg{}, discussed in section\,\ref{sect:ghostlumi}, except that here the interactions happen on the TAS and not with residual gas.
This confusion of the two beams does not appear in LUCID, where the signal from the upstream detector is used which means that 
the time-window for the incoming beam-1 bunch is more than 120\,ns before the passage of the beam-2 bunch. 

\begin{figure}[t]
\centering 
\includegraphics[width=\textwidth]{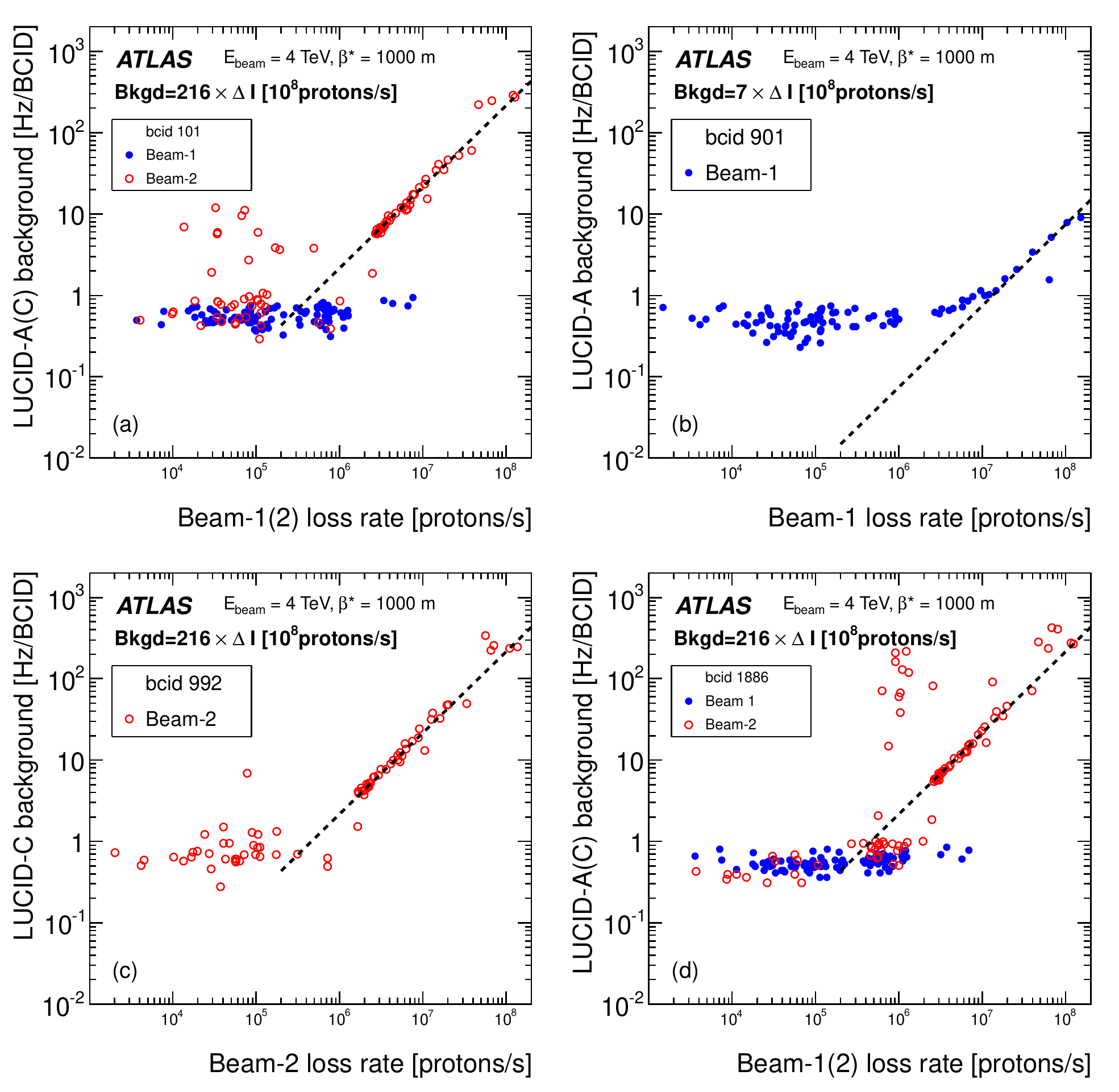}
\caption{Correlation between beam loss rate and the background rates measured by LUCID in the two paired 
BCIDs 101 (a) and 1886 (d) and two unpaired BCIDs 901 for beam-1 (b) and 992 for beam-2 (c). 
The data-points falling on the horizontal at small loss rates represent the normal beam-gas and noise level. The points on the
diagonal are associated to large losses due to cleaning in IR7. The dashed line is not a fit to the data,
but just gives an indication of the slope. Especially in BCID 1886 some large background is seen also for small beam intensity loss. 
This is due to cleaning in IR3.
}
\label{fig:alfaPlot_correlation}
\end{figure}

Figure\,\ref{fig:alfaPlot_correlation} shows the correlation between the bunch intensity loss rate and the BIB rate observed by LUCID.
The dashed lines in the plots are only to guide the eye, but they indicate that for a given beam the correlation is the same for 
each BCID, while it differs for the two beams. In particular, a comparison of figure\,\ref{fig:alfaPlot_correlation}(b) with the
other three plots of figure\,\ref{fig:alfaPlot_correlation} confirms that the ratio between background and beam loss 
is about a factor of 30 smaller for beam-1 than for beam-2. A value, perfectly consistent with that derived above from 
height-differences of individual peaks in figure\,\ref{fig:alfaPlot-all}.

The large group of outlier points with high background $(\sim\! 10^{2}$\,Hz/BCID) but moderate loss rate $(\sim\! 10^{6})$\,protons/s, seen in 
figure\,\ref{fig:alfaPlot_correlation}(d) for BCID 1886 in beam-2 direction, corresponds to the high background seen in 
figure\,\ref{fig:alfaPlot-all}(d) between 20:15 and 20:25\,UTC. At this time IR7 collimators were still open, so it seems evident 
that this background is due to scraping away off-momentum tails when the IR3 collimators close.

The background seen by the BCM was found to exhibit very similar behaviour to that observed in figure\,\ref{fig:alfaPlot_correlation}.

In conclusion, it is observed that at least in the 1000\,m optics ATLAS is much more sensitive to losses in IR3 than in IR7.
In fact the sensitivity is larger to beam-2 losses in IR7, a fraction of which leak to IR3 and from there to ATLAS, than to direct 
debris from beam-1 losses in IR7. This suggests that protons leaking out from IR7 are not intercepted to a significant degree by ATLAS (beam-1) 
or CMS (beam-2), but predominantly by IR3. 

Finally, it should be remarked that despite the very different optics, the results in section\,\ref{sect:ghostBG} suggest that this larger sensitivity to IR3 losses
could be present also in normal high-luminosity optics;  as discussed in section\,\ref{sect:ghostBG} ghost-BIB is observed only in beam-2 and
attributed to losses of ghost charge in IR3.
Even during abort-gap cleaning, which should increase losses in IR7, the ATLAS backgrounds in beam-2 increased more.
These indications together with the results obtained in this high-$\beta^*$ fill suggest that similar comparisons between cleaning losses and ATLAS 
backgrounds could be done in LHC Run-2 by recording background data during dedicated loss-map studies of the LHC. 

%% file: 25ns.tex
\subsection{Backgrounds during 25\,ns operation}

At the end of the 2012 $pp$ running the LHC operated for a few fills with a 25\,ns bunch separation.
The bunch pattern included 12 isolated bunches per beam, which were isolated 
by 88 BCIDs, i.e. much more than during 50\,ns operation.

\begin{figure}[h!]
\centering 
\includegraphics[width=0.7\textwidth]{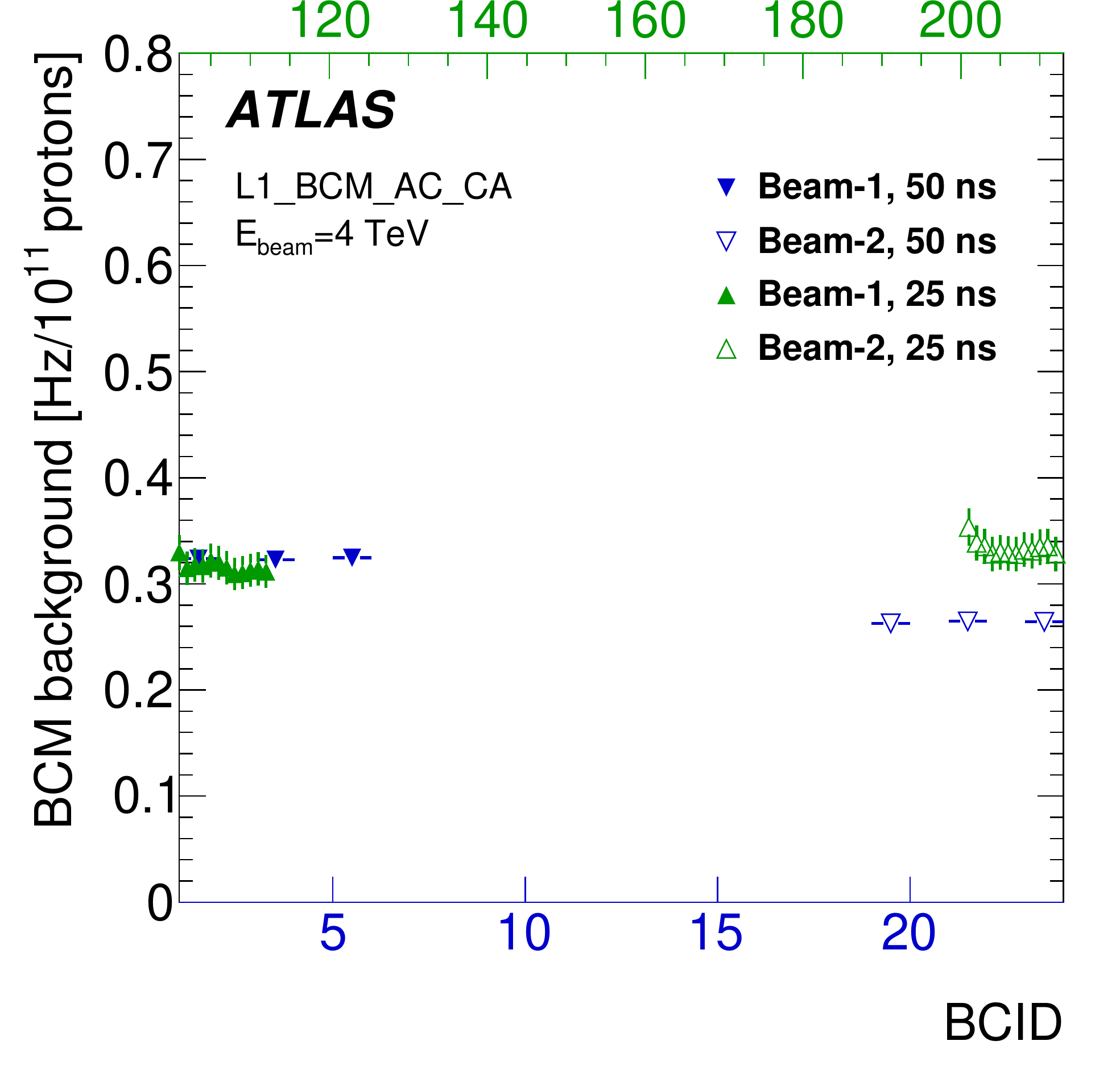}
\caption{Beam-gas rate per BCID in the events triggered by L1\_AC\_CA\_UNPAIRED\_ISO in the data with 50\,ns (downward triangles) and 25\,ns (upward triangles) bunch spacing.
The BCID values corresponding to the unpaired bunches in the 50\,ns (25\,ns) data are indicated by the bottom (top) $x$-axis.
For the 50\,ns data only LHC fills with 1368 colliding bunches taken before the September technical stop are considered.
Only the first well isolated unpaired bunches are shown for the bunch pattern used in the 25\,ns fill.
}
\label{fig:25beamgas}
\end{figure}

Figure\,\ref{fig:25beamgas} compares the L1\_BCM\_AC\_CA rates in the unpaired isolated bunches for the 25\,ns and 50\,ns operation. 
The observed beam-gas rates for beam-1 are at a similar level in both cases.
Unlike in the 50\,ns operation, where the beam-1 rate was about 25\% higher than that of beam-2, the 25\,ns operation exhibits an opposite 
trend with the background rate from beam-2 being approximately 5\% higher than that from beam-1.

%% file: conclusions.tex
\section{Conclusions}

A detailed analysis of beam-induced (BIB) and cosmic-ray backgrounds (CRB) in the ATLAS detector 
during the 2012 operation has been presented. The principal methods to monitor backgrounds in ATLAS
rely on the existence of unpaired bunches in the fill pattern. The BIB rates associated with these
are observed by the Beam Conditions Monitor (BCM) and fake jets with $\pt>10$\,\GeV{} in the calorimeters. 

The observation, made already in 2011 background data that beam backgrounds seen by the BCM are
strongly correlated with the residual pressure 22\,m from the interaction point, is re-established
and further elaborated. In particular, it is shown that a scaling of the BCM background with 
a linear combination of the pressures around the inner triplet results in a very constant rate
throughout 2012. Any deviations from this constant level are sensitive indications of abnormal beam-losses.

The background data collected in 2012 benefited from an optimised bunch pattern with 
dedicated unpaired bunches, well separated from the colliding ones.
Largely due to these cleaner conditions, the analysis of 2012 background data revealed features, 
related to ghost charge, which were not seen in the 2011 analysis. The most 
noteworthy observation is the high sensitivity of ATLAS to off-momentum ghost charge being lost in IR3.
The debris from these cleaning losses is seen both by the BCM and as fake jets in the calorimeters.
An analysis of these rates revealed a significant jump of BIB from beam-2 ghost charge in early August, coincident
with the LHC changing chromaticities. BIB, due to ghost charge losses, could 
be observed in the $3\,\mu$s long abort gap and for beam-2 was found to exhibit an unexpected hump-structure following the
chromaticity changes, but not before.
During periods of dedicated abort gap cleaning
by the LHC these losses increased. Immediately after the cleaning a reduction of background was observed,
but only for the order of  one minute. This suggests that the abort gap is filled by de-bunched beam
with a time constant of the order of one minute.

BIB from ghost charge can also be observed in form of fake jets. This analysis benefits from the
excellent time resolution of the ATLAS calorimeters, which allows use of the jet time to determine the 
direction of the BIB muon producing the fake jet.

Further analyses of the fake jet data show that after an initial decrease in early 2012 the rates
remained  constant throughout most of the year.
Fake-jet data were also collected during dedicated CRB data-taking and the energy and
multiplicity distributions of the reconstructed jets compare well with Monte Carlo results.
After proper normalisation, the rates of fake jet rates due to BIB and CRB could be compared, showing
that beyond $\pt\sim1\,\TeV$ the rates are similar, while at $\pt\sim 200$\,\GeV, the CRB-muons produce only 10\% of the total fake jet rate.

Two LHC fills of particular interest were subject of more detailed investigation. One fill, with an abnormally high
background in beam-1, revealed that an earlier quench of the inner triplet had caused 
adsorption of residual gas on the beam-screen. At the start of the next fill the collision debris hitting the
beam screen caused desorption of this gas, which resulted in the observed background excess. The evolution of the
background excess as a function of time provided information about the cleaning of the beam screen by the
collision debris.

Another fill of interest is the one with high $\beta^*$ for total cross-section measurements by the ALFA and TOTEM detectors,  
characterised by a sparse bunch pattern and very low luminosity. Since ALFA and TOTEM 
approached with their detectors very close to the beam, the halo was periodically cleaned down to
two nominal $\sigma$. 
During these cleaning periods significant fractions of the beam intensity were lost. At the same time
background spikes were observed in ATLAS, establishing a correlation between
cleaning losses and BIB in ATLAS, although only for this very special optics configuration.

The experience gained and observations made during this analysis allows for further improvements and better targeting
of background monitoring methods and analysis tools for LHC Run-2. In particular monitoring of BIB from 
ghost charge was shown to provide useful insights. The results presented in this paper indicate that with the new analysis methods ATLAS 
can resolve low enough levels of BIB to facilitate participation in dedicated loss-map measurements of the accelerator in Run-2. 
Such measurements promise to grant further insights into sources of background and should allow better predictions of background 
levels for different collimator configurations.

On the physics side, an improved understanding of the properties of BIB and CRB provides the basis for further 
optimisation of background rejection techniques. With the 25ns bunch spacing and higher collision energies in LHC Run-2, such improved 
background rejection will be crucial in order to enhance sensitivity to beyond the Standard Model phenomena in various channels, such as 
unconventional signatures of non-prompt decays, as well as measurements involving high $\pt$ jets.

%% file: app_TS3fitsBCM.tex
\section{BCM trigger configurations in 2012}
\label{app:TS3fitsBCM}

Until TS3, in mid-September, the BCM trigger was provided by the same read-out driver (ROD) as in 2011,
called BCM PRO with the L1\_BCM\_Wide trigger window in bins 28--48. During TS3 the trigger was moved 
to use a new BCM DEV ROD with the L1\_BCM\_Wide window re-aligned to bins 36--50, i.e. the 
same as the in-time window of L1\_BCM\_AC\_CA. Table\,\ref{tab:triggerwindows} summarises the
windows of the BCM level-1 and luminosity triggers.

\begin{table}
\centering
\begin{tabular}{l|c|c|c|l}
Trigger         & Type                   & Early window & In-time Window & comment \\ \hline
L1\_BCM\_AC\_CA & Early \& in-time       & $-6.25\,\pm\,2.73$ & $+6.25\,\pm\,2.73$ & All 2012 \\
L1\_BCM\_Wide   & A\&C coincidence       & ---                & $+4.29\,\pm\,3.90$ & Before TS3 \\
L1\_BCM\_Wide   & A\&C coincidence       & ---                & $+6.25\,\pm\,2.73$ & After TS3 \\
BCM\_TORx       & Single side A or C     & ---                & $+6.25\,\pm\,6.25$ & All 2012  \\ \hline
\end{tabular}
\caption{Summary of triggers used and their time windows (ns) with respect to the nominal collision time. 
The BCM\_TORx is not a L1 trigger item, but the count rate provided for luminosity determination, which is 
used for some analyses of this paper.}
\label{tab:triggerwindows}
\end{table}

An additional feature was that the collision-time in BCM DEV was shifted by roughly 6 bins
with respect to BCM PRO. While triggers after TS3 were given by BCM DEV, the data used
for the analysis of this paper was extracted from BCM PRO throughout 2012. 
The consequence is that in the data after TS3 the trigger windows appear shifted to bins 42--56 and 10--24. 

\begin{figure}[h!]
\centering 
\mbox{
\subfigure[]{
\includegraphics[width=0.49\textwidth]{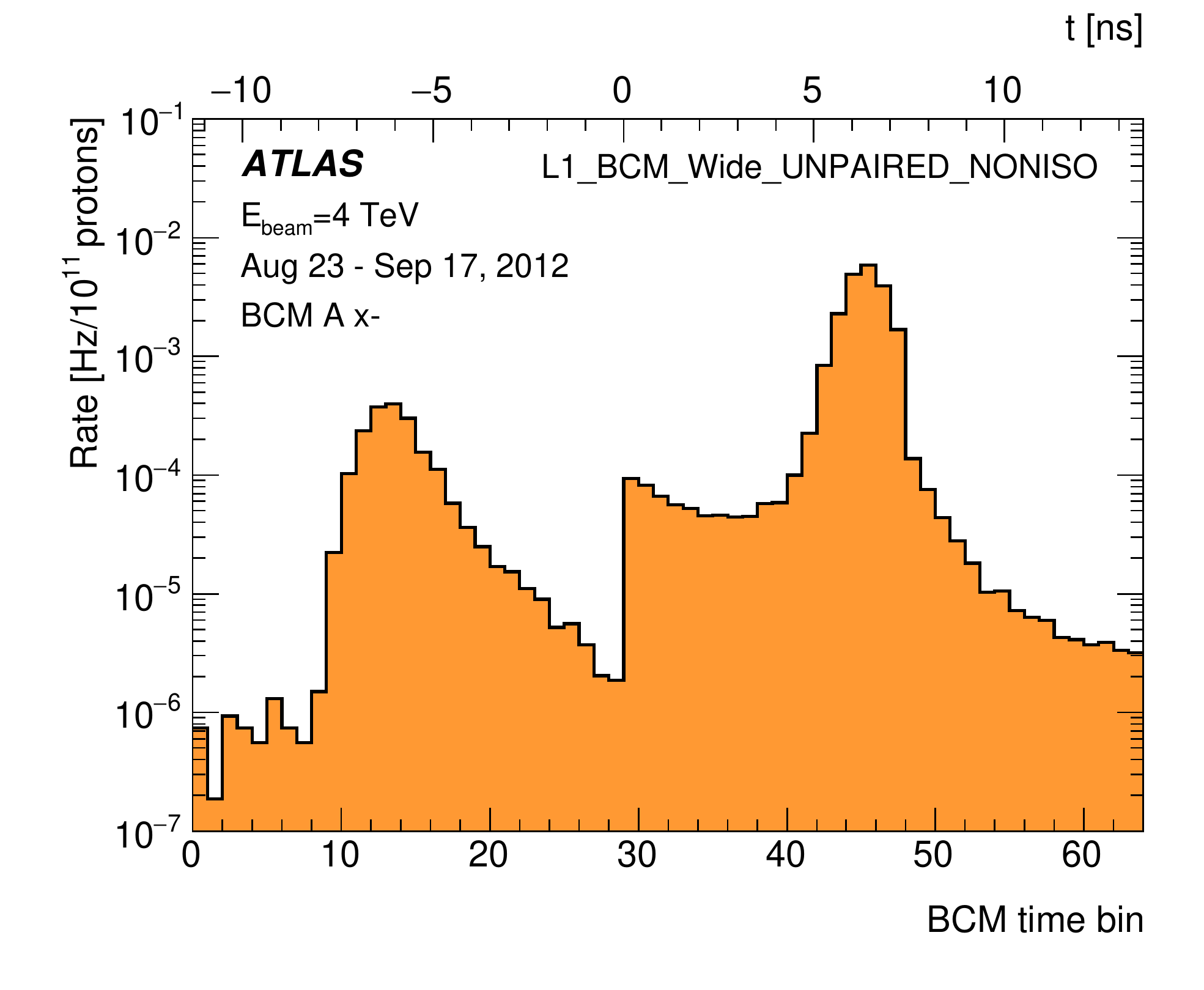}
\label{fig:ghost_trigchangea}
}
\subfigure[]{
\includegraphics[width=0.49\textwidth]{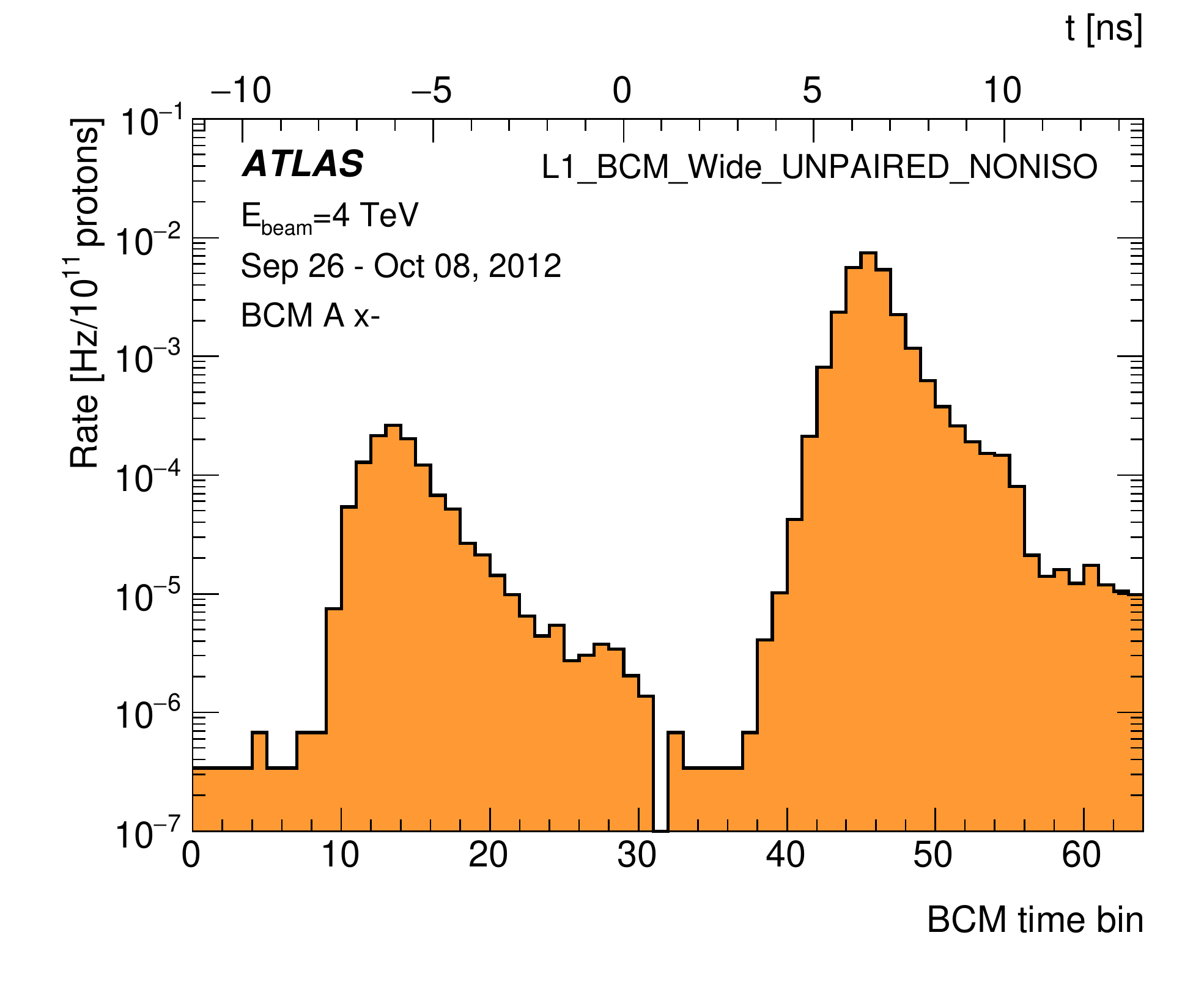}
\label{fig:ghost_trigchangeb}
}
}
\caption{Response of the BCM $x-$ station on side A in the events triggered by L1\_BCM\_Wide\_UNPAIRED\_NONISO
before (a) and after (b) the September technical stop.}
\label{fig:ghost_trigchange}
\end{figure}

The time distributions of the events before and after TS3 are illustrated in figure\,\ref{fig:ghost_trigchange}.
Prior to TS3 a sharp lower edge in bin 28 is seen, corresponding to the start of the trigger window. The upper edge
is after bin 48 and appears to cut the tail of the signal.
After TS3 the lower edge is well aligned with the rising signal and thus not visible in the plot. The upper edge is
seen after bin 56, which is far in the signal tail. This suggests that the signal is better contained in the
trigger window after TS3. The impact on the efficiency cannot be exactly quantified, but is at most 10\% (based
on the size of the tail apparently cut off in the left plot). Since the effect cannot be quantified accurately, no correction 
has been explicitly applied, but an implicit accounting for it is described below.

A major difference between the two RODs is that the trigger in the old BCM PRO was treating
the horizontal and vertical BCM modules independently - like two independent detectors, while 
the trigger in BCM DEV combined all modules together, which has a significant effect on
the trigger acceptance.

Assuming that back-to-back correlations between the secondaries produced in minimum-bias $pp$ collision events 
are negligible, the BCM-Wide rate after the trigger change should increase by exactly a factor of
two: if the (small) probability to produce a hit on one side in horizontal modules is denoted 
by $p$, then the probability for a coincidence is approximately $p^2$. Since the vertical modules
have independently the same probability, the total is $2p^2$. After combining the modules in the BCM DEV, the
probability for a hit on one side is $2p$ and the probability for coincidence is $4p^2$.

\begin{figure}[t]
\centering 
\mbox{
\subfigure[]{
\includegraphics[width=0.49\textwidth]{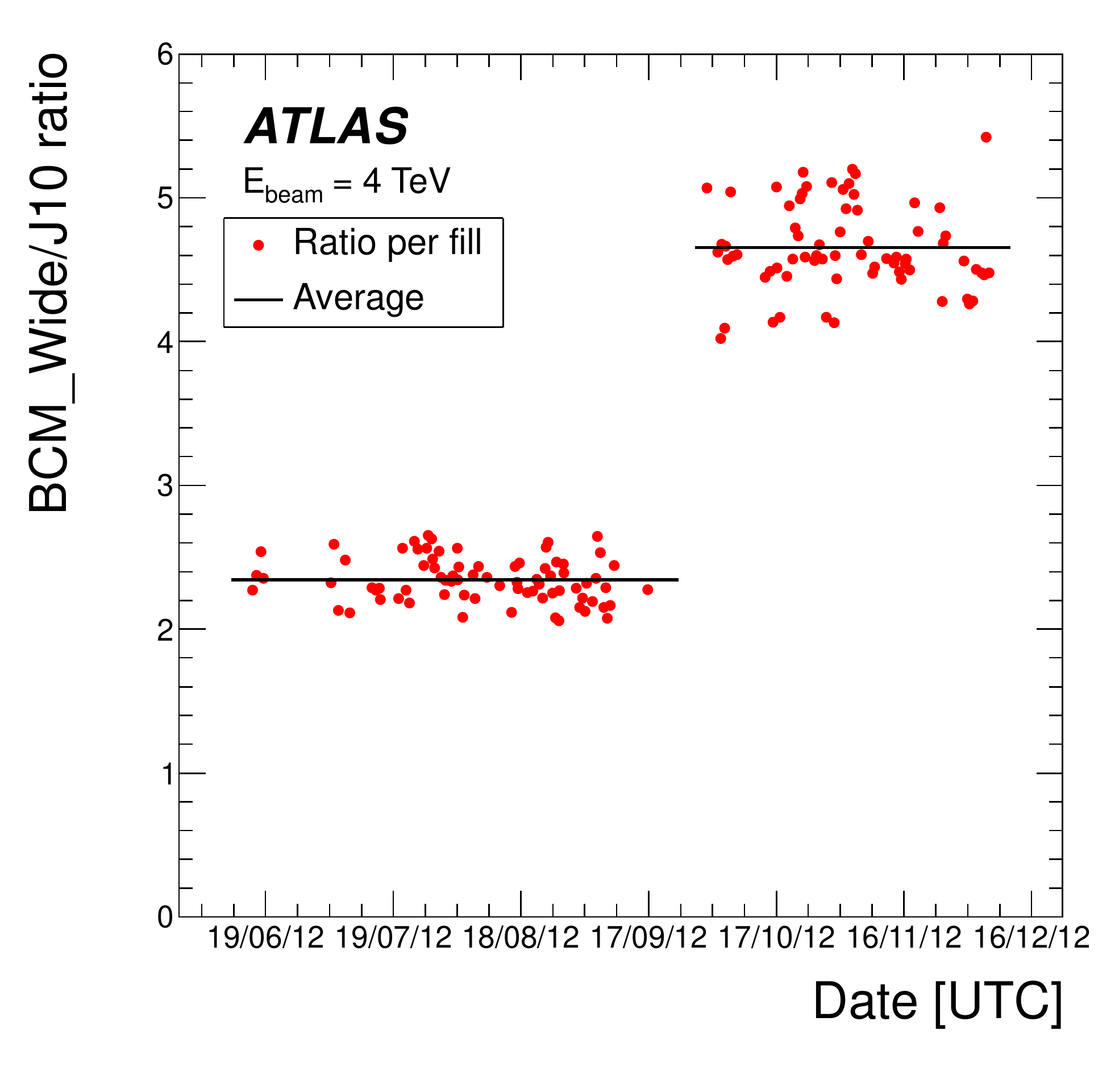}
\label{fig:bcmWideTS3jump}
}
\subfigure[]{
\includegraphics[width=0.49\textwidth]{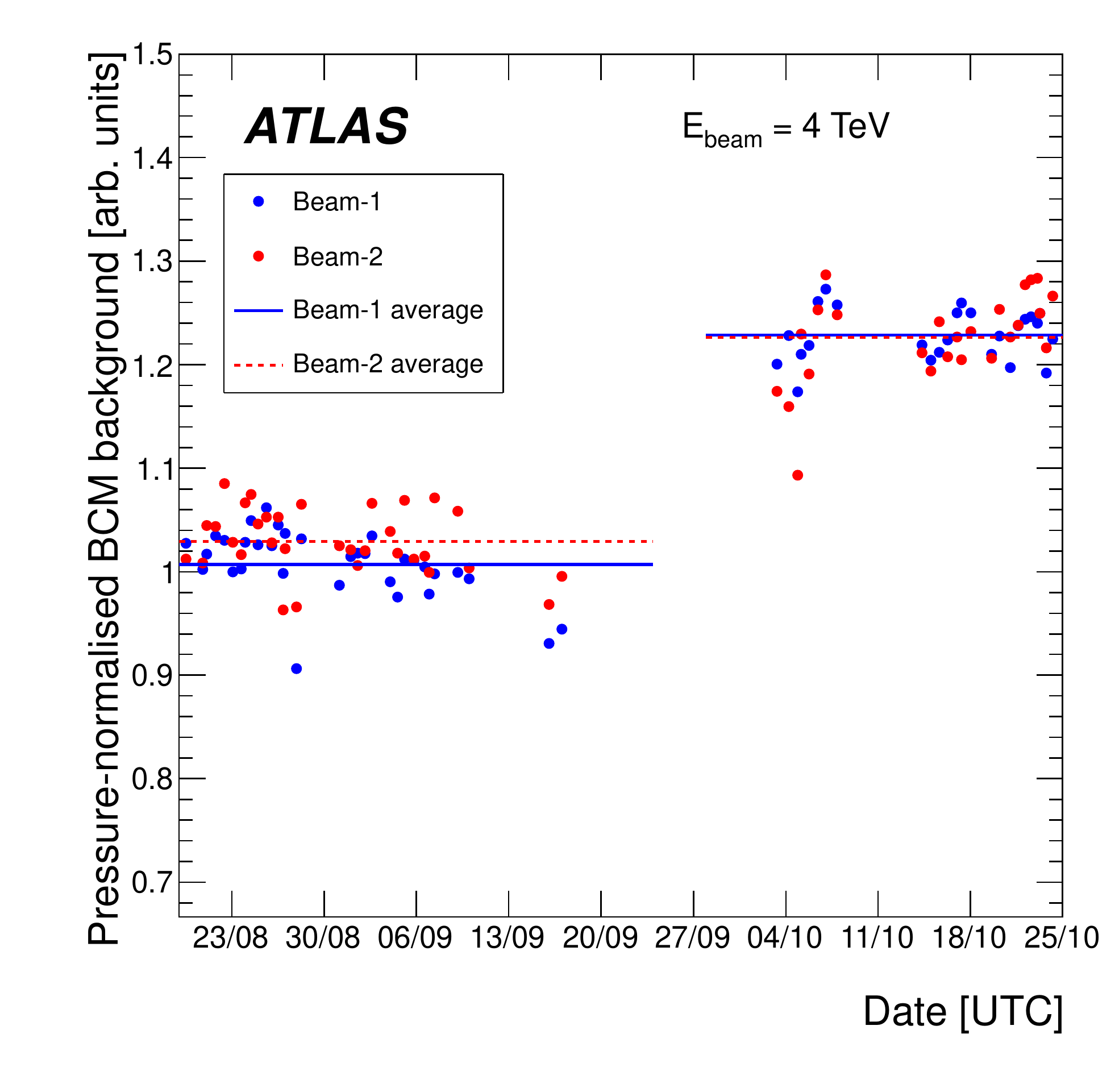}
\label{fig:bcmACCATS3jump}
}
}
\caption{Ratio between BCM\_Wide and J10 trigger rates before and after TS3 (a) and
trend of BCM\_AC\_CA rates after scaling with the 22\,m pressure (b).}
\end{figure}

In order to verify that back-to-back correlations are negligible, and also to assess the
possible efficiency impact of the trigger window re-alignment, the ratio between L1\_BCM\_Wide and L1\_J10 rates 
before and after TS3 were compared. The result is shown in figure\,\ref{fig:bcmWideTS3jump}.
Since no other changes to the system were made and beam conditions were comparable, the difference in this 
ratio allows to quantify the shift. Averaging the ratios before and after TS3, respectively, they 
are found to differ by a factor of $1.985\pm 0.021$. This confirms that a factor of two has to be applied to 
L1\_BCM\_Wide rates before TS3, in order to compensate for the effect of the trigger change.

It is more complicated to estimate the effect of the change on the BIB trigger L1\_BCM\_AC\_CA,
since it is designed to trigger on a correlation between early an in-time hits and there is
no information how often such hits are produced by the same particle traversing upstream and downstream 
modules in the same azimuth. Any cross-talk between horizontal and vertical modules, e.g. by two different 
secondaries from the same event, would lead to an increase of the trigger rate in BCM DEV. The upper limit
of the increase is a factor of 2, corresponding to fully uncorrelated hits in the modules.
Unlike for the collision-dominated L1\_BCM\_Wide, there is no other rate to which the BCM background
trigger could be compared.
The raw L1\_BCM\_AC\_CA data, even after bunch intensity normalisation, are found to have a scatter much larger
than the effect of the trigger change.
The method employed to reduce this scatter by taking into account the vacuum conditions is described in 
section\,\ref{sect:vacuum}. The normalization with the residual pressure produces a reasonably
flat distribution before and after TS3, and allows deriving the change in L1\_BCM\_AC\_CA rate to be a 
factor of $1.22\pm 0.01$ for beam-1 and $1.19\pm 0.01$ for beam-2. This is illustrated in figure\,\ref{fig:bcmACCATS3jump}.
For simplicity a unique factor of 1.2 was applied to L1\_BCM\_AC\_CA rates prior to TS3 for both beams.

In section\,\ref{sect:ghostlumi} rates extracted from the luminosity data are compared to L1\_BCM\_Wide trigger
rates. When doing this, the different window widths shown in table\,\ref{tab:triggerwindows} have to be taken into account.
Figure\,\ref{fig:ghost_trigchange} indicates that the signal from collisions or in-time BIB (figure\,\ref{fig:bcmresponse}) is 
almost fully contained in the L1\_BCM\_Wide window -- and of course in the wider BCM-TORx.
The count rates for hits distributed randomly in time, e.g. \agpp{} and noise, depend on the width of the windows and
if these rates are to be compared an appropriate correction has to be applied. 
The BCM-TORx rates correspond to an acceptance window of 12.5\,ns width, while the BCM\_Wide window is only 5.1\,ns wide, which implies
that a correction factor of 0.41 has to be applied to pedestal hits distributed uniformly in time.  Since no correction is needed 
for in-time BIB, the correction factor for random BIB+pedestal coincidences amounts to this 0.41.
Pedestal+pedestal coincidences have to be corrected by $0.41^2 = 0.17$ when comparing rates derived from BCM-TORx to L1\_BCM\_Wide
rates.

%% file: Acknowledgements.tex

We thank CERN for the very successful operation of the LHC, as well as the
support staff from our institutions without whom ATLAS could not be
operated efficiently.

We acknowledge the support of ANPCyT, Argentina; YerPhI, Armenia; ARC, Australia; BMWFW and FWF, Austria; ANAS, Azerbaijan; SSTC, Belarus; CNPq and FAPESP, Brazil; NSERC, NRC and CFI, Canada; CERN; CONICYT, Chile; CAS, MOST and NSFC, China; COLCIENCIAS, Colombia; MSMT CR, MPO CR and VSC CR, Czech Republic; DNRF and DNSRC, Denmark; IN2P3-CNRS, CEA-DSM/IRFU, France; GNSF, Georgia; BMBF, HGF, and MPG, Germany; GSRT, Greece; RGC, Hong Kong SAR, China; ISF, I-CORE and Benoziyo Center, Israel; INFN, Italy; MEXT and JSPS, Japan; CNRST, Morocco; FOM and NWO, Netherlands; RCN, Norway; MNiSW and NCN, Poland; FCT, Portugal; MNE/IFA, Romania; MES of Russia and NRC KI, Russian Federation; JINR; MESTD, Serbia; MSSR, Slovakia; ARRS and MIZ\v{S}, Slovenia; DST/NRF, South Africa; MINECO, Spain; SRC and Wallenberg Foundation, Sweden; SERI, SNSF and Cantons of Bern and Geneva, Switzerland; MOST, Taiwan; TAEK, Turkey; STFC, United Kingdom; DOE and NSF, United States of America. In addition, individual groups and members have received support from BCKDF, the Canada Council, CANARIE, CRC, Compute Canada, FQRNT, and the Ontario Innovation Trust, Canada; EPLANET, ERC, FP7, Horizon 2020 and Marie Sk{\l}odowska-Curie Actions, European Union; Investissements d'Avenir Labex and Idex, ANR, R{\'e}gion Auvergne and Fondation Partager le Savoir, France; DFG and AvH Foundation, Germany; Herakleitos, Thales and Aristeia programmes co-financed by EU-ESF and the Greek NSRF; BSF, GIF and Minerva, Israel; BRF, Norway; Generalitat de Catalunya, Generalitat Valenciana, Spain; the Royal Society and Leverhulme Trust, United Kingdom.

The crucial computing support from all WLCG partners is acknowledged
gratefully, in particular from CERN and the ATLAS Tier-1 facilities at
TRIUMF (Canada), NDGF (Denmark, Norway, Sweden), CC-IN2P3 (France),
KIT/GridKA (Germany), INFN-CNAF (Italy), NL-T1 (Netherlands), PIC (Spain),
ASGC (Taiwan), RAL (UK) and BNL (USA) and in the Tier-2 facilities
worldwide.

%% file: atlas_authlist.tex
\begin{flushleft}
{\Large The ATLAS Collaboration}

\bigskip

G.~Aad$^\textrm{\scriptsize 87}$,
B.~Abbott$^\textrm{\scriptsize 114}$,
J.~Abdallah$^\textrm{\scriptsize 65}$,
O.~Abdinov$^\textrm{\scriptsize 12}$,
B.~Abeloos$^\textrm{\scriptsize 118}$,
R.~Aben$^\textrm{\scriptsize 108}$,
M.~Abolins$^\textrm{\scriptsize 92}$,
O.S.~AbouZeid$^\textrm{\scriptsize 138}$,
N.L.~Abraham$^\textrm{\scriptsize 150}$,
H.~Abramowicz$^\textrm{\scriptsize 154}$
% ...
\bigskip
\\
% ...
$^{12}$ Institute of Physics, Azerbaijan Academy of Sciences, Baku, Azerbaijan\\
% ...
$^{65}$ University of Iowa, Iowa City IA, United States of America\\
% ...
$^{87}$ CPPM, Aix-Marseille Universit{\'e} and CNRS/IN2P3, Marseille, France\\
% ...
$^{92}$ Department of Physics and Astronomy, Michigan State University, East Lansing MI, United States of America\\
% ...
$^{108}$ Nikhef National Institute for Subatomic Physics and University of Amsterdam, Amsterdam, Netherlands\\
% ...
$^{114}$ Homer L. Dodge Department of Physics and Astronomy, University of Oklahoma, Norman OK, United States of America\\
% ...
$^{118}$ LAL, Univ. Paris-Sud, CNRS/IN2P3, Universit{\'e} Paris-Saclay, Orsay, France\\
% ...
$^{138}$ Santa Cruz Institute for Particle Physics, University of California Santa Cruz, Santa Cruz CA, United States of America\\
% ...
$^{150}$ Department of Physics and Astronomy, University of Sussex, Brighton, United Kingdom\\
% ...
$^{154}$ Raymond and Beverly Sackler School of Physics and Astronomy, Tel Aviv University, Tel Aviv, Israel\\
% ...
\end{flushleft}
